\documentclass{aa}
\usepackage{graphicx}
\usepackage[varg]{txfonts}
\usepackage{natbib,twoopt}
\usepackage[breaklinks=true]{hyperref} 
\bibpunct{(}{)}{;}{a}{}{,} 
\makeatletter
\newcommandtwoopt{\citeads}[3][][]{\href{http://adsabs.harvard.edu/abs/#3}%
{\def\hyper@linkstart##1##2{}%
\let\hyper@linkend\@empty\citealp[#1][#2]{#3}}}
\newcommandtwoopt{\citepads}[3][][]{\href{http://adsabs.harvard.edu/abs/#3}%
{\def\hyper@linkstart##1##2{}%
\let\hyper@linkend\@empty\citep[#1][#2]{#3}}}
\newcommandtwoopt{\citetads}[3][][]{\href{http://adsabs.harvard.edu/abs/#3}%
{\def\hyper@linkstart##1##2{}
\let\hyper@linkend\@empty\citet[#1][#2]{#3}}}
\newcommandtwoopt{\citeyearads}[3][][]%
{\href{http://adsabs.harvard.edu/abs/#3}
{\def\hyper@linkstart##1##2{}%
\let\hyper@linkend\@empty\citeyear[#1][#2]{#3}}}
\makeatother

\usepackage{lscape}
\usepackage{enumerate}
\usepackage{mathabx}
\newcommand{\vsin}{\hbox{$v \sin i$}} 
\newcommand{\vsini}{\hbox{$v \sin i$}} 
\newcommand{\cargto}{\hbox{CARMENES $N_{\rm GTO}$25}}
\newcommand{\FoIthree}{\hbox{RV$_{\rm corr}$3}}
\newcommand{\kms}{\hbox{km\,s$^{-1}$}}
\usepackage{graphicx}
\usepackage{xcolor}
\usepackage{soul}
\usepackage{siunitx}
\def\imagetop#1{\vtop{\null\hbox{#1}}}

\makeatletter
\renewcommand*\aa@pageof{, page \thepage{} of \pageref*{LastPage}}
\makeatother

\newcolumntype{M}{>{\hfil$\displaystyle}X<{$\hfil}} 
\newcolumntype{L}{>{\collectcell\AddLabel}r<{\endcollectcell}}

\usepackage{booktabs}
\usepackage{tabularx}
\usepackage{amsmath,array}
\usepackage{siunitx}

\usepackage{{makecell}}
\setcellgapes{3pt}\makegapedcells

\usepackage{array,collcell}

\begin{document} 
   \title{The CARMENES search for exoplanets around M dwarfs}
   \subtitle{Understanding the wavelength dependence of radial velocity measurements}

\titlerunning{Wavelength dependence of radial velocity measurements}

\author{S.\,V.~Jeffers\inst{1}, J.\,R.~Barnes\inst{2},  P.~Sch\"ofer\inst{3},   S.~Reffert\inst{4}, V.\,J.\,S.~B\'ejar\inst{5,6}, A.~Quirrenbach\inst{4}, A.~Reiners\inst{7}, Y.~Shan\inst{7,8}, M. R. Zapatero Osorio\inst{9}, B.~Fuhrmeister\inst{10}, P.\,J.\,Amado\inst{3}, J.\,A.~Caballero\inst{9}, I.~Ribas\inst{11,12}, C.~Cardona~Guill\'en\inst{5,6}, 
F.~Del~Sordo\inst{11,12},
M.~Fern\'andez \inst{3},
A.~Garc\'ia-L\'opez\inst{9}
A.~Guijarro\inst{13},
A.\,P.~Hatzes \inst{1}, 
M.~Lafarga \inst{14,15},
N.~Lodieu\inst{5,6},  
M.~K\"urster \inst{16}, 
K.~Molaverdikhani\inst{17},
D.~Montes\inst{18}, 
J.\,C.~Morales \inst{11,12}}

\authorrunning{Jeffers et al.}

\institute{$^{1}$ Thüringer Landessternwarte Tautenburg, Sternwarte 5, D-07778 Tautenburg, Germany \\
    \email{sandrajeffers.astro@gmail.com}\\
	 $^{2}$ School of Physical Sciences, The Open University, Walton Hall, MK7 6AA, Milton Keynes, United Kingdom\\
	 $^{3}$ Instituto de Astrof\'isica de Andaluc\'ia (CSIC), Glorieta de la Astronomía s/n, E-18008 Granada, Spain\\
	 $^{4}$ Landessternwarte, Zentrum für Astronomie der Universit\"at Heidelberg, K\"onigstuhl 12, D-69117 Heidelberg, Germany \\
	 $^{5}$  Instituto de Astrof\'isica de Canarias, c/ V\'ia L\'actea s/n, E-38205 La Laguna, Tenerife, Spain\\
     $^{6}$ Departamento de Astrof\'isica, Universidad de La Laguna, E-38206 Tenerife, Spain\\
	 $^{7}$ Institut f\"ur Astrophysik and Geophysik, Georg-August-Universit\"at, Friedrich-Hund-Platz 1, D-37077 G\"ottingen, Germany  \\
  $^{8}$  Centre for Planetary Habitability, Department of Geosciences, University of Oslo, Sem Saelands vei 2b, 0315 Oslo, Norway \\  
  $^{9}$ Centro de Astrobiolog\'ia (CSIC-INTA), ESAC, Camino Bajo del Castillo s/n, E-28692 Villanueva de la Ca\~nada, Madrid, Spain \\
     $^{10}$ Hamburger Sternwarte, Universit\"at Hamburg, Gojenbergsweg 112, D-21029 Hamburg, Germany  \\
 	 $^{11}$ Institut de Ci\`encies de l’Espai (CSIC), Campus UAB, c/ de Can Magrans s/n, E-08193 Bellaterra, Barcelona, Spain \\
     $^{12}$ Institut d’Estudis Espacials de Catalunya, E-08034 Barcelona, Spain\\
	 $^{13}$ Centro Astron\'omico Hispano en Andaluc\'ia (CAHA), Observatorio de Calar Alto, Sierra de los Filabres, E-04550 G\'ergal, Almer\'ia, Spain  \\
     $^{14}$ Department of Physics, University of Warwick, Gibbet Hill Road, Coventry CV4 7AL, UK \\
     $^{15}$ Centre for Exoplanets and Habitability, University of Warwick, Coventry CV4 7AL, UK \\
     $^{16}$ Max-Planck-Institut f\"ur Astronomie, K\"onigstuhl 17, D-69117 Heidelberg, Germany\\
     $^{17}$ Fakultät für Physik, Universitäts-Sternwarte, Ludwig-Maximilians-Universität München, Scheinerstr. 1, D-81679 München, Germany \\
	 $^{18}$ Facultad de Ciencias F\'isicas, Departamento de F\'isica de la Tierra y Astrof\'isica \& IPARCOS-UCM (Instituto de F\'isica de Part\'iculas y del Cosmos de la UCM), Universidad Complutense de Madrid, E-28040 Madrid, Spain\\
 }


\newcommand{\TN}[1]{\textcolor{red}{#1}}

   \date{Received 20 July 2023 / Accepted 20 October 2024}
 
  \abstract
   {Current exoplanet surveys are focused on detecting small exoplanets orbiting in the liquid-water habitable zones of their host stars.  Despite the recent significant advancements in instrumental developments, the current limitation in detecting these exoplanets is the intrinsic variability of the host star itself. 
     }
   {Our aim is to use the full CARMENES guaranteed time observations (GTO) data set spanning more than 8 years of observations of over 350 stars to investigate the wavelength dependence of high-precision radial velocities (RV), as stellar activity features should exhibit a wavelength dependence while the RV variation due to an orbiting planet will be wavelength independent.}
   {We use the chromatic index (CRX) to quantify the slope of the measured RVs as a function of logarithmic wavelength.  We investigate the dependence of the CRX in the full Carmenes GTO sample on 24 stellar activity indices in the visible and near-infrared channels of the CARMENES spectrograph and each star's stellar parameters.  We also present an updated convective turnover time scaling for the calculation of the stellar Rossby number for M dwarfs. }
   {Our results show that approximately 17\% of GTO stars show a strong or a moderate correlation between CRX and RV.  We can improve the measured RVs by a factor of up to nearly 4 in rms by subtracting the RV predicted by the CRX-RV correlation from the measured RVs.  Mid M dwarfs with moderate rotational velocities, moderate CRX-gradients and quasi-stable activity features have the best rms improvement factors.  }
  {We conclude that the CRX is a powerful diagnostic in mitigation of stellar activity and the search for low mass rocky planets. }


   \maketitle

\section{Introduction}

In recent years there has been exceptional progress in wavelength calibration and the development of high-precision, highly-stable instrumentation with the aim of detecting small rocky exoplanets orbiting in the liquid-water habitable zones of their host stars using the RV technique. These include ESPRESSO \citep{Pepe2021A&A...645A..96P}, EXPRES \citep{Jurgenson2016SPIE.9908E..6TJ}  and MAROON-X \citep{Bean2020AAS...23522504B} among others with tens of cm s$^{-1}$ precision.  Recently,  \cite{SurezM2020A&A...639A..77S} reported the achievement of a precision of 30 cm s$^{-1}$ for ESPRESSO observations of Proxima Centauri. 
The current limitation in detecting small rocky planets is not instrumental capability, but the intrinsic variability of the host stars themselves \citep[see recent works such as][and references therein]{Crass2021arXiv210714291C,Zhao2022AJ....163..171Z}.  

All stars with masses just above and lower than the Sun possess an internal convection zone and exhibit some degree of magnetic activity from very low \citep[e.g. the early M dwarf GJ\,887:][]{Jeffers2020Sci...368.1477J} to very high \citep[e.g. the early and mid M dwarfs AU  Mic and EV Lac:][]{Cale2021AJ....162..295C, Klein2022MNRAS.512.5067K,JeffersEVLac2022A&A...663A..27J, Bellotti2022A&A...657A.107B}.  The presence of this activity induces an asymmetry in the shape of the stellar photospheric absorption lines which limits our ability to determine the star's centre-of-mass motion via the position of the line's centre \citep{Dravins1981A&A....96..345D, Saar1997ApJ...485..319S, Lagrange2010A&A...512A..38L, Meunier2010A&A...512A..39M,Reiners2010ApJ...710..432R, Barnes2011MNRAS.412.1599B, Dumusque2014ApJ...796..132D,Haywood2014MNRAS.443.2517H}.  This consequently results in an additional activity-induced RV contribution which impacts the detection and the measurement of the masses of small planets orbiting close to their host stars.   There are also a large number of methods that have been developed to remove the activity-induced component to the RVs.  However, the conclusion is, from the extensive testing of different activity-removal methods by  \cite{Zhao2022AJ....163..171Z}, that no method succeeds in routinely achieving a reduction in the RV rms to sub-meter-per-second levels.

In this work we aim to understand the wavelength dependent activity diagnostics of the full data set secured as part of the CARMENES  GTO survey (Calar Alto high-Resolution search for M dwarfs with Exo-earths with Near-infrared and optical Échelle Spectrographs; \cite{Quirrenbach2018SPIE10702E..0WQ, Quirrenbach2020SPIE11447E..3CQ}.  The CARMENES GTO sample was selected to contain the brightest stars in the $J$ band for each M dwarf sub-spectral type bin that are observable from Calar Alto Observatory in southern Spain and have no stellar companion within 5 arcsec.  This selection procedure for the GTO sample ensures that it includes both magnetically active and inactive stars, and now after over 8 years of operation, and with more than 20\,000 spectra in hand \citep{Ribas2023A&A...670A.139R}, ensures that we have a statistically significant sample to understand the impact of stellar activity on high-precision RV measurements. One important diagnostic parameter in the search for exoplanets orbiting stars showing stellar activity is an analysis of the wavelength dependence of the measured RV values.  This is because features of stellar activity such as dark cool starspots are typically wavelength-dependent because the contrast between starspots and the stellar photosphere is smaller at longer wavelengths.  In contrast, the RV signal from an orbiting exoplanet is wavelength independent.  

The chromatic index (CRX) is a measure of the RV variation as a function of wavelength.  It is defined by taking the slope of the RVs in every spectral order as a function of logarithmic wavelength \citep{Zechmeister2018A&A...609A..12Z}. The CRX-index has been shown by \cite{JeffersEVLac2022A&A...663A..27J}, using a low-resolution Doppler-imaging technique, to be directly correlated with starspot coverage for the mid M-dwarf EV Lac. This is consistent with (i) the simulations of \cite{Baroch2020A&A...641A..69B}, who combined the CRX-index with  photometric  observations of YZ CMi to constrain the spot-filling factor and YZ CMi's spot and photosphere temperature contrast, and (ii) the previous work of ~\cite{TalOR2018A&A...614A.122T}, who investigated the CRX-index for only the most active stars in the CARMENES GTO sample using the first 1.5 years of CARMENES data.  More recently, \cite{Lafarga2020A&A...636A..36L} investigated periodicities found in RVs and nine activity indicators in the CARMENES GTO sample. They concluded that the CRX is as efficient as the line bisector \citep{Queloz2001A&A...379..279Q} as indicator of activity for active M dwarfs.    Other recent work by \cite{Cameron2021MNRAS.505.1699C} (and also \citealt{Lisogorskyi2019MNRAS.485.4804L}) has exploited the wavelength dependence of RV induced by stellar activity to separate RVs with dynamical origin from stellar activity induced variability.

In this paper we aim to empirically understand the wavelength variations of RVs using the CRX-index, and to quantify how to remove this contribution from the RVs.   In Section~2 we define the CRX-index and in Section~3 we describe the observations and define how the CRX stars were selected from the full CARMENES GTO sample. Section~4 we investigate the dependence of CRX-index on fundamental stellar parameters.
In Section~5, we use the wavelength information contained in the CRX-index to remove stellar activity from the RVs. We discuss the implications of our results in Section~6.  

\section{CRX: dependence of RVs with wavelength}

\begin{figure}
\centering
\includegraphics[angle=270,width=0.45\textwidth]{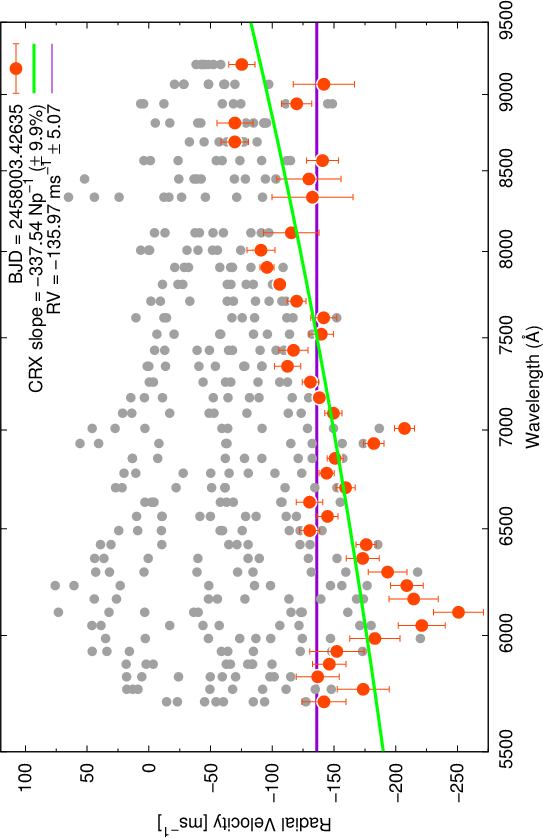}
\caption{Illustration of the RV variation with wavelength, where the central wavelength is given for each order.   Shown are the same data as in Figure ~\ref{fig:EVLac_CRX}, with the steepest positive slope shown by red points.  The green line indicates the slope of these points, which is the CRX-index for each individual spectrum.  The weighted RV measurement is shown by the purple line. The wavelength spacing is logarithmic.} 
\label{fig:CRX_Mathias}
\end{figure}

\begin{figure*}
\centering
\includegraphics[angle=270,width=0.8\textwidth]{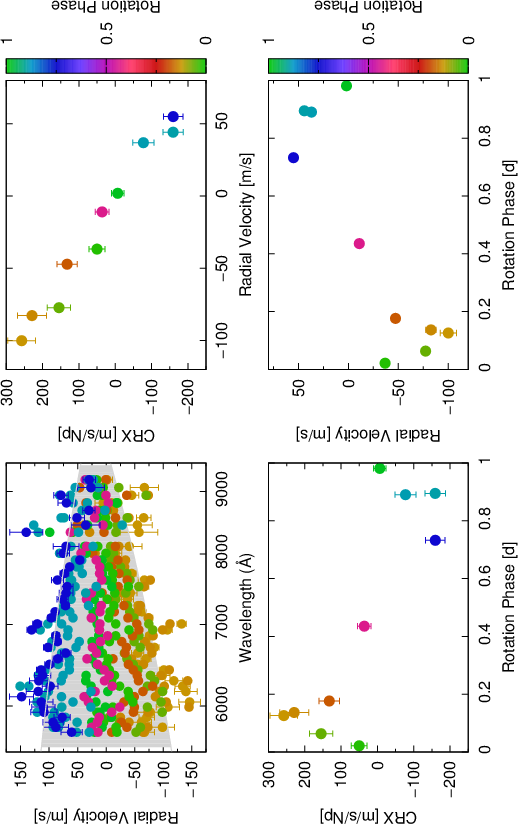}
\caption{{Illustration of the behaviour of the CARMENES visible channel CRX for the mid-M dwarf EV~Lac. Top Left: Variation of RV as a function of wavelength (spacing is logarithmic).  The CRX-index is the slope of RV vs log($\lambda$) for each spectrum. A total of 10 CARMENES spectra are shown covering approximately one stellar rotation. The upper and lower boundaries are the fits to the most positive and negative RV values.  Top right: The CRX-gradient, illustrating the CRX-index as a function of {\em mean} RV. Bottom Left and Bottom Right: CRX vs rotation phase and RV vs rotation phase illustrating the anti-correlation of CRX-index (i.e. the negative CRX-gradient shown in the top right panel). The plotted points in each panel are colour coded according to stellar rotation phase.}}
\label{fig:EVLac_CRX}
\end{figure*}

\subsection{Definition of the CRX}
The CRX-index is defined as the slope of the RV as a function of logarithmic wavelength \citep{Zechmeister2018A&A...609A..12Z} for each spectrum.  The RV of each spectrum is computed using weighted averages to account for the wavelength dependence of the RVs (more details later in Section 3.1).  This is illustrated in Figure~\ref{fig:CRX_Mathias}, which shows the wavelength dependent RVs of 10 spectra of the mid-M dwarf EV Lac. The spectrum exhibiting the steepest positive gradient is illustrated with red points and a linear fit in green. The RV value computed by {\tt{serval}} is shown as a horizontal purple line.

Figure~\ref{fig:EVLac_CRX} shows the CRX calculation in more detail for the same 10 spectra that are illustrated in Figure~\ref{fig:CRX_Mathias}. Because the spectra span less than two complete rotations of EV Lac (\hbox{$P_{\textrm {rot}}=4.349$~d;} \citealt{JeffersEVLac2022A&A...663A..27J}), the effects of evolution due to stellar activity should be minimal.
Each spectrum is coloured according to the stellar rotation phase. The CRX-index measured for each of the spectra in Figure~\ref{fig:EVLac_CRX} (top left panel) is shown as a function of RV in Figure~\ref{fig:EVLac_CRX} (top right panel), where the colour coding is the same in both panels.  

The gradient of the CRX-index versus RV distribution for a series of spectra is referred to as the CRX-gradient in the work that follows. For each star, we define the CRX-length as  RV$_{\rm max}$ - RV$_{\rm min}$, where the RV$_{\rm min}$ and the RV$_{\rm max}$ are the minimum and maximum RV values from all of the available spectra (using 90 and 10 percentiles).  We prefer this definition since CRX variation (CRX$_{\rm max}$ - CRX$_{\rm min}$) and CRX are directly related, while CRX-length is most readily calculated directly from the RVs, even in the absence of CRX-index measurements.  The large variations in the CRX-index as a function of rotation phase (as illustrated in Figure~\ref{fig:EVLac_CRX}) demonstrate how the  CRX-index evolves over the stellar rotation period due to the presence of activity features on the stellar surface.  The CRX-index for each individual spectrum, and consequently the CRX-gradient for a series of spectra, has been directly linked to the distribution of dark starspots on the stellar surface using low resolution Doppler imaging techniques by \cite{JeffersEVLac2022A&A...663A..27J}. The lower two panels of Figure~\ref{fig:EVLac_CRX} show the variation of CRX-index and RV as a function of stellar rotational phase, with the same colouring as in the upper panels.    

The data shown in Figure~\ref{fig:EVLac_CRX} are for the VIS channel only as the errors in the NIR channel are too high for meaningful interpretation for the case of EV Lac.   Even though the NIR data 
are not shown, we do not expect to see a convergence of the slopes at longer wavelengths as is suggested by the outer edges of the light grey region in Figure~\ref{fig:EVLac_CRX} (top left panel).  As discussed by ~\cite{JeffersEVLac2022A&A...663A..27J}, a black-body scaling of the spot and photospheric flux ratios will remain constant continuing to NIR wavelengths, in agreement with PHOENIX models for the spot / photosphere ratio. A more detailed description of the full data set of EV~Lac was previously presented by ~\cite{JeffersEVLac2022A&A...663A..27J}. 

\subsection{Comparison of the CRX with different instruments}

The CRX will show different behaviour when investigated with different spectrographs with
varying wavelength coverage and spectral resolution.  The comparison of the CRX between data secured with HARPS and CARMENES has been previously investigated by \cite{Zechmeister2018A&A...609A..12Z} for the stars YZ CMi, and GJ\,3379, and 
\cite{Kossakowski2022A&A...666A.143K} for AD Leo.   In general the CRX shows a negative slope for both HARPS and CARMENES datasets of YZ CMi while for GJ\,3379, only the CARMENES data show a significant chromatic variation.  
Similarly, CARMENES M dwarf K2-18 has been reported by \cite{Radica2022MNRAS.517.5050R} to show chromatic behaviour at a level that is below our threshold of Pearson's $r>$ 0.5.  Similar to  GJ\,3379, this is not visible in the HARPS dataset of K2-18.  Surprisingly, the CRX of AD Leo shows a positive slope for archival HARPS data and a negative slope for CARMENES data \citep{Kossakowski2022A&A...666A.143K, Zechmeister2018A&A...609A..12Z} and significantly lower amplitude for much redder wavelengths \citep{Carmona2023A&A...674A.110C}.  The comparison of chromaticity of RVs across a very broad wavelength regimes has the potential to give a crucial insight into the activity of these stars.  However, simultaneous observations across a broad wavelength range are required to truly compare the wavelength dependence as features of stellar activity can evolve on timescales of a few stellar rotation periods.  

\subsection{Modelling the CRX}

The effect of dark starspots on the wavelength dependence of RVs has previously been investigated by \cite{Desort2007A&A...473..983D,Barnes2011MNRAS.412.1599B, Cameron2021MNRAS.505.1699C} among many others.  In particular, the CRX has been simulated by \cite{Baroch2020A&A...641A..69B} using 
spectroscopic and photometric data for the mid-M dwarf YZ CMi.  They demonstrated that the CRX-index, computed for each spectrum, can be used to determine, and break the degeneracy between, the spot-filling factor and temperature contrast between the starspot and unspotted photosphere.

Recent work has shown that the CRX-index versus RV relation can have an inclined lemniscate ($\infty$) or figure-of-eight shape, with a typically negative slope, which results from a slight phase offset between the phase curves of the RVs and the CRX values \citep[examples include][for YZ CMi and EV Lac respectively]{Baroch2020A&A...641A..69B, JeffersEVLac2022A&A...663A..27J}.  This is comparable to the inclined-lemniscate shape established for RV versus Bisector Inverse Slope (BIS) relations for co-rotating cool spots \citep{Desort2007A&A...473..983D,Saar1997ApJ...485..319S,Boisse2011A&A...528A...4B}. For the BIS, \cite{Boisse2011A&A...528A...4B} attribute this to result from foreshortening and limb-darkening effects, while  
for the CRX \cite{Baroch2020A&A...641A..69B} consider that the offset results from the combined effects of limb darkening and convective blue shift.  The example of the CRX-index versus RV relation shown Figure 2 for the mid-M dwarf EV\,Lac is almost a straight line due to the dominance of high-latitude spot features (see the surface brightness maps reconstructed by ~\cite{JeffersEVLac2022A&A...663A..27J}) and to the limited time sampling of the data which covers only a few rotation periods.   We would expect a higher amplitude lemniscate shape if there was a higher degree of spot coverage located at equatorial latitudes as has been modelled for the BIS ~\citep{Boisse2011A&A...528A...4B}.   

\subsubsection{Spot to photosphere contrast}

\begin{figure}
\centering
\imagetop{\includegraphics[angle=270,width=0.43\textwidth]{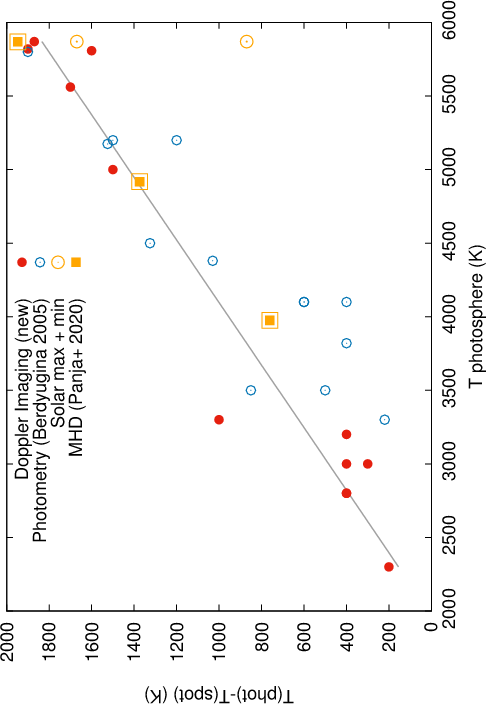}}
\caption{Revised photospheric to spot temperature. Original photometric data points from \cite{berdyugina05starspots} are shown as light blue open circles.  Doppler imaging values are shown as filled red circles where the values are taken from Table~\ref{tab:temp_contrast_DI} and the values for the Sun are shown as two large open light blue circles with a temperature of 5800\,K.}
\label{fig:Spot_temp_contrast}
\end{figure}

The spot to photospheric temperature contrast decreases towards redder wavelengths because the contrast between dark starspot and the photosphere is smaller at longer wavelengths.   As this is one of the key diagnostic parameters of the CRX, we first update the temperature contrast relation originally published by \cite{berdyugina05starspots} who only used contrasts derived from photometric data sets for spectral types earlier than M3V.  Here, we additionally include values used in Doppler imaging studies covering the full spectral type range from early-G to late-M (Table~\ref{tab:temp_contrast_DI}), as well as the MHD simulations of starspots for different spectral types \citep{Panja2020ApJ...893..113P}.  Our revised spot to photosphere temperature contrast (using Doppler Imaging results only) is equated as follows:     

\begin{equation}
\Delta {T} = a\,T_{\rm phot} + b
\end{equation}

\noindent where $a=0.47$ and $b=-926.97$\,K.  The fit to all Doppler Imaging data is plotted in Figure~\ref{fig:Spot_temp_contrast}.

\begin{table}[]
     \caption{\label{tab:temp_contrast_DI} Empirical data used to derive spot to photospheric temperature contrasts using the Doppler imaging technique.}
    \centering
    \begin{tabular}{lccccc}
    \hline
    \hline
    \noalign{\smallskip}
Star  &  Sp.  &  $T_{\rm phot}$ &  $T_{\rm spot}$ &  $\Delta T$ &  Ref. \\
  & type & (K) & (K) & (K) & \\
    \noalign{\smallskip}
\hline
    \noalign{\smallskip}
HD 106506 & G1.0\,V & 5870 & 4000 & 1870 & 1 \\
EK Dra & G1.5\,V & 5561 & 3861 & 1700 & 2 \\
HD 171488 & G2.0\,V & 5808 & 4200 & 1600 & 3 \\
HD 29615 & G3.0\,V & 5820 & 3920 & 1900 & 4 \\
AB Dor & K0.0\,V & 5000 & 3500 & 1500 & 5 \\
EV Lac & M3.5\,V  & 3300 & 2300 & 1000 & 7 \\
V394 Peg & M4.0\,V &  3200 & 2800 & 400 &  8 \\
HU Del & M4.5\,V & 3000 & 2700 & 300 &  9 \\
GJ 65A & M5.5\,V & 2800 & 2400 & 400 &  10 \\
GJ 65B & M6.0\,V  & 2800 & 2400 & 400 &  10 \\
LP 944--20 & M9.5\,V & 2300 & 2100 & 200 &  9  \\
    \noalign{\smallskip}
\hline
    \end{tabular}
\tablebib{{1: \citealt{Waite2017MNRAS.465.2076W};
2: \citealt{Waite2017MNRAS.465.2076W}; 
3: \citealt{Jeffers2008MNRAS.390..635J}; 
4: \citealt{Waite2015MNRAS.449....8W};
5: \citealt{Jeffers2005MNRAS.359..729J};
6: \citealt{JeffersEVLac2022A&A...663A..27J};
7: \citealt{Morin2008MNRAS.384...77M};
8: \citealt{Barnes2015ApJ...812...42B};
9: \citealt{Barnes2017MNRAS.471..811B}}}
\end{table}

\begin{figure*}
\centering
\includegraphics[angle=270,width=0.43\textwidth]{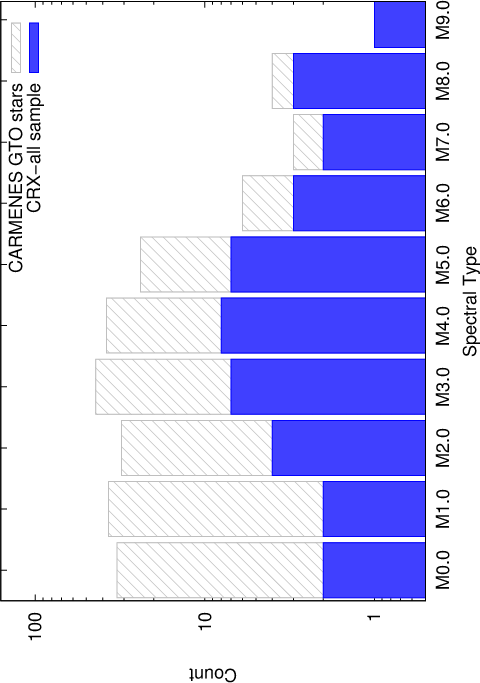}
\includegraphics[angle=270,width=0.43\textwidth]{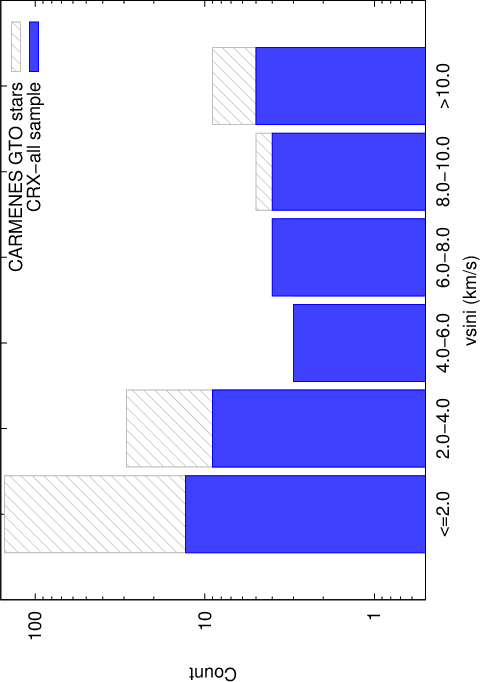}
\caption{Spectral type distribution (left panel) and the distribution of \vsini\ of the CRX-all sample, coloured in blue.  Also shown are all of the CARMENES GTO stars with more than 25 observations as indicated by shaded grey bars.}
\label{fig:CRX_SPT_vsini_hist}
\end{figure*}

\subsubsection{CRX-gradient and length}

The simulations of \cite{Baroch2020A&A...641A..69B} show that the gradient of the CRX-RV relation is an important diagnostic as it provides a key insight into how quickly the RV variations induced by a cool starspot decrease as a function of increasing (logarithmic) wavelength.   In the simulations of \cite{Baroch2020A&A...641A..69B}, the CRX-gradient primarily depends on the spot to photosphere temperature contrast ratio for the case of one large spot located at a latitude of 78$^\circ$.  In addition to the CRX-gradient, \cite{Baroch2020A&A...641A..69B} also reported that the length of the CRX-index versus RV relation depends on both the spot-to-photosphere contrast and the spot filling factor. This is expected because for a given single spot size (i.e. a star with a fixed filling factor), the wavelength dependent RV amplitude and hence gradient measured by CRX-index is larger for a larger spot-to-photosphere contrast. Similarly, for a fixed spot to photospheric contrast, increasing a spot's size, and thus the spot filling factor, will also lead to a greater CRX-length.    

The models of \cite{Baroch2020A&A...641A..69B} considered the impact of one starspot located at high latitudes on the surface of the mid-M dwarf YZ CMi. However results from Doppler imaging studies (examples are listed in Table~\ref{tab:temp_contrast_DI}) show that in reality, the surfaces of M dwarfs are likely to be much more densely spotted.  In this work we use the CRX-gradient and the CRX-length as the two main diagnostic parameters of our investigation.   We investigate how the CRX-gradient and CRX-length of stars in the CARMENES GTO sample depend on their fundamental stellar parameters and levels of stellar activity.

\begin{figure}
\centering
\includegraphics[angle=270,width=0.43\textwidth]{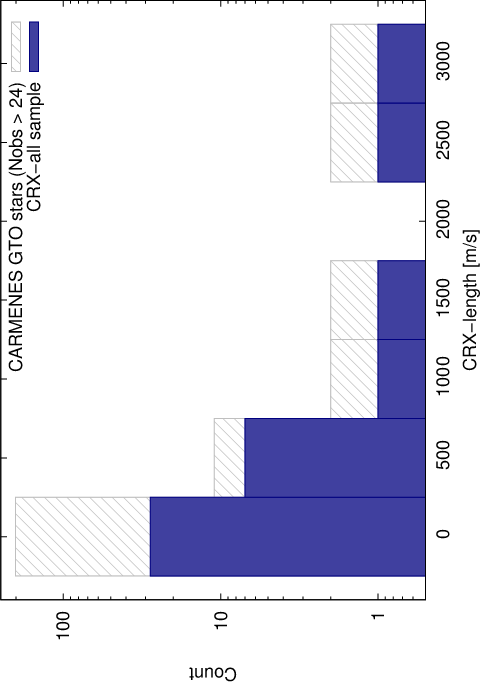}
\caption{Distribution of CRX-length for the CARMENES GTO sample (light grey) and the CRX-all sample (navy).  For clarity,  the CRX-length values of the CARMENES GTO sample are only shown up to a value of 3000 m\,s$^{-1}$}.
\label{fig:CRX_length_hist}
\end{figure}

\begin{figure}
\centering
\includegraphics[angle=270,width=0.43\textwidth]{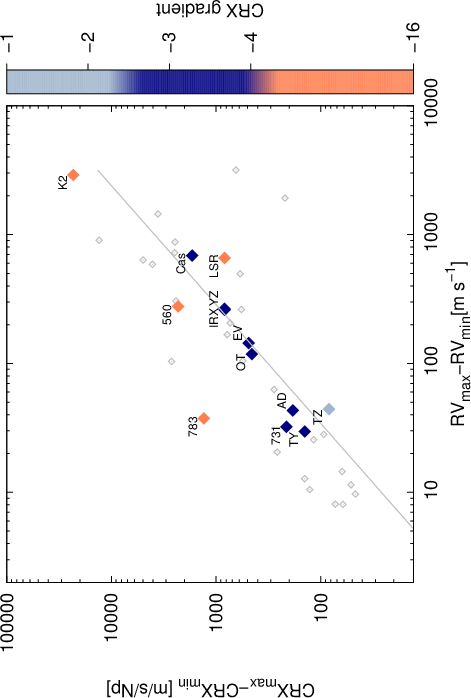}
\caption{The variation of CRX (CRX$_\textrm{max}$\,-\,CRX$_\textrm{min}$) as a function of RV variation (CRX-length = RV$_\textrm{max}$\,-\,RV$_\textrm{min}$). Fitted is a straight line to moderate slopes (coloured navy). Points are coloured by CRX-gradient. 
}
\label{fig:Var_CRX_RV}
\end{figure}

\section{Data}
\subsection{Observations and data processing}

The data investigated in this work are the full set of CARMENES Guaranteed Time Observations (GTO) of more than 350 M dwarfs which have been observed since the start of the CARMENES GTO survey in 2016.  The CARMENES spectrograph is installed at the 3.5\,m telescope at Calar Alto Observatory in Spain. It comprises two fibre-fed and cross-dispersed spectrographs, one at visible wavelengths (VIS, 520--960\,nm) and the other at near-infrared wavelengths (NIR, 960--1710 nm) with a resolution of $\sim$94\,600 in the VIS and $\sim$80\,400 in NIR and an average sampling of 2.8 pixels \citep{Quirrenbach2018SPIE10702E..0WQ}.  The data were processed in the usual manner for CARMENES data, i.e. with {\tt{caracal}} \citep[CARMENES reduction and calibration software;][]{Zechmeister2014A&A...561A..59Z, Caballero2016csss.confE.148C}, and the RVs were calculated using {\tt{serval}} \citep[Spectrum radial velocity analyzer][]{Zechmeister2018A&A...609A..12Z} using weighted averages to account for the wavelength dependence of the RVs, where  the individual orders are weighted.  The impact of telluric lines on the measured RVs was removed \citep{Nagel2023A&A...680A..73N} using the template division telluric modeling technique.    The nightly-zero-point (NZPs) were also subtracted from the serval RV measurements (see \cite{Ribas2023A&A...670A.139R} section 4.4 for a detailed explanation). 

\begin{figure}
\centering
\includegraphics[angle=270,width=0.43\textwidth]{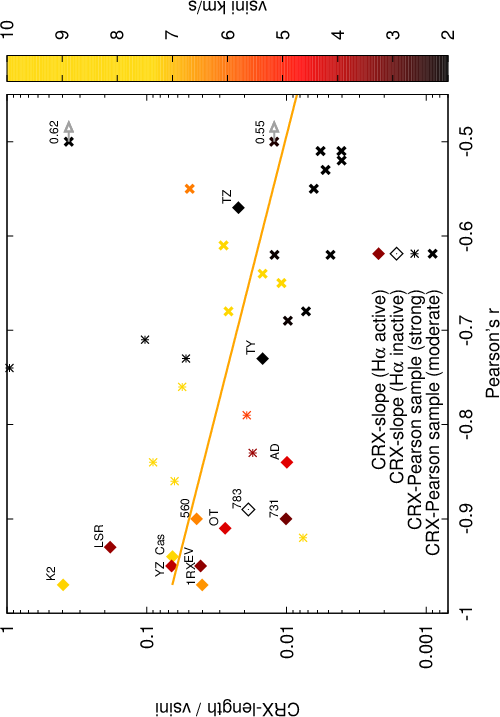}
\includegraphics[angle=270,width=0.43\textwidth]{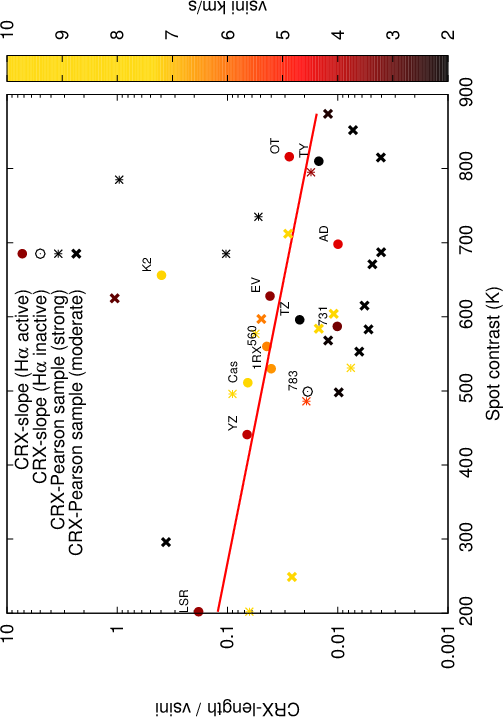}
\includegraphics[angle=270,width=0.43\textwidth]{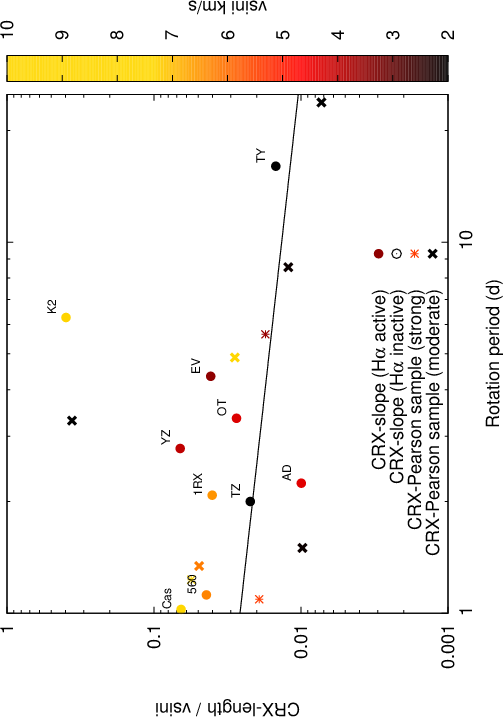}
\caption{Dependence of the CRX-length as a function of Pearson's $r$ value (upper panel) and spot contrast (middle panel) and rotation period (lower panel) for the CRX-all stars sample.   Closed symbols indicate H$\alpha$ active stars and open symbols H$\alpha$ inactive stars.  Points are coloured by projected stellar rotational velocity (\vsini).  The CRX-slope stars are labelled as indicated in Table~\ref{tab:CRX-param}.  In the top panel, the two arrows indicate two stars with positive Pearson's r correlation.  Their true Pearson's value is shown in the label for each point. }
\label{fig:CRXlength_Pearson}
\end{figure}

\subsection{CARMENES NIR channel}

In this work we only use data from the CARMENES VIS channel and not from the CARMENES NIR channel because the NIR channel yields RV measurements of less precision than the VIS channel. This is partly because of the diminished RV information content in the observed NIR domain ~\cite{Reiners2018A&A...612A..49R}.  But also, the CARMENES NIR channel suffered from thermo-mechanical instability since the start of operations and until some corrective measures were applied a few years later. Consequently, the NIR RV data were in general not used during the exploitation of Guaranteed Time Observations (but see ~\cite{Morales2019Sci...365.1441M} and ~\cite{Bauer2020A&A...640A..50B}). Subsequent instrument upgrades have dramatically improved the precision of the NIR channel (Schaefer et al. 2024; Varas et al. 2024, in prep) and a careful analysis of the newly collected NIR data is undergoing

\subsection{Activity Indices and CCF parameters}

Given the large wavelength coverage of CARMENES we have a large number of activity indicators available.  We use a total of 26 indicators as follows

\begin{itemize}
    \item Six RV, CRX and differential line widths (dLW) calculated by {\tt{serval}} - i.e. RV$_{\rm VIS}$ and  RV$_{\rm NIR}$, CRX$_{\rm VIS}$ and CRX$_{\rm NIR}$, dLW$_{\rm VIS}$ and dLW$_{\rm NIR}$.

    \item Sixteen activity indices at VIS and NIR wavelengths computed following the procedure described by \cite{Schoefer2019A&A...623A..44S}.  The full list of spectral line indices comprise: 
$\log( L_{{\rm H}\rm\alpha}/L_{\rm bol})$, pEW(H$\rm \alpha$),  HeD3, NaD, Ca IRT-a,-b,-c,	He\,I\,10830, Pa $\beta$, CaH2, CaH3, TiO\,7050, TiO\,8430,	TiO\,8860, VO\,7436, VO\,7942, and FeH Wing-Ford.   We define stars with positive pEW(H$\alpha$), where H$\alpha$ is in absorption, to be H$\alpha$ inactive stars and with negative values, where H$\alpha$ is in emission, to be H$\alpha$ active stars.  The minimum detectable emission in the H$\alpha$ line depends on the resolution of the spectrograph and the signal-to-noise of the data and has been shown by \cite{Schoefer2019A&A...623A..44S} to be $-$0.3\,\AA\ for CARMENES.

    \item Four cross correlation function (CCF) parameters, CCF-Contrast, CCF-RV, CCF-FWHM, and CCF-BIS were computed following \cite{Lafarga2020A&A...636A..36L}. \\
\end{itemize}

\subsection{CRX-all sample}

From the full CAMRENES GTO sample of 350 stars, firstly, we select stars with a minimum number of observations $N_{\rm obs} >$ 25  as this is the minimum number of observations required to give a reliable CRX-index versus RV relation, and to provide a statistically significant correlation with activity indicators.   The CARMENES GTO sample with $N_{\rm obs} >$ 25 comprises a total of 215 stars and is referred to as CARMENES $N_{\rm GTO}$25 sample in the rest of this work.  Parameters for all 215 \cargto~stars are listed in Table B.1 (Appendix).

We then computed the statistical significance of the correlation of CRX-index with RV for each of the 215 stars using Pearson's $r$ coefficient. We define a strong correlation (or anti-correlation)
where Pearson's $r >|0.7|$ , and a moderate correlation to be $0.5 < |r| < 0.7$ .  The statistical significance is computed by the Student's t-test probability value,  $p$.  We assumed values of $p < 0.03$ imply no strong evidence to reject the (no correlation) null hypothesis.  There are a total of 39 stars with a strong or a moderate correlation and these stars are referred to as CRX-all in the work that follows and are listed in Table~\ref{tab:CRX-param}.  From this, 21 stars have strong correlations between the CRX-index and RV and 18 stars have a moderate correlation.

\subsection{CRX-slope sample}

The CRX-all sample is further classified based on a visual inspection of the CRX-index versus RV relation for each star.  A total of 13 stars exhibit a tight linear CRX-index versus RV relationship and comprise both strong and moderate statistical CRX-index versus RV correlations. This sample is referred to as the CRX-slope sample in the rest of this work. To avoid confusion we refer to the actual measured value of the CRX slope as the CRX-gradient in the rest of this work. The CRX-index versus RV relation is shown for each of these stars in Figure~\ref{fig:CRX-sample}.  

\subsection{CRX-Pearson sample}
The 26 remaining stars are refereed to as the CRX-Pearson sample in the rest of this work and are subdivided into Pearson-strong sample comprising 9 stars and Pearson-moderate sample comprising 17 stars. The CRX-index versus RV for these two samples are respectively shown in Figures~\ref{fig:CRX-Pearson-strong} and ~\ref{fig:CRX-Pearson-mod}.   The RV-CRX correlation deviates from a straight line in this sample, where the deviations can take the form of a high degree of scatter or a generally extended cloud shape. The reasons for these deviations are not well understood and are beyond the current scope of this work.\\

\noindent {\underline{Sample Summary:}}
\begin{itemize}
\item CRX-all: 39 stars with Pearson's strong, $|r| \geq 0.7$ (21 stars), or moderate, $0.5 < |r| < 0.7$ (18 stars) correlations
    \item CRX-slope: 13 stars with a tight linear relationship, as inspected visually; they are a subset of the the CRX-all sample
    \item CRX-Pearson: 26 stars with non-linear but moderate (17 stars) or strong (9 stars) Pearson's $r$ values
\end{itemize}
\subsection{Comparison to the \cargto-sample}

The distribution of the CRX-all stars compared to the \cargto-sample in terms of spectral type and projected stellar rotational velocity, or \vsin\ is shown in Figure~\ref{fig:CRX_SPT_vsini_hist}.  The 39 stars in the CRX-all sample cover all spectral types and broadly follows the spectral type distribution of the \cargto-sample.  The distribution of \vsini\ values for the CRX-all stars and the \cargto-sample are shown in Figure~\ref{fig:CRX_SPT_vsini_hist} (right panel).  Approximately one third of the CRX-all stars have \vsini\ values $<$ 2.0\,km\,s$^{-1}$, one third have \vsini\ values between 2.0--4.0 \kms\ and the remaining third have values 4.0--10.5 \kms. The wavelength dependence of the RVs depends on many fundamental stellar parameters and in the following sections we will investigate the  correlations and trends in the activity indices available from the CARMENES spectra.   

\section{Results: CRX: gradient and length}

As previously discussed, preliminary models indicate that the two most useful parameters to characterise the CRX-index versus RV relation are its gradient and length.   For the CRX-slope sample, the measured CRX-gradient values typically have slopes ranging from $-$2.65 Np$^{-1}$ to $-$3.70 Np$^{-1}$ (8 stars), with 4 stars having slopes in the range $-$4.6 Np$^{-1}$ to $-$17 Np$^{-1}$, and 1 star having a very shallow slope of $-$1.16 Np$^{-1}$.  The colouring of points in Figure~\ref{fig:Var_CRX_RV} and subsequent Figures is chosen to reflect these broad groupings.   We caution the direct interpretation of the steepest slopes for the 3 stars with the latest M-dwarf spectral types as for these stars there is less flux in the bluer orders and may potentially bias the results.  Since our aim is to apply uniform analysis to all stars we have included these stars but do not draw any conclusions from them.  In contrast, the steep slope of the young M\,3.0 dwarf K2-33 results from very high levels of stellar activity given its young age.    

CRX-length values range from 30 ms$^{-1}$ also for the  M\,2.5 dwarf TYC 3529-1437-1 to 2897 ms$^{-1}$ for the M\,3.0 dwarf K2-33.  The distribution of CRX-length values for the CRX-all sample compared to the CARMENES GTO sample is shown in Figure~\ref{fig:CRX_length_hist}. In the work that follows we normalise the CRX-length by dividing the measured length by the stellar \vsini.

Furthermore, in Figure~\ref{fig:Var_CRX_RV} we show the CRX variation, CRX$_\textrm{max}$\,--\,CRX$_\textrm{min}$, as a function of the total RV variation, RV$_\textrm{max}$\,--\,RV$_\textrm{min}$, which we define as the CRX-length. The CRX-slope stars with moderate gradients show a linear increase with increasing RV variation.  \\

Figure~\ref{fig:CRXlength_Pearson} illustrates the correlation between the CRX-length and the Pearson's $r$ value where longer CRX-length / \vsini\ values have a higher Pearson's $r$ value indicating a stronger statistical correlation.  We note that low CRX-length values occur at low \vsini\ values. There was no correlation noted between the CRX-gradient and the Pearson's $r$ value. 
We further investigated the correlation of CRX-gradient with the CRX-length.  We find no correlation between these two parameters (plots not shown).   We also examined the dependence of the CRX-gradient on the inferred spot to photospheric temperature contrast (see Section 2.2.1).  While the results do not show a correlation with CRX-gradient, the CRX-slope stars lower contrasts for stars with higher CRX-length / \vsini\ values as shown in the middle panel of Figure~\ref{fig:CRXlength_Pearson}.  Stars with low \vsini\ values show the lowest CRX-length / \vsini\ values over a range of medium to high contrasts.  Increasing the \vsini\ value for a given contrast often results in a higher CRX-length / \vsini\ value.  The lower panel of Figure~\ref{fig:CRXlength_Pearson} shows that there is a tendency of stars with short stellar rotation periods  to show slightly higher CRX-length / \vsini\ values.  

Our initial results indicate that the interpretation and understanding the nature of the CRX is more challenging than suggested by the models \citep{Baroch2020A&A...641A..69B}.  However, it should be noted that only one starspot was modelled on the surface of YZ CMi and the reality is certainly more complicated.  M dwarfs in general are likely to have more complicated distributions of starspots, where the latitude of the starspot will also have an impact on the CRX-gradient and length \citep{Barnes2011MNRAS.412.1599B}, as well as plage regions \citep{Meunier2010A&A...512A..39M, Jeffers2014MNRAS.438.2717J}.  Given that M dwarfs typically have much lower levels of convective blueshift compared to the more massive G and K dwarfs, its impact on the RV precision is unlikely to be a large effect \citep{Liebing2021A&A...654A.168L} compared to more massive solar-type stars.  The dependence of the CRX gradient and length as a function of stellar activity parameters will be investigated in later sections of this work.   

\subsection{Dependence on activity indicators}

In this section we investigate how the CRX gradient and CRX-length of each of the 13 CRX-slope stars depend on their measured activity indices in relation to the CRX-Pearson and \cargto-samples.  The general patterns of activity in the CARMENES GTO sample have previously been presented and discussed in detail by \cite{Jeffers2018A&A...614A..76J}.

In the following Figures ~\ref{fig:Vsini_SPT}  to~\ref{fig:Bfield_SPT}, the stars in the CRX-slope sample are shown as diamond symbols, CRX-Pearson strong stars with an asterisk symbol, CRX-Pearson moderate stars as a cross, and the remaining stars in the \cargto-sample are shown as small grey circular points. For each figure, the left and right panels are colour-coded based on the CRX gradient and the CRX-length / \vsini, respectively.  

The colouring for the CRX gradient and the CRX-length plots were chosen to represent the sample in terms of global groups rather specific stars.  In the text that follows we discuss these plots in terms of the shallowest to the steepest slopes, and the shortest to the longest CRX-length values.  The stellar activity indices that we investigate are stellar (i) rotational velocity or \vsini\, (Figure~\ref{fig:Vsini_SPT}) (ii) chromospheric emission via H$\alpha$ (Figure~\ref{fig:Halpha_SPT}), (iii) diagnostic of spot coverage via TiO7050 \AA\ band (Figure~\ref{fig:SPT_TiO}), and finally (iv) average magnetic field strength (Figure~\ref{fig:Bfield_SPT}).

\begin{figure*}
\centering
\mbox{
\imagetop{\includegraphics[angle=270,width=0.43\textwidth]{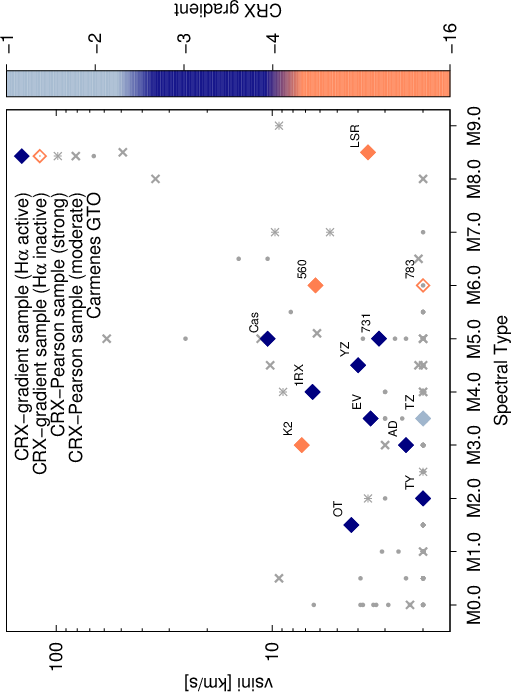}}
\imagetop{\includegraphics[angle=270,width=0.43\textwidth]{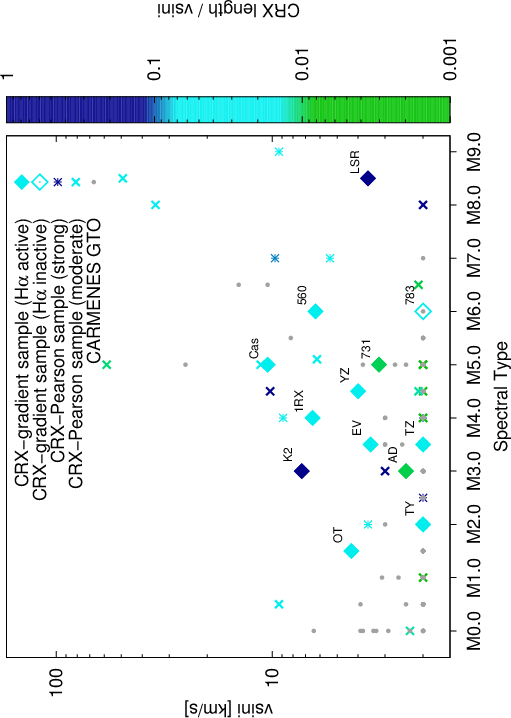}}}
\caption{Variation of projected rotational velocity (\vsini) as a function of spectral type for the CARMENES GTO sample with $N_{\rm obs}$ $>$ 25.  Stars in the CRX-slope sample are coloured according to the measured gradient(left panel) and CRX-length (right panel). Stars in the CRX-slope sample that are H$\alpha$ active are indicated by closed symbols and H$\alpha$ inactive stars as open symbols. The minimum \vsini\ value plotted is 2 \kms. The stars are labelled as indicated in Table~\ref{tab:CRX-param} and also whether their large-scale magnetic field geometry is SS: Stable + Strong or CW: Complex Weak.}
\label{fig:Vsini_SPT}
\end{figure*}

{\renewcommand{\arraystretch}{1.25}
\begin{table*}
\caption{\label{tab:CRXgrad-activity} CRX-gradient with activity.  }
 \centering
 \begin{tabular}{l|c|c|c|c}
 \hline
 \hline
 Parameter  & Moderate gradients & Steep gradients & Figure No\\
 \hline
 Spectral Type  & early to mid M & late-M & ~\ref{fig:Vsini_SPT}\\  
 \vsini\ &  $\leq$ 11 \kms\ & low to moderate & ~\ref{fig:Vsini_SPT} \\
 $\log( L_{{\rm H}\rm\alpha}/L_{\rm bol})$  & active (high) & active (moderate) & \ref{fig:Halpha_SPT} \\
  H$\alpha$ sat / unsat  & saturated & saturated & \ref{fig:Halpha_vsini} \\
 TiO 7050  & lowest value per SpT bin & late M hook & \ref{fig:SPT_TiO} \\
 log($L_{\rm{x}}$/$L_{\rm{bol}}$) & saturated & ... & not shown$^1$ \\
 Ave mag. field  & highest & moderate & \ref{fig:Bfield_SPT}\\
Ave mag. field sat / unsat  & saturated & saturated & \ref{fig:Bfield_SPT}\\

 \hline
 \noalign{\smallskip}
 \end{tabular}
\tablefoot{The shallow gradients are not shown as this group only comprises only 1 star.  \tablefoottext{1} {The} 
log($L_{\rm{x}}$/$L_{\rm{bol}}$) 
plot is not shown as many of the stars in the CRX-all sample do not have X-ray values.}
\end{table*}

{\renewcommand{\arraystretch}{1.25}
\begin{table*}
 \caption{\label{tab:CRXlength-activity} CRX-length with activity}
 \centering
 \begin{tabular}{l|c|c|c|c}
 \hline
 \hline
 Parameter & Shortest lengths & Moderate lengths & Longest lengths & Figure\\
 \hline
 Spectral Type & mid M & early to mid M & mid to late M & ~\ref{fig:Vsini_SPT}\\  
 \vsini\ & $\leq$2-3.5 \kms\ & moderate to high & moderate & ~\ref{fig:Vsini_SPT} \\
 $\log( L_{{\rm H}\rm\alpha}/L_{\rm bol})$ & active: low to moderate & active: moderate to high & active: moderate to high & \ref{fig:Halpha_SPT} \\
 H$\alpha$ sat / unsat & unsaturated & both & saturated & \ref{fig:Halpha_vsini} \\
 TiO 7050 &  lowest value per SpT bin  & lowest value per SpT bin & late M hook + & \ref{fig:SPT_TiO} \\
 log($L_{\rm{x}}$/$L_{\rm{bol}}$) & ... & saturated & saturated & not shown$^1$ \\
 Ave mag. field & moderate & highest & ... & \ref{fig:Bfield_SPT}\\
Ave mag. field sat / unsat  & unsaturated & both & ... \\
 \hline
 \noalign{\smallskip}
 \end{tabular}
  \tablefoot{\tablefoottext{1}{{The} log($L_{\rm{x}}$/$L_{\rm{bol}}$) plot is not shown as many of the stars in the CRX-all sample did not have X-ray values.}}
 \end{table*}
}

\subsection{CRX-gradient}

In Table~\ref{tab:CRXgrad-activity} we have summarised the information contained in the left-hand panels of Figures ~\ref{fig:Vsini_SPT}  to~\ref{fig:Bfield_SPT} to understand how the CRX gradient varies as a function of a range of stellar activity parameters.    Comparing the gradients of the CRX-slope sample with the full \cargto-sample \citep[see][]{Jeffers2018A&A...614A..76J}, we conclude that the stars with the shallowest CRX-gradients are among the least active stars.  Since there is only one star in this category we do not include the shallowest slopes in Table~\ref{tab:CRXgrad-activity}.  Stars with moderate CRX gradients occur at early to mid-M spectral types, have high levels of stellar activity (Figures~\ref{fig:Vsini_SPT},~\ref{fig:Halpha_SPT}) and show saturated $\log( L_{{\rm H}\rm\alpha}/L_{\rm bol})$ and average magnetic field strengths (Figures~\ref{fig:Halpha_vsini},~\ref{fig:Bfield_SPT}).  Similarly, the steepest slopes occur at later M spectral types at moderate levels of activity (e.g. \vsini~and $\log( L_{{\rm H}\rm\alpha}/L_{\rm bol})$, Figures~\ref{fig:Vsini_SPT}, ~\ref{fig:Halpha_SPT}). These stars show saturation at lower $\log( L_{{\rm H}\rm\alpha}/L_{\rm bol})$ values compared to stars with mid-M spectral types \citep[][]{Mohanty2003ApJ...583..451M} (Figures~\ref{fig:Halpha_vsini},~\ref{fig:Bfield_SPT}).

\subsection{CRX-length}
The correlations of the CRX-length / \vsini\ with the stellar activity indicators as shown in the right-hand panels of Figures~\ref{fig:Vsini_SPT}  to~\ref{fig:Bfield_SPT} are summarised in Table~\ref{tab:CRXlength-activity} for ease of comparison.  Generally, the shortest CRX-lengths are found for stars with low to moderate H$\alpha$ activity levels, with early to mid-M spectral types and moderate average magnetic field strengths (Figures~\ref{fig:Vsini_SPT}, ~\ref{fig:Halpha_SPT},~\ref{fig:Bfield_SPT}).  The moderate CRX-length values occur for very active stars and the longest lengths are for very active stars at late M spectral types.  As noted in for the CRX gradient, activity in the latest M dwarfs saturates at lower $\log( L_{{\rm H}\rm\alpha}/L_{\rm bol})$ values compared to mid M dwarfs (Figures~\ref{fig:Halpha_vsini},~\ref{fig:Bfield_SPT}).

\subsection{CRX-Pearson}

The remaining stars in the CRX-Pearson sample follow the same trends.  Stars with a strong Peason's correlation typically have longer lengths and are located at mid to late M spectral types.  The correlation of the CRX-length with various parameters was previously discussed in more detail in Section 3.  

\begin{figure*}
\centering
\mbox{
\imagetop{\includegraphics[angle=270,width=0.43\textwidth]{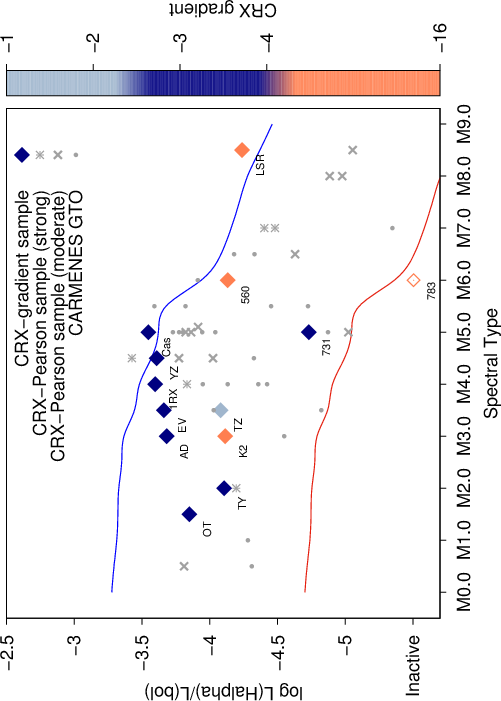}}
\imagetop{\includegraphics[angle=270,width=0.43\textwidth]{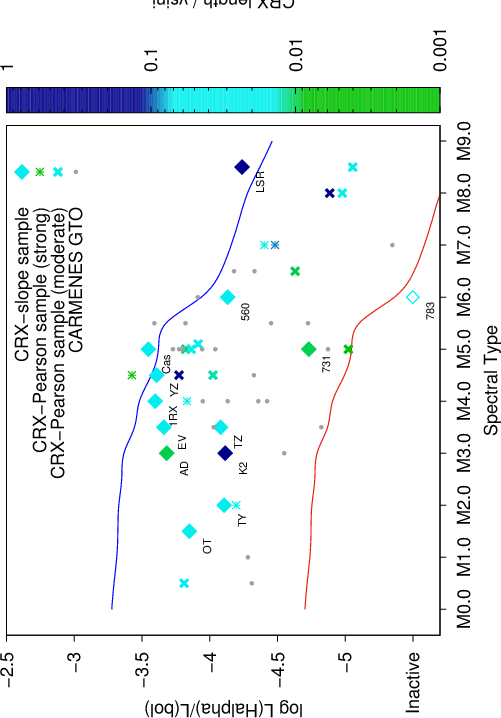}}
}
\caption{Normalised H$\alpha$ luminosities as a function of spectral type for CARMENES GTO sample with $N_{\rm obs}$ $>$ 25.  The solid orange line indicates the minimum values of emission in the H$\alpha$ line that our CARMENES observations can detect, which is defined to be pEW (H$\alpha$) = $-$0.3 \AA\ \citep{Schoefer2019A&A...623A..44S}.  A population of extremely active stars with pEW(H$\alpha$) < $-$8.0 are indicated by the solid blue line as previously noted by \cite{Jeffers2018A&A...614A..76J}. H$\alpha$ inactive stars are shown at the base of each plot as indicated by the y-axis. The stars in the CRX-slope sample are indicated by the diamond shapes which are coloured based on the CRX gradient of the data points in the left panel and CRX-length in the right panel. The stars are labelled as indicated in Table~\ref{tab:CRX-param}.}
\label{fig:Halpha_SPT}
\end{figure*}

\begin{figure*}
\centering
\mbox{
\imagetop{\includegraphics[angle=270,width=0.43\textwidth]{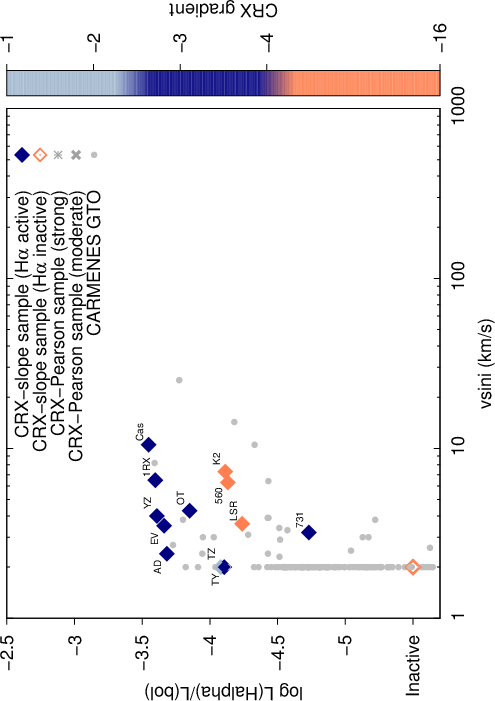}}
\imagetop{\includegraphics[angle=270,width=0.43\textwidth]{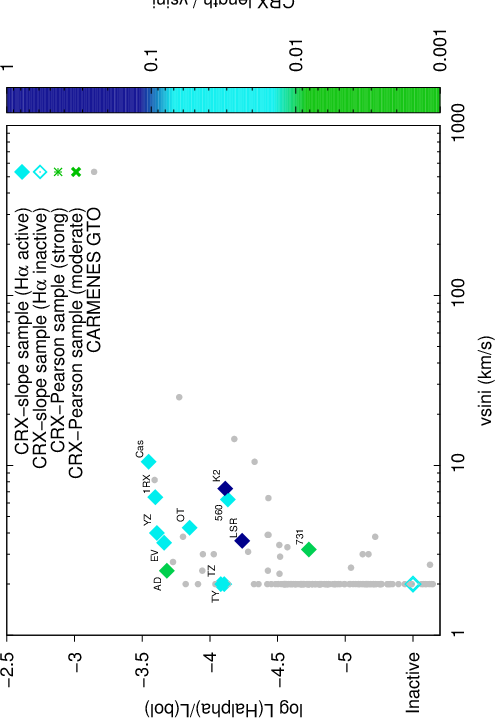}}
}
\caption{Normalised H$\alpha$ luminosities as a function of projected rotational velocities (\vsini) for CARMENES GTO sample with $N_{\rm obs}$ $>$ 25.  Stars in the CRX-slope sample are shown by diamond shaped point which are open for H$\alpha$ inactive stars and filled for H$\alpha$ active stars.  The CRX-slope sample stars are coloured based on CRX-gradient  (left panel) and CRX-length (right panel). The stars are labelled as indicated in Table~\ref{tab:CRX-param}. }
\label{fig:Halpha_vsini}
\end{figure*}

\begin{figure*}
\centering
\mbox{
\imagetop{\includegraphics[angle=270,width=0.43\textwidth]{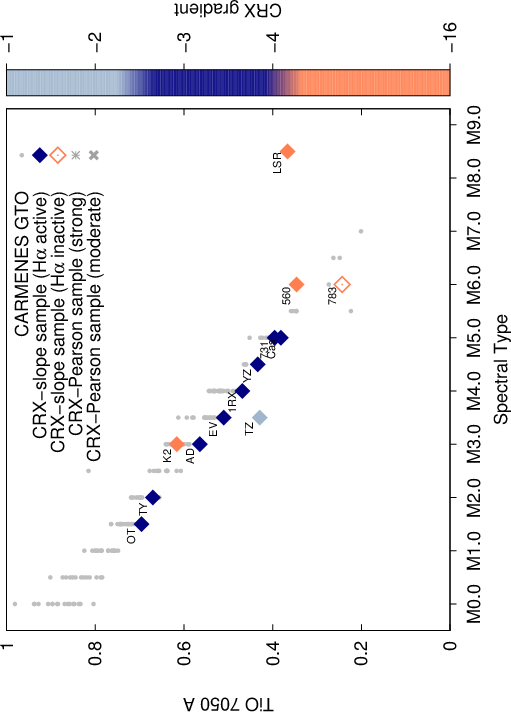}}
\imagetop{\includegraphics[angle=270,width=0.43\textwidth]{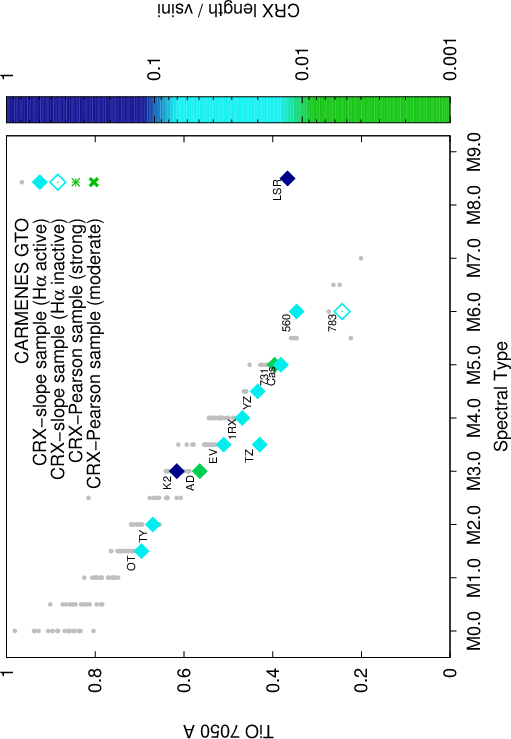}}
}
\caption{Variation of TiO 7050 index as a function of stellar spectral type for the CARMENES GTO sample with $N_{\rm obs}$ $>$ 25.  The CRX-slope sample of stars is indicated by diamonds which are filled for H$\alpha$ active stars and open for H$\alpha$ inactive stars.  The CRX-slope sample stars are coloured depending on CRX-gradient (left panel) and CRX-length (right panel). The stars are labelled as indicated in Table~\ref{tab:CRX-param}.}
\label{fig:SPT_TiO}
\end{figure*}

\subsection{Noteworthy points}

Our aim has been to present the results of the CRX-gradient and length in the context of stellar activity in a clear and concise manner.  In addition to the points already discussed, there are a few additional noteworthy points

\begin{itemize}
    \item {\underline{Ca~{\sc ii} H\&K:}} We also investigated the dependence of the CRX on the Ca~{\sc ii} H\&K lines as derived for the CARMENES GTO sample by \cite{Perdelwitz2021A&A...652A.116P}.  However, there were only a few measurements available for the stars in the CRX-all sample from archival observations and it was not possible to determine any distinct correlations or trends; therefore these plots are not shown.
    \item {\underline{X-ray:}} As summarised in Tables~\ref{tab:CRXgrad-activity}, and ~\ref{tab:CRXlength-activity}, X-ray data were only available for a small number of stars and it was not 
    possible to derive any trends or correlations.  As part of determining the log($L_{\rm{x}}$/$L_{\rm{bol}}$) relation as a 
    function of Rossby number, we noticed that the fit parameters 
    provided by \cite{Wright2018MNRAS.479.2351W} in their equations 5 and 6 contain typos and are not the values derived from fitting the data in their Table 3. Notably, \cite{Reiners2022A&A...662A..41R} independently found that there was an offset with the $\log{\tau}$ values from \cite{Wright2018MNRAS.479.2351W} for very low mass stars.  The correct scaling for the empirical calibration data of \cite{Wright2018MNRAS.479.2351W} is as follows (and is an update to their Equations 5 and 6):

\begin{equation}
\begin{array}{ll}
\log{\tau} = (0.58 \pm 0.04) + (0.28 \pm 0.01) (V-Ks)
\label{eqn:Updated_log_tau}
\end{array}
\end{equation}

\begin{equation}
\begin{array}{ll}
\log{\tau} &= (2.30 \pm0.06) - (1.38 \pm 0.245) (M/M_\odot) \\
& + (0.22 \pm 0.19) (M/M_\odot)^2
\label{eqn:Updated_log_tau_mass}
\end{array}
\end{equation}

where log\,$\tau$ is the convective turnover time and $(V-Ks)$ is the magnitude difference between the $V$ and $Ks$ bands.  The difference in the Rossby numbers derived using the fit parameters directly taken from \cite{Wright2018MNRAS.479.2351W} and Equation~\ref{eqn:Updated_log_tau} is illustrated in Figure~\ref{fig:Xray_diff} for the CARMENES GTO sample.
\end{itemize}

\begin{figure}
\includegraphics[angle=270,width=0.43\textwidth]{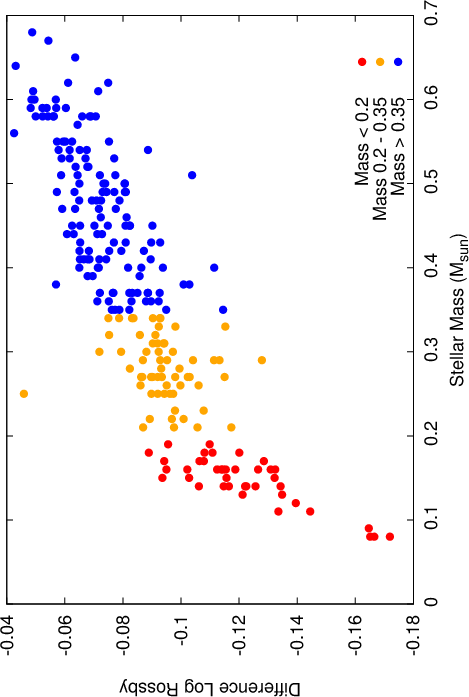}
\caption{Difference in Rossby number as a function of stellar mass derived using the fit in Equation~\ref{eqn:Updated_log_tau} and the parameters published in \cite{Wright2018MNRAS.479.2351W} (their equations 5 and 6).  Points are coloured by stellar mass following ~\cite{Shan2024A&A...684A...9S}.}
\label{fig:Xray_diff}
\end{figure}

\begin{figure*}
\centering
\mbox{
\imagetop{\includegraphics[angle=270,width=0.43\textwidth]{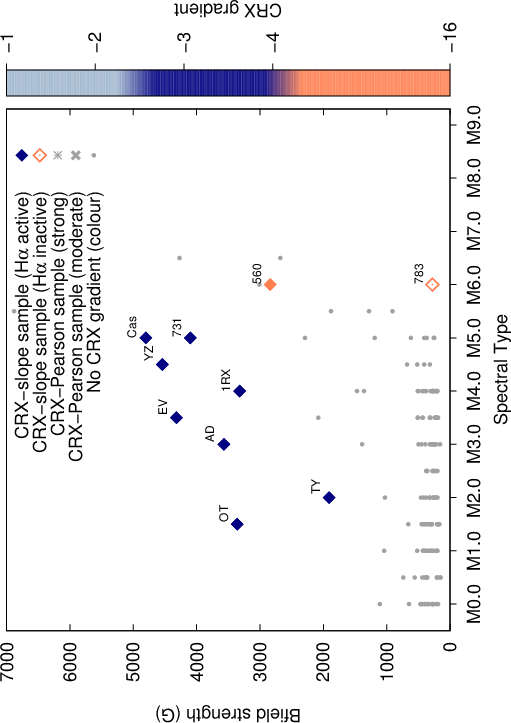}}
\imagetop{\includegraphics[angle=270,width=0.43\textwidth]{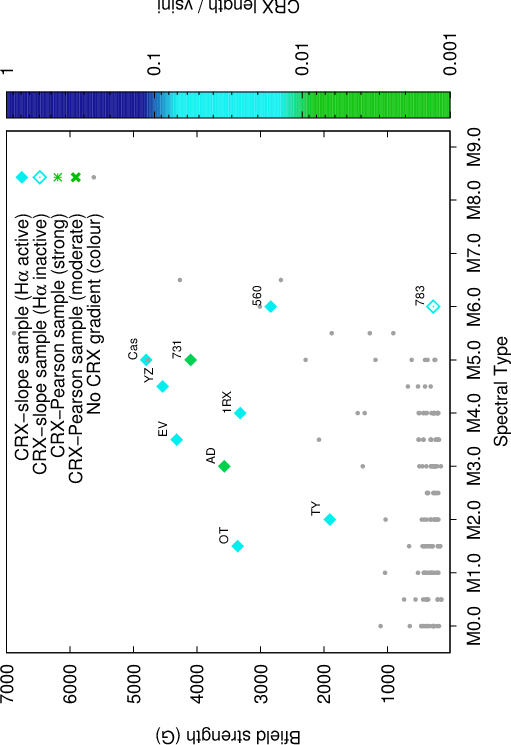}}}
\imagetop{\includegraphics[angle=270,width=0.43\textwidth]{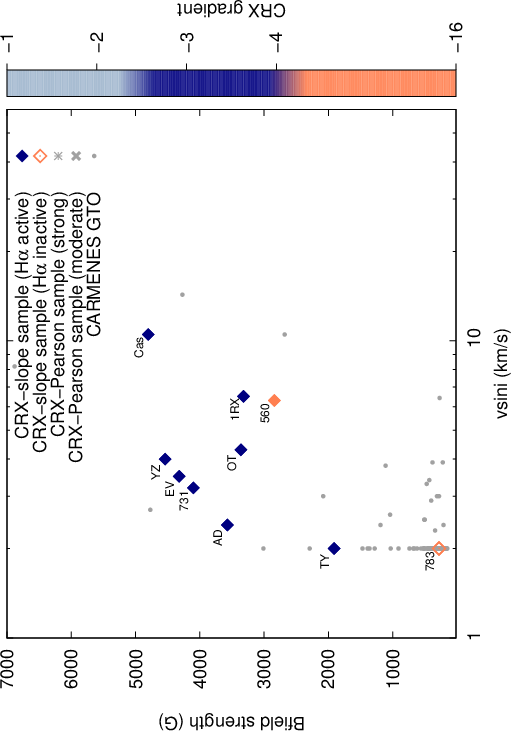}}
\imagetop{\includegraphics[angle=270,width=0.43\textwidth]{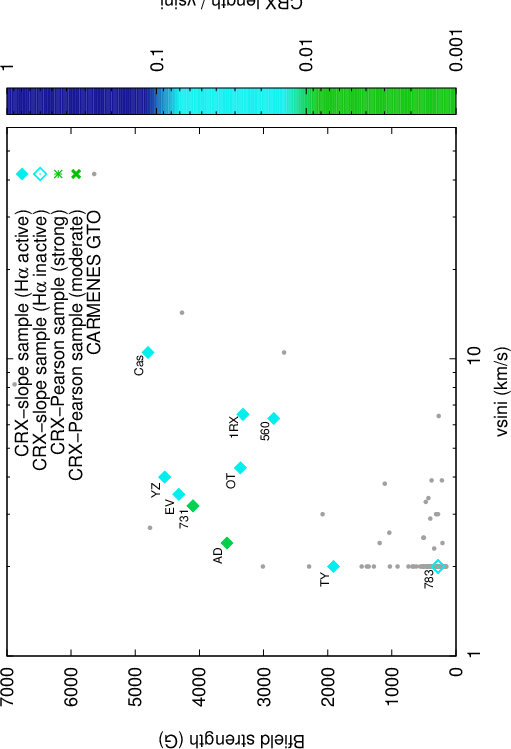}}
\caption{Average magnetic field strength as a function of spectral type (upper panels) and \vsini\ (lower panels) for the CARMENES GTO sample with $N_{\rm obs}$ $>$ 25.  Stars in the CRX-slope sample are indicated by diamonds and are coloured according to CRX-gradient (left panel) and CRX-length (right panel). The minimum \vsini\ value plotted is 2 \kms. The stars are labelled as indicated in Table~\ref{tab:CRX-param}.}
\label{fig:Bfield_SPT}
\end{figure*}

\subsection{Stars not showing CRX-gradient}

To understand why some stars show a clear CRX slope, or have a statistically significant correlation between CRX and RV, and other stars not, we inspected the CRX-index versus RV relations of seven stars that are not part of the CRX-all sample.  These stars were selected from regions that did not contain any stars from the CRX-all sample in (i) the \vsini~versus spectral type and (ii) the H$\alpha$ versus spectral type plots.  The stars are listed in Table~\ref{tab:noCRX_param} and their CRX-index versus RV relations are shown in Appendix ~\ref{fig:nonCRX_stars}.  The two stars, Teegarden's star and K2-18 are exoplanet hosts \citep{Zechmeister2019A&A...627A..49Z, Sarkis2018AJ....155..257S, Plavchan2020Natur.582..497P,Martioli2021A&A...649A.177M,Wittrock2022AJ....164...27W, Dreizler2024A&A...684A.117D}. For both Teegarden's star and K2-18, the planet-induced RV amplitude is a large fraction of the CRX-length value. In particular, K2-18 has been reported by \cite{Radica2022MNRAS.517.5050R} to exhibit a wavelength dependence in RV based on the line-by-line method, but this below the Pearson's $r$ threshold of 0.5 that we consider in this work.  

\begin{table*}
 \caption{\label{tab:noCRX_param}Sample of stars not showing a CRX slope.}
 \centering
 \begin{tabular}{ll c r r c}
 \hline
 \hline
\noalign{\smallskip}
Karmn & Name & Sp. & $v \sin{i}$ & $P_{\rm rot}$ & $\log{L_{{\rm H}\alpha} / L_{\rm bol}}$ \\
& & type & (km\,s$^{-1}$) & (d) & \\
\noalign{\smallskip}
\hline
\noalign{\smallskip}
J02530+168 & Teegarden's Star & M7.0\,V & $<$2.0 & 96.00 & --5.40 \\ 
J05084-210 & 2MASS J05082729--2101444 & M5.0\,V & 25.2 & 0.28 & --3.28 \\ 
J06024+498 & G 192-015 & M5.0\,V & $<$2.0 & 105.00 & Inact. \\
J07361-031 & BD-02 2198 & M1.0\,V & 3.1 & 12.2 & --4.28 \\
J11302+076 & K2-18 & M2.5\,V & $<$2.0 & 38.83 & Inact. \\
J14257+236 & BD+24 2733 & M0.5\,V & $<$2.0 & 17.60 & Inact. \\
J19346+045 & BD+04 4157 & M0.0\,V & 3.9 & 12.90 & Inact.\\
\noalign{\smallskip}
\hline
\end{tabular}
    \tablebib{{The full list of stellar parameters for each of these stars is listed in the Appendix Table B.1.  The rotation period of Teegarden's star is taken from \cite{Dreizler2024A&A...684A.117D}. CRX-RV relations for each of these stars are plotted in Figure~\ref{fig:nonCRX_stars}.}}
\end{table*}

\section{Results: Removing stellar activity from RVs using CRX}

The RV signatures resulting from presence of stellar activity such as spots and plage regions is wavelength dependent, due to the decrease of the photospheric / spot contrast, whereas a planetary companion would induce a RV signal that is wavelength independent.  In this section we use the wavelength dependent information parameterised by the CRX to remove the signatures of stellar activity from the CRX-all sample of 39 stars.    The subtraction of the CRX, via a linear fit to the CRX-RV anti-correlation, from the RV results in a decrease in the RV rms by a factor of up to 3.89 in the case of 1RXS J050156.7+4233 (labeled: 1RX).  The results for the CRX-all sample are listed in Table~\ref{tab:CRX-param}. We find a total of 9 stars or 25\% of the CRX-all sample have a factor of improvement greater than 2.0.

There is a strong dependence of the factor of improvement (FoI) with Pearson's $r$ value as shown in the top left-hand panel of Figure~\ref{fig:Factor_Imp}.  A linear fit shows that this relation can be equated as follows:

\begin{equation}
\begin{array}{ll}
  \mathrm{FoI} = -18.51 r -14.48   & (r < -0.80) \\
  \mathrm{FoI} = -1.98 r + 0.04   & (r > -0.80)
\end{array}
\end{equation}

where FoI is the factor of improvement and r is the Pearson's r value.  Only 3 stars show an RV rms reduction that is greater than a factor of 3. These are 1RXS J050156.7+010845, YZ CMi and EV Lac, which will be referred to as the RV$_{\rm corr}$3} stars in the rest of this work.  All three stars have mid-M spectral types and are located on the saturated part of the rotation-activity correlation (Figure~\ref{fig:Halpha_vsini}).  For mid-M spectral types, this occurs at higher normalised H$\alpha$ values than for the latest M dwarf spectral types.   

The correlation of the Factor of Improvement with stellar rotation period is shown in Figure~\ref{fig:Factor_Imp} (upper right panel).  The highest Factor of Improvement values are shown for fast to moderately rotating stars with rotation periods of approximately a few days to just less than 10 days.  This is likely because such moderately rotating stars have more stable spot patterns.  

The TiO indicator can be considered as an indicator of dark starspot coverage, even on slowly rotating and less active stars
~\cite{Vogt1979PASP...91..616V,Amado2002A&A...381..517A,Oneal2004AJ....128.1802O}.
To remove the spectral type dependence that we previously showed in Figure~\ref{fig:SPT_TiO}, we plot the maximum minus the minimum value for each star (Figure~\ref{fig:Factor_Imp} middle left panel).  The \FoIthree stars show moderate values, while the cluster of low \vsini\ stars with very small Factor of Improvement values show very small TiO 7050 variations.

Furthermore, we show the factor of improvement as a function of normalised H$\alpha$ in Figure~\ref{fig:Factor_Imp} (middle row right panel).  Notably, the \FoIthree\ stars have very high levels of H$\alpha$ activity.  The remaining stars show H$\alpha$ values spanning the full range from low to high levels of H$\alpha$ activity.    

Finally in the lower panels of Figure~\ref{fig:Factor_Imp} we investigate the dependence of the Factor of Improvement value on the CRX-length / \vsini\ and spot contrast ratio (left and right panels respectively).  The factor of improvement increases with increasing  CRX-length / \vsini\ values, though importantly the \FoIthree\ stars don't follow this trend with the highest factor of improvement values and moderate CRX-length / \vsini\ values.  The lowest 
\vsini\ stars have the lowest factor of improvement values and the lowest CRX-length / \vsini\ measurements.   The variation of the factor of improvement with spot contrast ratio shows a general trend of increasing factor of improvement values with decreasing spot contrast ratios.  The highest factor of improvement values occurring for moderate spot coverage ratios that are typical for mid-M spectral types.  

\begin{figure*}
\centering
\mbox{
\imagetop{\includegraphics[angle=270,width=0.43\textwidth]{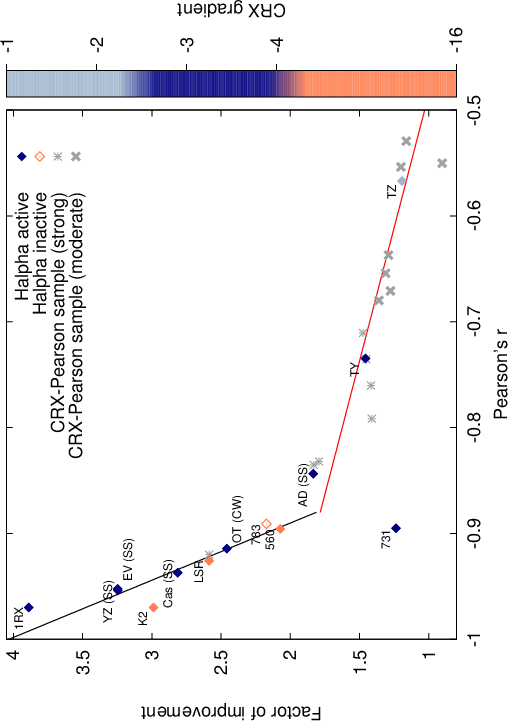}}
\imagetop{\includegraphics[angle=270,width=0.43\textwidth]{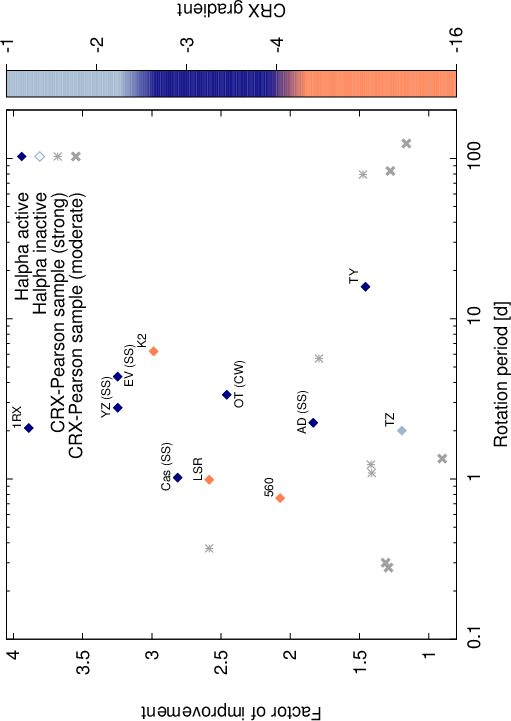}}
}
\mbox{
\imagetop{\includegraphics[angle=270,width=0.43\textwidth]{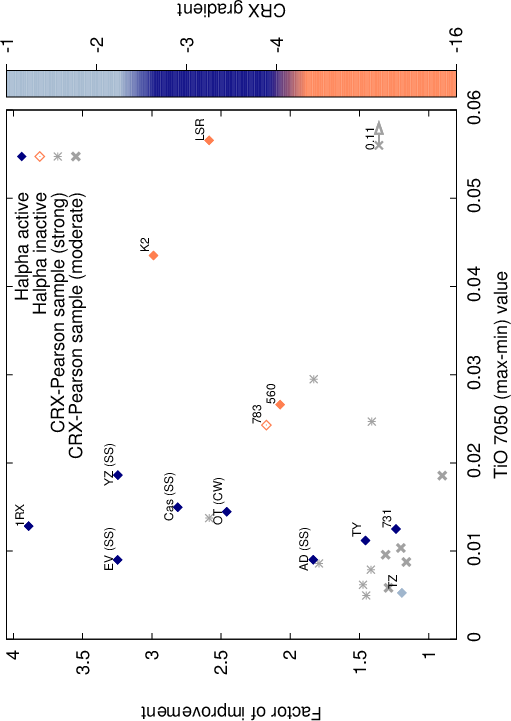}}
\imagetop{\includegraphics[angle=270,width=0.43\textwidth]{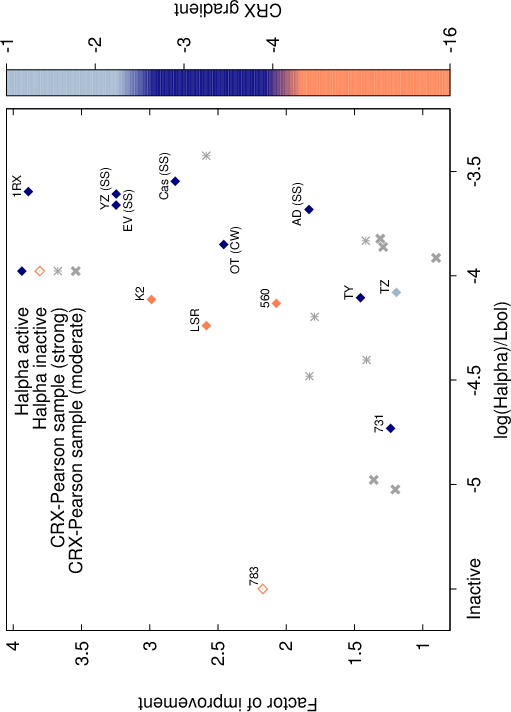}}
}
\mbox{
\imagetop{\includegraphics[angle=270,width=0.43\textwidth]{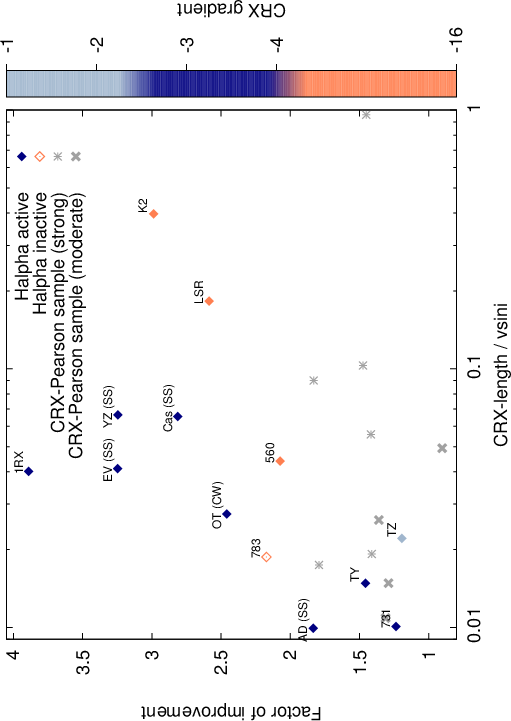}}
\imagetop{\includegraphics[angle=270,width=0.43\textwidth]{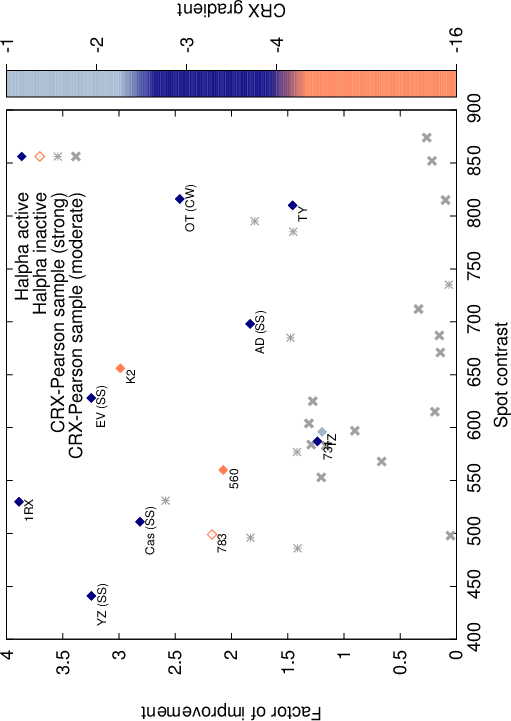}}
}
\caption{Factor of improvement in the RV rms from the subtraction of the RV estimated from the RV-CRX correlation for each star in the CRX-all sample. Filled points indicate H$\alpha$ active stars while open points show H$\alpha$ inactive stars.  Symbols are coloured based on CRX-gradient value or none (grey points).  The factor of improvement is shown as a function of Pearson's $r$, stellar rotation period (upper row, left and right), stellar TiO7050 max-min value (using 95\% percentiles) and normalised H$\alpha$ luminosity (middle row row left and right) and CRX-length / \vsini\, and (zoomed) spot contrast ratio (lower row left and right).  The stars are labelled as indicated in Table~\ref{tab:CRX-param}.}
\label{fig:Factor_Imp}
\end{figure*}

\section{Discussion}
 
The advantage of the CARMENES spectrograph is its large wavelength coverage and high RV precision.  In this work we have investigated the wavelength dependence of RVs for the \cargto\ sample of 215 stars. In this section we discuss our results.

\subsection{CRX-gradient and CRX-length as a function of stellar parameters}

Our results show that from the \cargto\ sample of 215 stars that 39 stars show a strong or a moderate correlation between the CRX and the RV (CRX-all stars).  Furthermore, a subset of 13 of these stars show a tight linear relationship (CRX-slope stars).  

We demonstrate that the larger the variation is in CRX-length / \vsini\, the larger the variation is in CRX (i.e.$\Delta CRX$).  The highest variations in CRX-length and CRX tend to occur for the least massive stars with the steepest CRX-gradients where their activity levels are saturated.  The caveat with this is that these stars have very little flux at bluer wavelenghts and may be biased.  However, the very active star K2-22 also shows a very steep slope but has a spectral type of M 3.0 V.  This shows that our methods can detect a CRX-RV correlation even for more rapidly rotating stars with high levels of stellar activity.   

There is also a strong dependence of the CRX-length / \vsini\ on the Pearson's r value with the slowest rotating stars having the lowest  CRX-length / \vsini\ values and at most moderate Pearson's r values for the CRX-RV relation.  This is in agreement with the previous work of \cite{TalOR2018A&A...614A.122T}.  There is also a correlation of the  CRX-length / \vsini\ with spot to photosphere temperature contrasts where stars with lower contrasts (e.g. at late M spectral types) show higher CRX-length /\vsini\ values.    While these results are valid for our sample of stars, a true determination of the inter-dependency between CRX-length / \vsini\ and the stellar spot to photosphere temperature contrast is only possible knowing the distribution and evolution of spots and other features of stellar activity on the stellar surface.  As noted by \cite{Barnes2011MNRAS.412.1599B}, the latitude of a starspot (or other activity signature) will impact the activity induced RV, with higher latitude spots having a lower RV than equatorial spots.  Similarly, the latitude of spots will impact the lemniscate shape ~\citep{Boisse2011A&A...528A...4B}. We already know from our extensive investigation of stellar activity the CARMENES consortium, that every M dwarf is an individual when it comes to its activity patterns and evolution timescales. 

\subsubsection{Dependence of CRX-gradient on stellar activity}

Since there is only one star with a classification of having a shallow CRX-gradient, we restrict the discussion of our results to the stars with moderate and steep slopes.   The stars in the CRX-slope sample with moderate CRX-gradients cover a range in spectral type from early-M to mid M as indicated by the navy points in Figures~\ref{fig:Vsini_SPT} to ~\ref{fig:Bfield_SPT}.  These stars are active stars as evidenced by: (i) higher \vsini\ values of up to 5 \kms\ for stars with early-M spectral types and a range of \vsini\ values from 2 to 11 \kms\ at mid-M spectral types (Figure~\ref{fig:Vsini_SPT}), (ii) a range in rotation periods from 1 to 11 days,  (iii) the highest $\log( L_{{\rm H}\rm\alpha}/L_{\rm bol})$ values (Figure~\ref{fig:Halpha_SPT}) in the CRX-all sample of 39 stars (iv) located in the upper part of the $\log( L_{{\rm H}\rm\alpha}/L_{\rm bol})$ as a function of \vsini\ (Figure~\ref{fig:Halpha_vsini}), (v)  lowest TiO7050 values per spectral type bin (Figure~\ref{fig:SPT_TiO}), (vi) saturation in X-rays (not shown), (vi) highest average magnetic field strengths as a function of spectral type and \vsini\ (Figure~\ref{fig:Bfield_SPT}, upper and lower panels respectively).  They also have, for our M-dwarf sample, comparatively high to moderate $\Delta T$ spot/photosphere temperature contrasts.

There are only a few stars with high CRX-gradient values and the latest spectral types of the CRX-slope sample.  Despite their high-CRX-gradients, these stars exhibit moderate levels of activity (i) they have moderate to high \vsini\ levels (Figure~\ref{fig:Vsini_SPT}), (ii) are the most H$\alpha$ active stars in the \cargto-sample for the latest spectral types (Figure~\ref{fig:Halpha_SPT}), (iii) are located in the active part of the $\log( L_{{\rm H}\rm\alpha}/L_{\rm bol})$ as a function of \vsini\ (Figure~\ref{fig:Halpha_vsini}), (iv) are located in the `hook' part of the TiO7050 per spectral type bin (Figure~\ref{fig:SPT_TiO}), (v) show moderate average magnetic field strengths  as a function of spectral type and \vsini\ (Figure~\ref{fig:Bfield_SPT}, upper and lower panels respectively).   Their moderate activity is because late-M dwarfs saturate at lower activity levels compared to early- and mid-M dwarfs \citep[for more details see][]{Mohanty2003ApJ...583..451M}.    These stars have the smallest $\Delta T$ spot/photosphere temperature contrasts and previously explained, we advise caution in the literal interpretation of these steep slopes as they may be biased due to the lack of flux at blue orders compared to the rest of the sample.

\subsubsection{Dependence of CRX-length on stellar activity}
The stars with the shortest CRX-length values have mid-M spectral types and show low to moderate levels of stellar activity as evidenced by (i) their low \vsini\ values (Figure~\ref{fig:Vsini_SPT}) and a range of moderate rotation periods (Figure~\ref{fig:CRXlength_Pearson}, (ii)  are the least H$\alpha$ active stars (Figure~\ref{fig:Halpha_SPT}), (iii) are located in the unsaturated part of the rotation-activity relation (Figure~\ref{fig:Halpha_vsini}), (iv) have low to moderate average magnetic field strengths and are located on the unsaturated part of the magnetic field - rotation relation (Figure~\ref{fig:Bfield_SPT}, lower and upper panels respectively).  

Stars with moderate CRX-length values typically are early- to mid- M dwarfs with moderate to high levels of stellar activity.  This is shown by (i) their increased \vsini\ levels and short rotation periods (Figure~\ref{fig:Vsini_SPT}), (ii) generally the H$\alpha$ activity levels for their spectral types (Figure~\ref{fig:Halpha_SPT}), (iii) that they are mostly located on the saturated part of the rotation-activity relation (Figure~\ref{fig:Halpha_vsini}) and X-ray distributions (not shown), (iv) highest average magnetic field strengths for their spectral types (Figure~\ref{fig:Bfield_SPT} upper and lower panels).   

As we know that the late-M dwarfs saturate in activity we already know what to expect.  The CRX-length is longest for the stars with mid- to late-M spectral types.  Compared to the \cargto\ sample, they are not the most active stars, in both H$\alpha$ and average magnetic field strengths, for their spectral type, but they do have moderate to fast \vsini\ values and short rotation periods (Figure~\ref{fig:Vsini_SPT}).  Their normalised H$\alpha$ levels are close to or above the blue line (indicating very high levels of H$\alpha$ activity) in Figure~\ref{fig:Halpha_SPT} and are on the saturated part of the rotation-activity relation Figure ~\ref{fig:Halpha_vsini} and ~\ref{fig:Bfield_SPT}.  They also cover the `hook' in the TiO distribution with spectral type.

\subsection{Average magnetic field strength}

The average magnetic field strengths were computed from the telluric corrected co-added templates with exceptional high signal to noise values \citep{Reiners2022A&A...662A..41R}.  The CRX-gradient shows moderate values for the stars with the highest average magnetic field strengths at mid-M spectral types.  The average magnetic field strength have been shown by \cite{Haywood2014MNRAS.443.2517H,Haywood2022ApJ...935....6H} to be a proxy of RV variations in the Sun and other stars in agreement with \cite{Meunier2010A&A...512A..39M}.   Recently, the work of ~\cite{Ruh24} investigated the RV jitter and its relation to the average magnetic field strength.  They showed that M dwarfs with excess activity-rotation induced RV jitter have magnetic field filling factors that are dominated by a component with a strength of 2 to 4 kG.

\subsection{Correlations with Bisector Inverse Span (BIS)}

All of the CRX-all stars have a strong or moderate (anti-) correlations between CRX and Moment 1 indicators such as RV.    Recent models of stellar activity, and its impact on the line moments ~\cite{Barnes2024MNRAS.534.1257B}, show that there is a high likelihood of having correlations between odd numbered moments such as Moment 1 (e.g. RV) and Moment 3 (e.g. the BIS).    While the CRX-index and BIS measure different aspects of the stellar spectra, it is important to also note that the BIS also includes a chromatic effect.  This is because the skewness of the line profile is caused by the different velocity patterns as a function of formation depth, with redder lines being weaker. 

Given the similarity of the lemniscate shape of the CRX and the BIS ~\citep[see for e.g.][]{Boisse2011A&A...528A...4B}, we additionally investigate the statistical correlation between the CRX and BIS (Moment 3 line moment).  We use Pearson's r for the 39 stars in the CRX-all sample.  Our results are shown in Table~\ref{tab:CRX-param} and confirm that Moment 1 and Moment 3 are linked with 13 stars having either a strong or a moderate Pearson's r value between the CRX and BIS.  The 3 stars that show the highest factor of improvement also show a strong correlation between CRX and BIS  (see Appendix Table A.1).   This is in agreement with our previous observational results for the mid M dwarf EV Lac ~\citep{JeffersEVLac2022A&A...663A..27J}, the results of \cite{Schoefer2022A&A...663A..68S} for four stars in the CARMENES sample.  

Recently, ~\cite{Lafarga2021A&A...652A..28L} investigated the optimal activity indicators to identify activity induced RV signals in M dwarf stars with a range of activity levels.  They report that the CRX and BIS are the most effective indicators for tracing activity in high-mass and high-activity  (measured by H$\alpha$) M dwarfs.  A total of 9 CRX-slope and 4 CRX-Pearson stars are also included in their sample, where for each of these star the stellar rotation period is detected in the RV (9 CRX-slope, 4 CRX-Pearson) and the CRX (9 CRX-slope, 1 CRX-Pearson).  A total of 6 CRX-slope stars and 3 CRX-Pearson stars have both the a periodic signal detected at the stellar rotation period in BIS.  

\begin{figure}
\includegraphics[angle=270,width=0.43\textwidth]{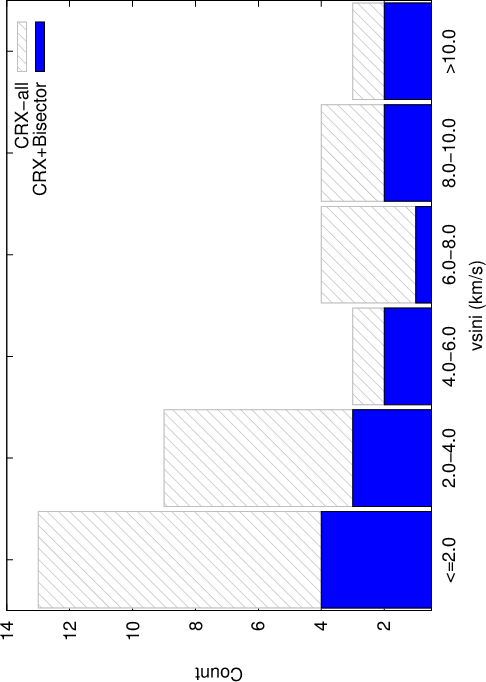}
\caption{The \vsini\ distribution of all 39 stars in the CRX-all sample (CRX-slope and CRX-Pearson) shown as grey shaded bars.  The stars with a strong or a moderate correlation of the CRX with BIS are shown in blue.    }
\label{fig:CRX_BIS_hist}
\end{figure}

Previously, the results of \cite{Saar1997ApJ...485..319S} reported that the BIS span varies as (\vsin)$^{3.3}$.
In Figure~\ref{fig:CRX_BIS_hist} we show the distribution of \vsini\ of the CRX-all sample where the distribution of the stars with a strong or a moderate correlation between the CRX-index and the BIS are highlighted. 
Our results show that BIS is a significant parameter, at moderate \vsini\ rates and even at low \vsini\ levels.  While models of fixed starspots are important, a comprehensive analysis also needs to take into account the variations in the latitude, longitude and lifetimes of the spot distributions over the time span of data collection.

\begin{table*} 
\centering
\caption{Summary of relations derived in this work \label{tab:equn}}    
\begin{tabular}{r@{~}c@{~}lcc}  
\hline
\multicolumn{3}{c}{Relation} & Figure & Equ No (text)\\
\hline
$\Delta {T} $ &=& $ 0.47\,T_{\rm phot} -926.97$ & 3 & 1 \\
$\log(CRX_{\rm max}-CRX_{\rm min}) $&=&$ 1.08 \log(RV_{\rm max}-RV_{\rm min}) + 0.336 $& 5 & ...\\ 
$\log(CRX_{\rm length}/\vsini) $&=&$ -1.71 r -2.846 $ &   6 & ...\\
$\log(CRX_{\rm length}/\vsini) $&=&$ 9.7 \times 10^{-4} 
\Delta T - 0.967$ &  6 &...\\
$\log(CRX_{\rm length}/\vsini) $&=&$ -0.283 \log P_{\rm rot} -1.585 $ &  6 & ...\\
$\log{\tau} $&=&$ (0.58 \pm 0.04) + (0.28 \pm 0.01) (V-Ks)$ & 12 & 2 \\
$\log{\tau} $&=&$ (2.30 \pm0.06) - (1.38 \pm 0.245) (M/M_\odot)  + (0.22 \pm 0.19) (M/M_\odot)^2$ & 12 & 3 \\
$\mbox{Factor of Improvement} $&=&$ -18.51 r -14.48 \hspace{1cm}  (r < -0.80) $& 14 & 4\\
\hspace{3.25cm} &=&$ -1.98 r + 0.04   \hspace{1.35cm} (r > -0.80) $ & 14 & 4\\
\hline
\end{tabular}
 \tablefoot{where $\Delta T$ is the Spot / Photospheric temperature contrast ratio and $r$ is Pearson's r}
\end{table*}

\subsection{Subtraction of the CRX}

The subtraction of a linear fit of the CRX-RV anti-correlation from the RV results in a reduction of the RV rms by a factor of up to 4.  There is a tight correlation of factor of improvement with Pearson's r for both the CRX-slope and Pearson-all samples, where the best stars typically have rotation periods of several days (Figure~\ref{fig:Factor_Imp} top panels), low levels of $\Delta$ TiO and very high H$_\alpha$ values (Figure~\ref{fig:Factor_Imp} middle panels)
and moderate CRX-length / \vsini\ values and low to moderate M dwarf spot contrasts.    The moderate rotation periods and low $\Delta$ TiO values point to stable spot feature, though the high H$_\alpha$ values could result from plage regions appearing dark at the limb centre as suggested by ~\cite{Johnson2021MNRAS.504.4751J}.

Whether the star has a moderate or a steep CRX-gradient does not influece the factor of improvement values.  However, in both the $\Delta$ TiO and the H$\alpha$ activity relations, the CRX-slope stars with the steepest gradients occupy a separate branch or region compared to the CRX-slope stars with moderate gradients.  Furthermore, we show that all of the CRX-slope stars with moderate gradients have CRX-length / \vsini\ values that are low to moderate (with the best factor of improvement values) and that the steepest gradients have the longest lengths.  The stars with the longest CRX-length values are in the saturated regime of the rotation-activity plot and have moderate CRX slopes.  These relations are summarised in Figure~\ref{fig:Factor_Imp} and Table~\ref{tab:equn}.

\subsubsection{Planet detection thresholds}

We have demonstrated that the improvement in the RV rms by removing the contribution of the CRX can be up to a factor of nearly four.  The implications that this has for improving the detectability of small exoplanets was simulated using the dataset of EV Lac.  An exoplanetary signal is first injected  in the uncorrected data and then the activity contribution is removed ~\citep{Cardona23}. With the linear correction factor it is possible to detect exoplanets that are approximately 2.5 times less massive compared to the same dataset but without the correction.  The exoplanetary signal can be identified even if it is not a significant signal in the original dataset.   We  will explore this further in forthcoming papers. 

Other techniques that exploit the wavelength dependence of activity induced RVs include the work by \cite{Cameron2021MNRAS.505.1699C}. They present test simulations by adding injected signals of low-mass planets to 5 years of HARPS Sun-as-a-Star observations of the solar spectrum where planets can be recovered with a precision of $\sim$ 6.6 cm s$^{-1}$.   

In addition to using the wavelength dependence of RVs, there are currently a large number of methods in the literature to remove the impact of stellar activity.  An excellent summary of these methods is presented by \cite{Meunier2021arXiv210406072M} (see their Table 1).  Also \cite{Zhao2022AJ....163..171Z} presented the results of a comparison of different methods (expanding from the work of \cite{Dumusque2017A&A...598A.133D}) and concluded that no method is yet performing better than classical linear decorrelation, further supporting the results we obtained in this work.

\subsubsection{Surface mapping}
  
The large-scale magnetic field geometry has been reconstructed for a total of 6 stars in the CRX-all sample.  From these 6 stars, YZ CMi and EV Lac are also part of the \FoIthree\ stars with the highest reduction in RV rms after the subtraction of the CRX. V388 Cas, AD Leo and OT Ser are CRX-slope stars and AU Mic is a CRX-Pearson (moderate) star.   
The mid-M dwarfs YZ CMi and EV Lac consistently show an axisymmetric, predominantly poloidal and  simple large-scale magnetic field geometry that does not change significantly over time periods of approximately 1 year \citep{Donati2008MNRAS.390..545D, Morin2008MNRAS.390..567M, Morin2010MNRAS.407.2269M}.  The long-term stability of YZ CMi enabled \cite{Baroch2020A&A...641A..69B} to fit CARMENES data with one stable starspot.   Recently, \cite{JeffersEVLac2022A&A...663A..27J} also reconstructed several low resolution surface brightness maps of EV Lac using CARMENES data, and also found a long-term stable component with small-scale variability which is consistent with EV Lac's TESS light curve.

The remaining 3 stars in the CRX-all sample with large-scale magnetic field maps show a variety of geometries. 
The mid- M dwarf V388~Cas exhibits a comparably simple large-scale magnetic field geometry as reconstructed for YZ~CMi and EV~Lac which is also axisymmetric and predominantly poloidal.  The large-scale magnetic field geometry of the mid-M dwarf AD~Leo shows strong and simple magnetic features, which are mostly poloidal and axisymmetric and with low level variations over the time span of 1 year.  On the other hand, the early-M dwarf OT~Ser shows latitudinal rings of mixed polarities that are axisymmetric and dominated by the poloidal component but with also a significant toroidal component \citep{Donati2008MNRAS.390..545D,Morin2008MNRAS.390..567M, Morin2010MNRAS.407.2269M}.

The active planet-hosting star AU~Mic, which is a CRX-Pearson moderate star in this work, has been the focus of an intensive observational campaign to reconstruct its surface activity patterns \citep{Klein2021MNRAS.502..188K,Klein2022MNRAS.512.5067K}.  AU~Mic has surface brightness features that evolve rapidly, which is consistent with its young age of 22 Myr and frequent flaring events \citep{Robinson2001ApJ...554..368R}.  It cannot be excluded that the reason why AU~Mic is not a CRX-slope star is because its stellar activity features evolved on  a much more rapid timescale compared to the observational cadence of the CARMENES data of AU~Mic and which may explain the curved nature of AU~Mic's CRX-RV relation (see Figure~\ref{fig:CRX-Pearson-mod}) and \citep{Cale2021AJ....162..295C}.

These Zeeman Doppler Imaging (ZDI) observations point to that (1) early-M dwarfs have weak but complex fields and consequently activity patterns, (2) mid-M dwarfs have less complex but strong magnetic fields and (3) late-M dwarfs can have either of these configurations.    The results from brightness (Doppler) imaging typically exhibit more complex spot patterns for all M spectral types \citep[][for example]{Barnes2001MNRAS.326..950B,barnes04hkaqr, barnes15mdwarfs} but these results are biased towards stars with \vsini\ values that are in excess of 15--20\kms. 

\subsection{Sampling of data}

One point to consider is that the observations of the CARMENES GTO stars have been secured with the main aim of detecting close-in orbiting exoplanets and not characterising the stellar activity of the target stars.  This means that there can be gaps in the observational timeseries of longer than several stellar rotation periods which is the timescale that we expect stellar activity features such as spots and plage region to evolve. 

If the stellar activity features, and consequently the CRX, have evolved over this time frame then determining correlations and removing the contribution of the CRX will be biased by the time sampling of the dataset.   For example, the very detailed analysis of EV Lac by \cite{JeffersEVLac2022A&A...663A..27J} showed that it is possible to see additional (anti-)correlations between the line moments or parameters with smaller data sets that cover only a few stellar rotation periods and where the features of stellar activity have not significantly evolved.   In this work we have shown that the subtraction of the CRX from the RVs works optimally for moderate rotation periods, which is likely to result from a combination of stable global starspot patterns which are more likely to be well sampled by our CARMENES data. 

For very active stars, with higher rotation periods, this could mean that the stellar activity features have evolved significantly over the timespan of the observations whereas the more stable activity patterns are less impacted by the cadence of the observations. In the future we will investigate if the more rapidly evolving stellar activity can be mitigated using densely sampled observations covering a large wavelength range, such is as used by the RedDots exoplanet search program \citep{Jeffers2020Sci...368.1477J}.

\section{Conclusions}

In this work we have investigated the wavelength dependence of high-precision RV measurements via the CRX.   We conclude the following:
\begin{itemize}
    \item Approximately 17\% of stars in the Carmenes GTO sample are CRX-all stars (i.e. stars with strong-moderate RV Pearson's $r$ correlations of CRX-index vs RV).
    \item Dependence on stellar activity diagnostics shows that CRX-all stars have low to moderate activity levels and cover the full range of M dwarf spectral types from M1.5 to M8.5.
    \item The variation of RVs with wavelength can be used to remove the impact of stellar activity with an improvement in RV rms of up to a factor of 4. The optimal targets are moderately active stars. 
    \item The correction effectiveness is related to how stable and structured the activity patterns are, though more active stars would be feasible with regular cadence or high density sampled high-precision RVs.
\end{itemize}
The power of using the wavelength dependent information contained in the spectra themselves is that it is an instantaneous decorrelation.

\section{Data availability}

The Figures contained in Appendix A are only available online at
https://zenodo.org/records/14013877.

\begin{acknowledgements}

We would like to thank the referee for their very insightful and constructive comments which helped to improve the clarity of the paper. 
  This publication is based on observations collected under the CARMENES Legacy+ project.
  
  CARMENES is an instrument at the Centro Astron\'omico Hispano en Andaluc\'ia (CAHA) at Calar Alto (Almer\'{\i}a, Spain), operated jointly by the Junta de Andaluc\'ia and the Instituto de Astrof\'isica de Andaluc\'ia (CSIC).
    
  CARMENES was funded by the Max-Planck-Gesellschaft (MPG), 
  the Consejo Superior de Investigaciones Cient\'{\i}ficas (CSIC),
  the Ministerio de Econom\'ia y Competitividad (MINECO) and the European Regional Development Fund (ERDF) through projects FICTS-2011-02, ICTS-2017-07-CAHA-4, and CAHA16-CE-3978, 
  and the members of the CARMENES Consortium 
  (Max-Planck-Institut f\"ur Astronomie,
  Instituto de Astrof\'{\i}sica de Andaluc\'{\i}a,
  Landessternwarte K\"onigstuhl,
  Institut de Ci\`encies de l'Espai,
  Institut f\"ur Astrophysik G\"ottingen,
  Universidad Complutense de Madrid,
  Th\"uringer Landessternwarte Tautenburg,
  Instituto de Astrof\'{\i}sica de Canarias,
  Hamburger Sternwarte,
  Centro de Astrobiolog\'{\i}a and
  Centro Astron\'omico Hispano-Alem\'an), 
  with additional contributions by the MINECO, 
  the Deutsche Forschungsgemeinschaft (DFG) through the Major Research Instrumentation Programme and Research Unit FOR2544 ``Blue Planets around Red Stars'', 
  the Klaus Tschira Stiftung, 
  the states of Baden-W\"urttemberg and Niedersachsen, 
  and by the Junta de Andaluc\'{\i}a.
  
  This work was based on data from the CARMENES data archive at CAB (CSIC-INTA).
  
  We acknowledge financial support from the Agencia Estatal de Investigaci\'on (AEI/10.13039/501100011033) of the Ministerio de Ciencia e Innovaci\'on and the ERDF ``A way of making Europe'' through projects 
  PID2021-125627OB-C31,		
  PID2019-109522GB-C5[1:4]	
and the Centre of Excellence ``Severo Ochoa'' and ``Mar\'ia de Maeztu'' awards to the Instituto de Astrof\'isica de Canarias (CEX2019-000920-S), Instituto de Astrof\'isica de Andaluc\'ia (CEX2021-001131-S), and Centro de Astrobiolog\'ia (MDM-2017-0737).
  This work was also funded by the Generalitat de Catalunya/CERCA programme, and the DFG through the priority program SPP 1992 `Exploring the Diversity of Extrasolar Planets' (Jeffers, JE 701/5-1).
  
  \end{acknowledgements}

\bibliographystyle{bibtex/aa.bst} 
\bibliography{bibtex/CRX_main_ALL.bib} 

\newpage
\onecolumn
\begin{appendix}
\section{Supplementary information}
 
\begin{table*}[b]
\hspace{-2.0cm}
 \caption{\label{tab:CRX-param}Derived CRX parameters of the CRX-all stars sample.}
 \centering
 \begin{tabular}{lll c rr c cc c}
 \hline
 \hline
 \noalign{\smallskip}
Karmn & Name & Label & Sp. & \multicolumn{2}{c}{CRX} & Pearson & Factor & Spot/phot. & BIS \\
& & (plots) & type & gradient & length & $r$ & RV red. & cont. & (mod / \\
& & & & (Np$^{-1}$) & (ms$^{-1}$)  & & & (K) & strong) \\

 \noalign{\smallskip}
\hline
 \noalign{\smallskip}
\multicolumn{9}{c}{CRX-slope sample} \\
 \noalign{\smallskip}
\hline
 \noalign{\smallskip}
J01033+623 & V388 Cas & Cas & M5.0\,V & --2.67 $\pm$ 0.20 & 686 & --0.94 & 2.81 & 511 & M \\
J02002+130 & TZ Ari & TZ & M3.5\,V & --1.16 $\pm$ 0.18 & 44 & --0.57 & 1.19 & 596 & ...\\
J04198+425 & LSR J0419+4233 & LSR$^\diamondsuit$ & M8.5\,V & --4.64 $\pm$ 0.33 & 658 & --0.93 & 2.59 & 202 & ...\\
J05019+011 & 1RXS J050156.7+010845 & 1RX & M4.0\,V & --3.03 $\pm$ 0.15 & 261 & --0.97 & 3.89 & 530 & S \\
J07403--174 & LP 783--002 & 783$^\diamondsuit$ & M6.0\,V & --16.98 $\pm$ 1.21 & 37 & --0.89 & 2.17 & 499 & ... \\
J07446+035 & YZ CMi & YZ & M4.5\,V & --2.92 $\pm$ 0.13 & 265 & --0.95 & 3.25 & 441 & S \\
J10196+198 & AD Leo & AD & M3.0\,V & --3.64 $\pm$ 0.25 & 43 & --0.84 & 1.83 & 698 & S \\
J10584--107 & LP 731--076 & 731 & M5.0\,V & --3.51 $\pm$ 0.69 & 32 & --0.90 & 1.24 & 587 & ...\\
J14321+081 & LP 560--035 & 560$^\diamondsuit$ & M6.0\,V & --8.68 $\pm$ 0.58 & 277 & --0.90 & 2.07 & 560 & ...\\
J15218+209 & OT Ser & OT & M1.5\,V & --3.65 $\pm$ 0.23 & 118 & --0.91 & 2.46 & 816 & S \\
J16102--193 & K2-- 33 & K2 & M3.0\,V & --8.49 $\pm$ 0.45 & 2897 & --0.97 & 2.99 & 656 & ... \\
J18174+483 & TYC 3529--1437--1 & TY & M2.0\,V & --2.97 $\pm$ 0.33 & 30 & --0.73 & 1.46 & 810 & M \\
J22468+443 & EV Lac & EV & M3.5\,V & --3.16 $\pm$ 0.10 & 144 & --0.95 & 3.25 & 628 & S \\
 \noalign{\smallskip}
\hline
 \noalign{\smallskip}
\multicolumn{9}{c}{CRX-Pearson strong} \\
 \noalign{\smallskip}
\hline
 \noalign{\smallskip}
J04173+088 & LTT 11392 & * & M4.5\,V  & ...\hspace{0.8cm}...  & 1447 & --0.92 & 2.43 & 531 & ... \\
J06396--210 & LP 780--032 & * & M4.0\,V  & ...\hspace{0.8cm}...  & 206 & --0.71 & 1.47 & 685 & M \\
J08536--034 & LP 666--009 & * & M9.0\,V  & ...\hspace{0.8cm}...  & 588 & --0.86 & ... & 202 & ...  \\
J09033+056 & NLTT 20861 & * & M7.0\,V  & ...\hspace{0.8cm}...  & 874 & --0.84 & 1.83 & 496 & S\\
J11044+304 & LSPM J1104+3027 & * & M3.0\,V  & ...\hspace{0.8cm}...  & 105 & --0.73 & ... & 735 & ...\\
J11201--104 & LP 733--099 & * & M2.0\,V  & ...\hspace{0.8cm}...  & 63 & --0.83 & 1.79 & 795 & S \\
J13536+776 & RX J1353.6+7737 & * & M4.0\,V  & ...\hspace{0.8cm}...  & 496 & --0.76 & 1.42 & 577 & ...\\
J16555--083 & VB 8 & * & M7.0\,V  & ...\hspace{0.8cm}...  & 104 & --0.79 & 1.41 & 486 & ...\\
J23556--061 & GJ 912 & * & M2.5\,V  & ...\hspace{0.8cm}...  & 1918 & --0.74 & 1.45 & 785 & ...\\
 \noalign{\smallskip}
\hline
 \noalign{\smallskip}
\multicolumn{9}{c}{CRX-Pearson moderate} \\
 \noalign{\smallskip}
\hline
 \noalign{\smallskip}
J02358+202 & BD+19 381 & $\times$ & M2.0\,V  & ...\hspace{0.8cm}...  & 8 & --0.51  & ...  & 815 & M \\
J04153--076 & omi02 Eri C & $\times$ & M4.5\,V  & ...\hspace{0.8cm}...  & 26 & --0.62  & ...  & 568 & ...  \\
J06318+414 & LP 205--044 & $\times$ & M5.0\,V  & ...\hspace{0.8cm}...  & 635 & --0.65 & 1.31 & 604 & ... \\
J06594+193 & GJ 1093 & $\times$ & M5.0\,V  & ...\hspace{0.8cm}...  & 13 & --0.55 & 1.2 & 553 & ... \\
J07582+413 & GJ 1105 & $\times$ & M3.5\,V  & ...\hspace{0.8cm}...  & 8 & --0.52  & ...  & 687 & ... \\
J10482--113 & LP 731--058 & $\times$ & M6.5\,V  & ...\hspace{0.8cm}...  & 21 & --0.69  & ...  & 498 & ... \\
J13450+176 & BD+18 2776 & $\times$ & M0.0\,V  & ...\hspace{0.8cm}...  & 28 & 0.55  & ...  & 874 & ... \\
J15412+759 & UU UMi & $\times$ & M3.0\,V  & ...\hspace{0.8cm}...  & 3165 & --0.67 & 1.28 & 625 & ... \\
J16581+257 & BD+25 3173 & $\times$ & M1.0\,V  & ...\hspace{0.8cm}...  & 15 & --0.68  & ...  & 852 & ...\\
J18022+642 & LP 071--082 & $\times$ & M5.0\,V  & ...\hspace{0.8cm}...  & 168 & --0.64 & 1.29 & 584 & ... \\
J18027+375 & GJ 1223 & $\times$ & M5.0\,V  & ...\hspace{0.8cm}...  & 11 & --0.53 & 1.16 & 583 & ... \\
J19216+208 & GJ 1235 & $\times$ & M4.5\,V  & ...\hspace{0.8cm}...  & 11 & --0.51  & ...  & 615 & ... \\
J19255+096 & LSPM J1925+0938 & $\times$ & M8.0\,V  & ...\hspace{0.8cm}...  & 903 & --0.68 & 1.36 & 249  \\
J19422--207 & 2MASS J19421282 \\
 & --2045477 & $\times$ & M5.1\,V  & ...\hspace{0.8cm}...  & 306 & --0.55  & ...  & 597 & ...\\
J20451--313 & AU Mic & $\times$ & M0.5\,V  & ...\hspace{0.8cm}...  & 262 & --0.61  & ...  & 712 & M  \\
J23064--050 & 2MUCD 12171 & $\times$ & M8.0\,V  & ...\hspace{0.8cm}...  & 722 & 0.62  & ...  & 296 & ...\\
J23216+172 & LP 662-027 & $\times$ & M4.0\,V & ...\hspace{0.8cm}... & 10 & --0.62 & ... & 671 & S \\
\noalign{\smallskip}
\hline
\end{tabular}
\vspace{-0.3cm}
\tablefoot{\tablefoottext{$\diamondsuit$} {Caution} should be exercised in the interpretation of these CRX-gradient values as for the later spectral types there is less flux in the bluer wavelengths and may bias the result.}
\end{table*}

\newpage

\begin{figure*}[b]
\centering
\includegraphics[angle=270,width=0.95\textwidth]{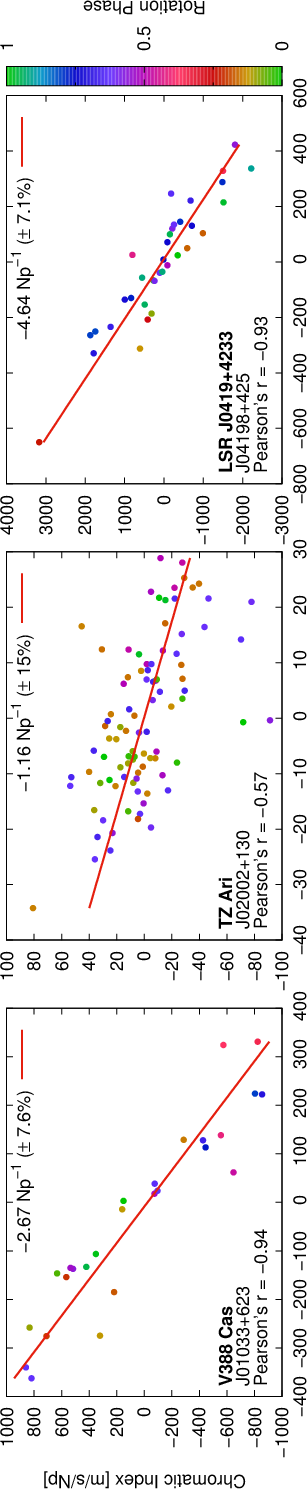}
\includegraphics[angle=270,width=0.95\textwidth]{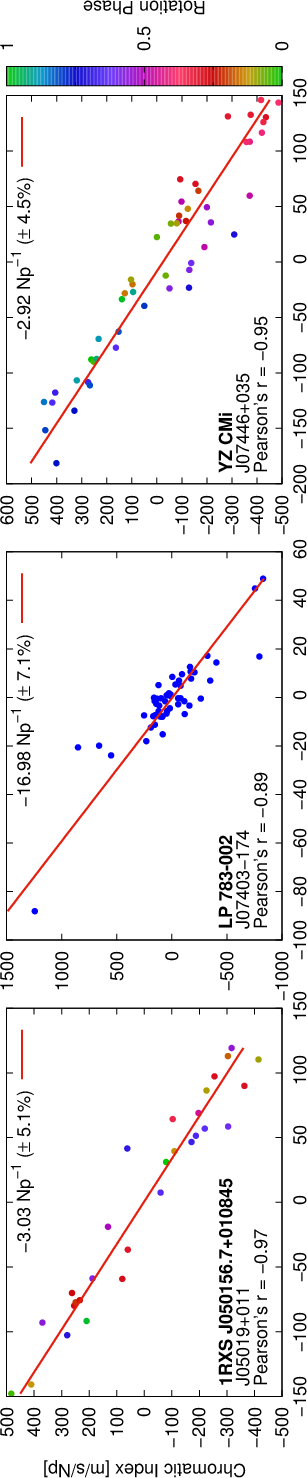}
\includegraphics[angle=270,width=0.95\textwidth]{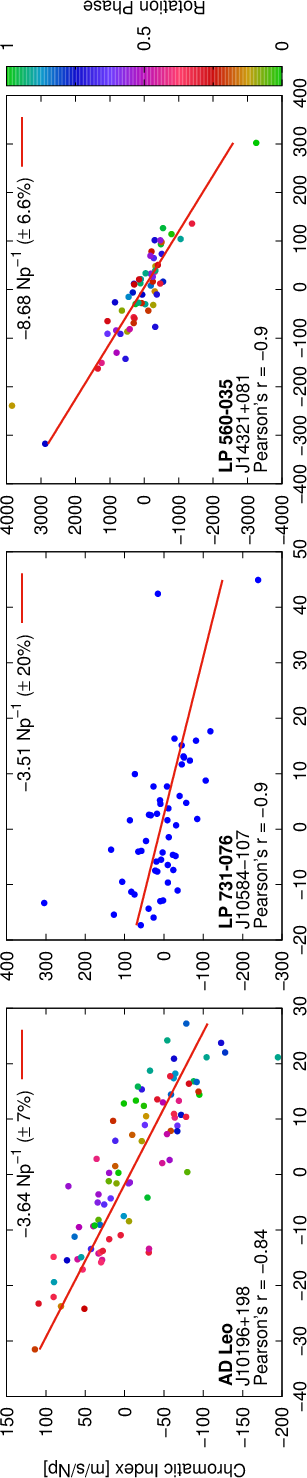}
\includegraphics[angle=270,width=0.95\textwidth]{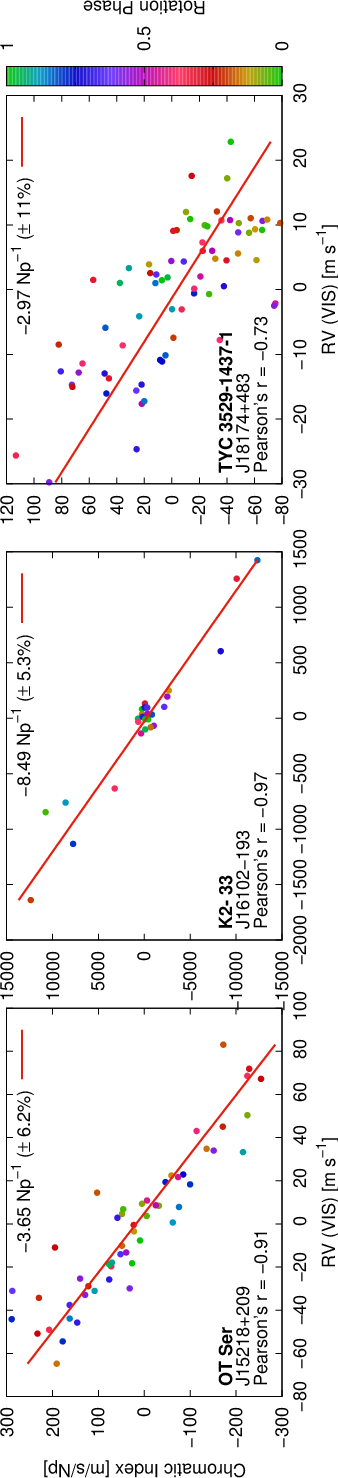}
=\includegraphics[angle=270,width=0.33\textwidth]{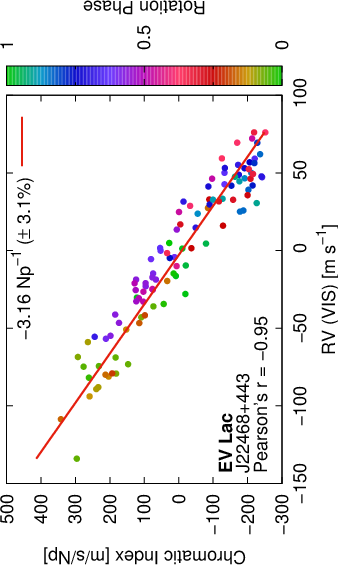}
\caption{CRX as a function of RV for each star in the CRX-slope sample.  Points are coloured by rotational phase derived from the rotation periods of ~\cite{Shan2024A&A...684A...9S}.  Stars with only blue coloured points do not have a known rotation period.}
\label{fig:CRX-sample}
\end{figure*}

\begin{figure*}
\centering
\includegraphics[angle=270,width=0.99\textwidth]{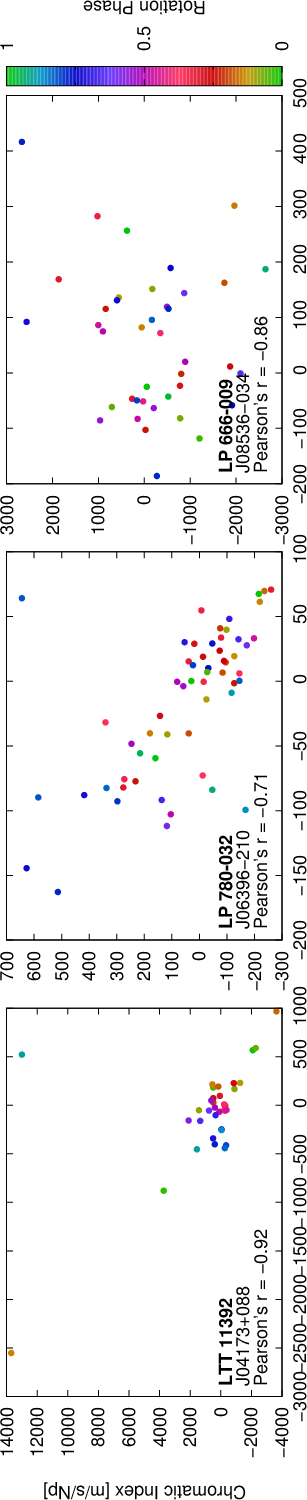}
\includegraphics[angle=270,width=0.99\textwidth]{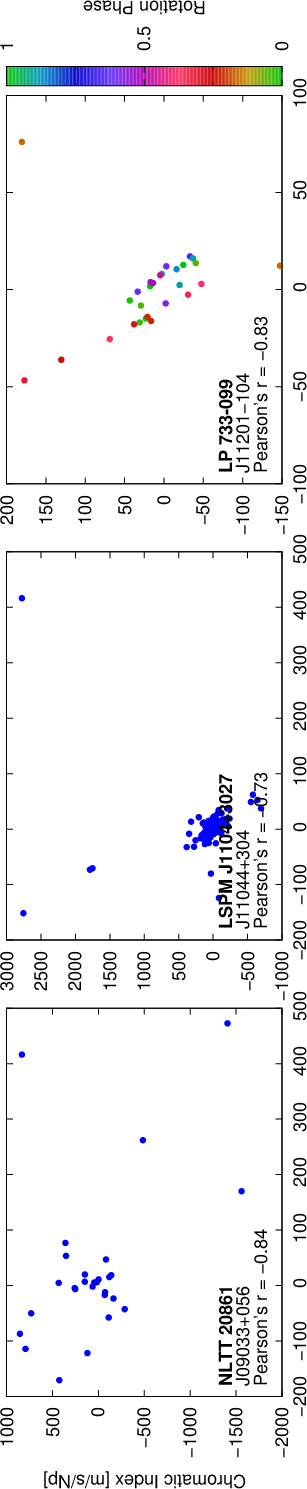}
\includegraphics[angle=270,width=0.99\textwidth]{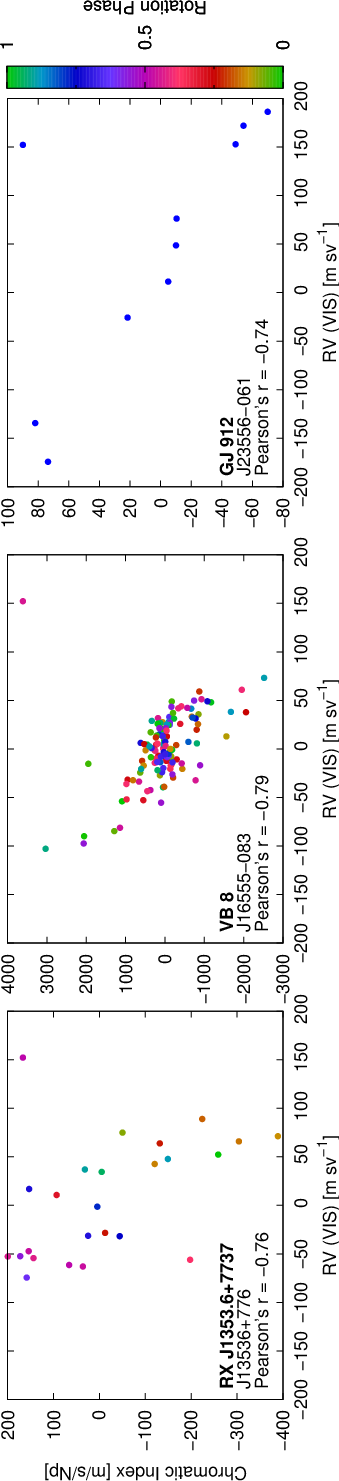}
\caption{CRX as a function of RV for each star in the CRX-Pearson (strong) sample. Points are coloured by rotational phase derived from the rotation periods of ~\cite{Shan2024A&A...684A...9S}.  Stars with only blue coloured points do not have a known rotation period.}
\label{fig:CRX-Pearson-strong}
\end{figure*}

\begin{figure*}
\centering
\includegraphics[angle=270,width=0.99\textwidth]{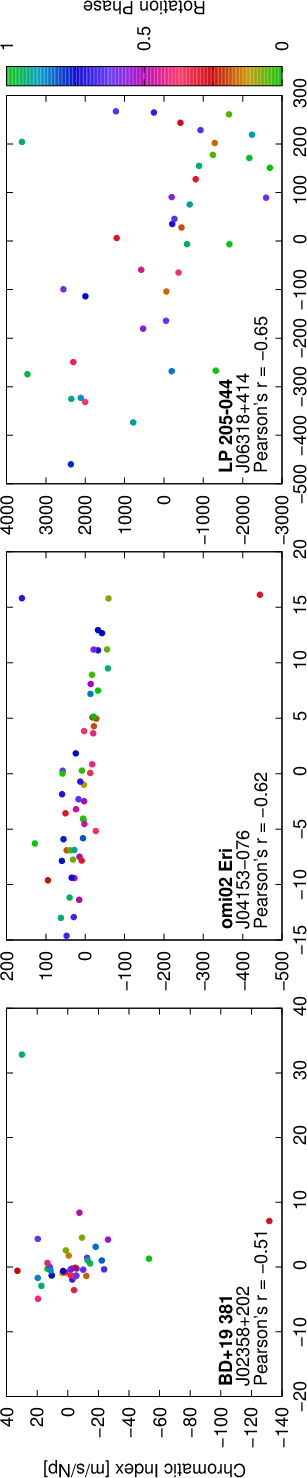}
\includegraphics[angle=270,width=0.99\textwidth]{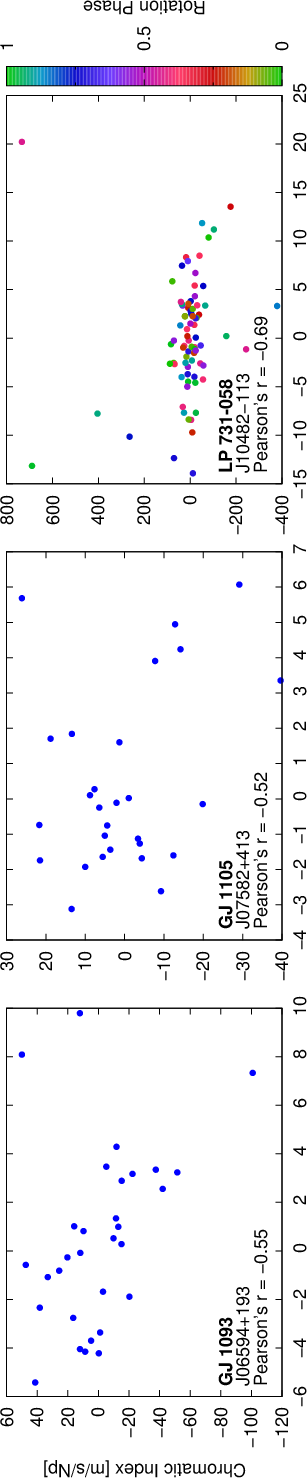}
\includegraphics[angle=270,width=0.99\textwidth]{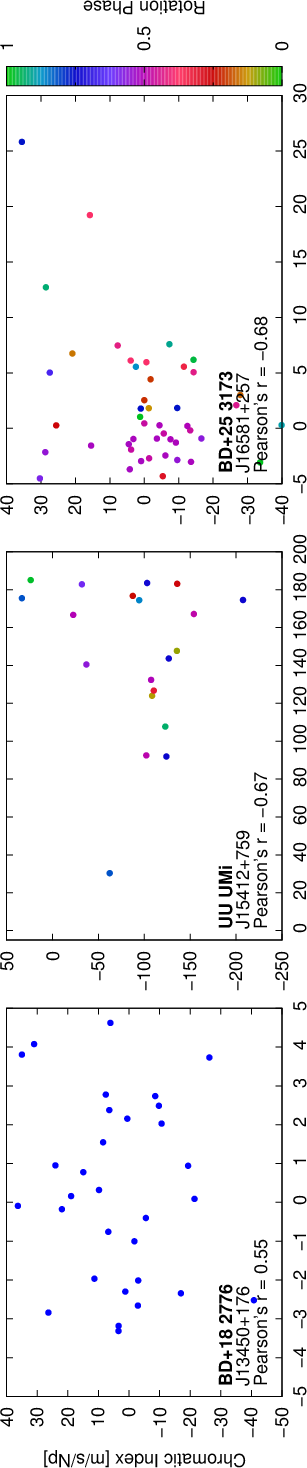}
\includegraphics[angle=270,width=0.99\textwidth]{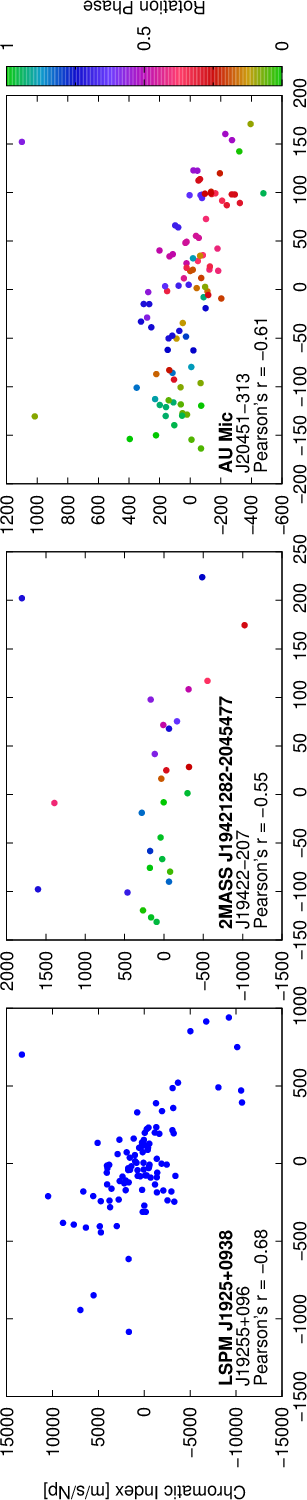}
\includegraphics[angle=270,width=0.66\textwidth]{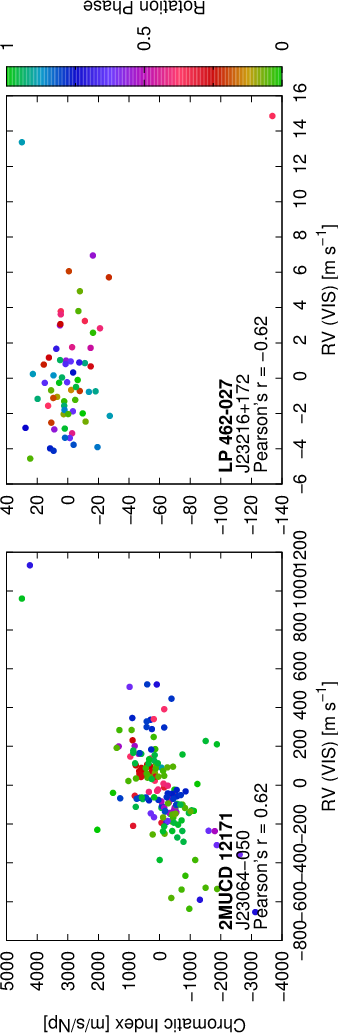}
\caption{CRX as a function of RV for each star in the CRX-Pearson (moderate) sample. Points are coloured by rotational phase derived from the rotation periods of ~\cite{Shan2024A&A...684A...9S}.  Stars with only blue coloured points do not have a known rotation period.}
\label{fig:CRX-Pearson-mod}
\end{figure*}

\begin{figure*}
\centering
\includegraphics[angle=270,width=0.99\textwidth]{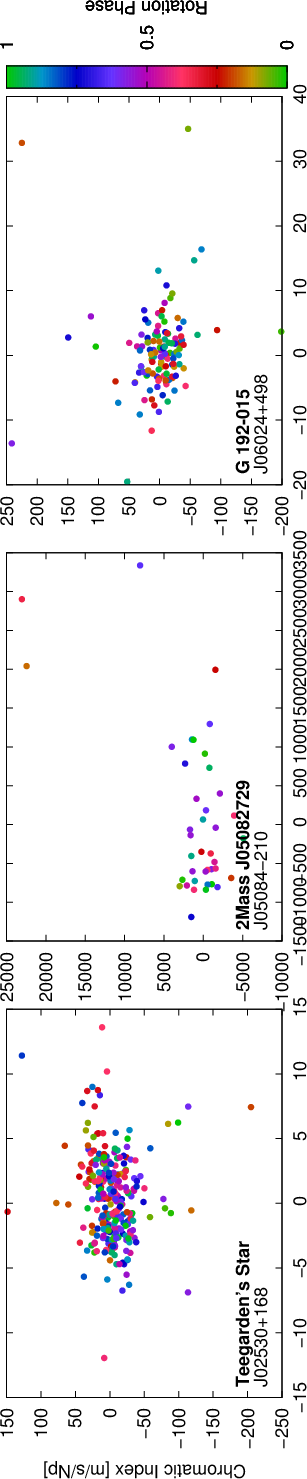}
\includegraphics[angle=270,width=0.99\textwidth]{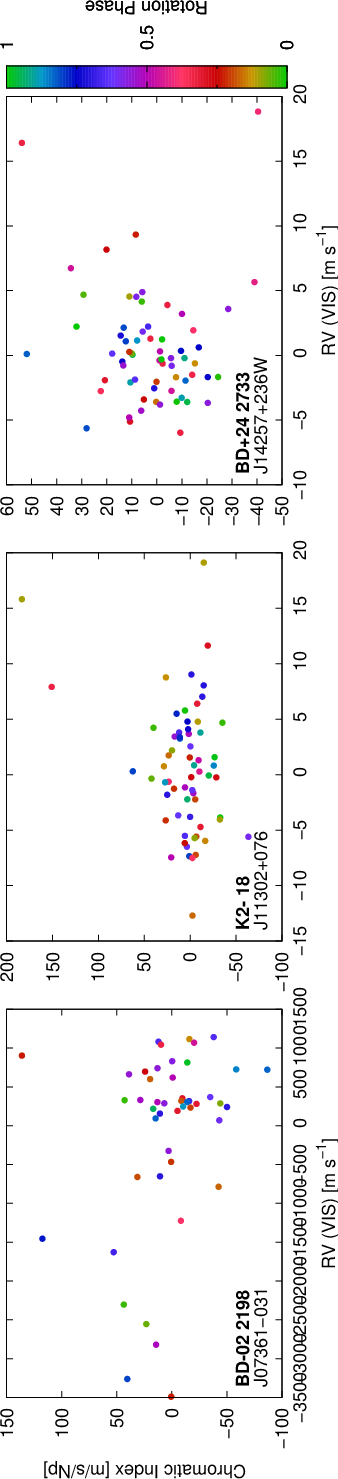}
\includegraphics[angle=270,width=0.33\textwidth]{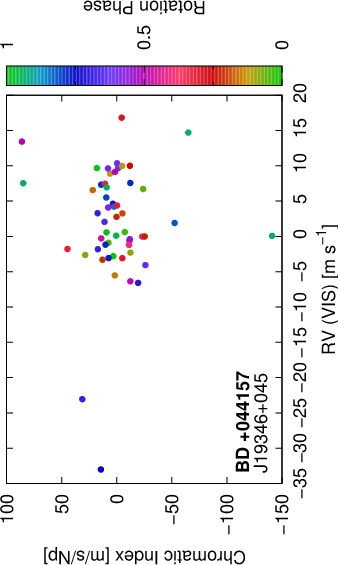}
\caption{A selection of stars not showing a clear CRX slope.  Their stellar parameters are shown in Table~\ref{tab:noCRX_param}. Points are coloured by rotational phase derived from the rotation periods of ~\cite{Shan2024A&A...684A...9S}.}
\label{fig:nonCRX_stars}
\end{figure*}


\begin{figure*}
\centering
\includegraphics[angle=270,width=0.95\textwidth]{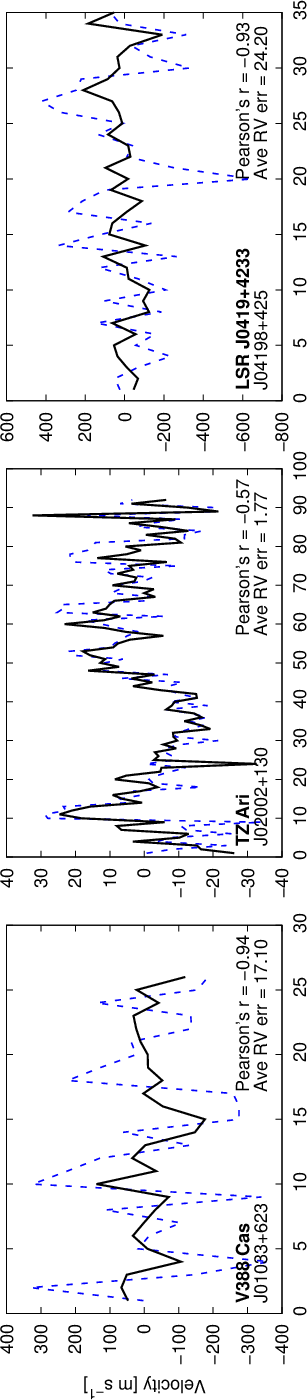}
\includegraphics[angle=270,width=0.95\textwidth]{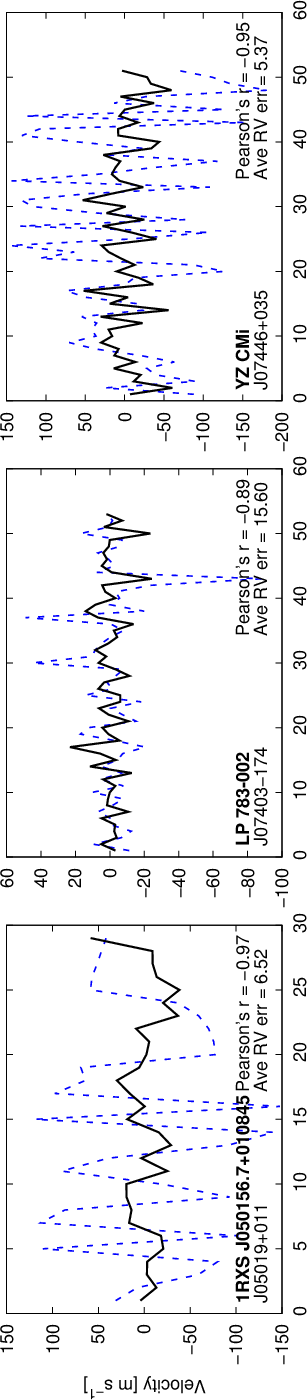}
\includegraphics[angle=270,width=0.95\textwidth]{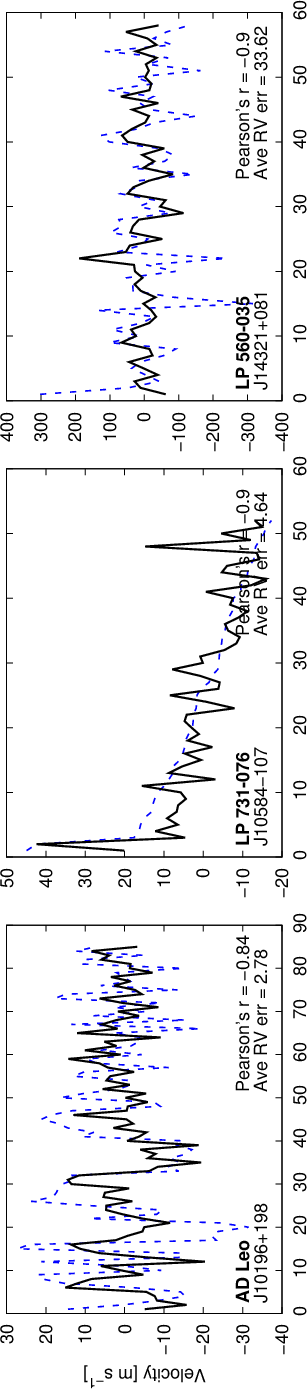}
\includegraphics[angle=270,width=0.95\textwidth]{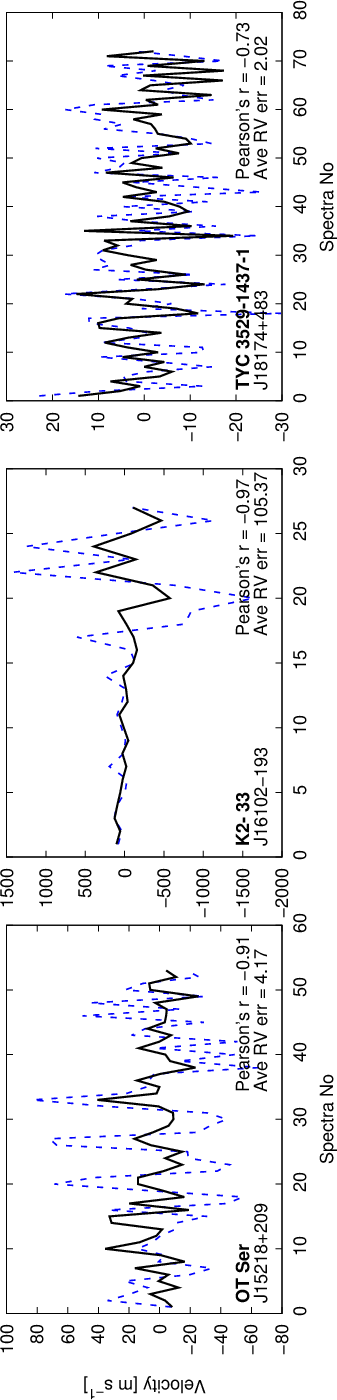}
\includegraphics[angle=270,width=0.33\textwidth]{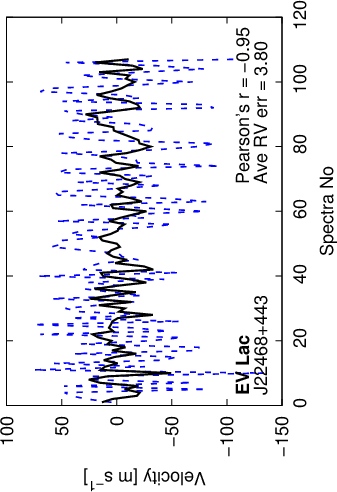}
\caption{Capability of a linear fit to the RV-CRX correlation to remove activity-induced RV variations.  For each of the 13 stars in the CRX-slope sample, the measured uncorrected RV for all CARMENES visible spectra is shown as a blue dashed line.  The subtracted CRX values are shown as the solid black line.  The rms values are shown in the key.}
\label{fig:Appendix_CRX_sub}
\end{figure*}

\begin{figure*}
\centering
\includegraphics[angle=270,width=0.95\textwidth]{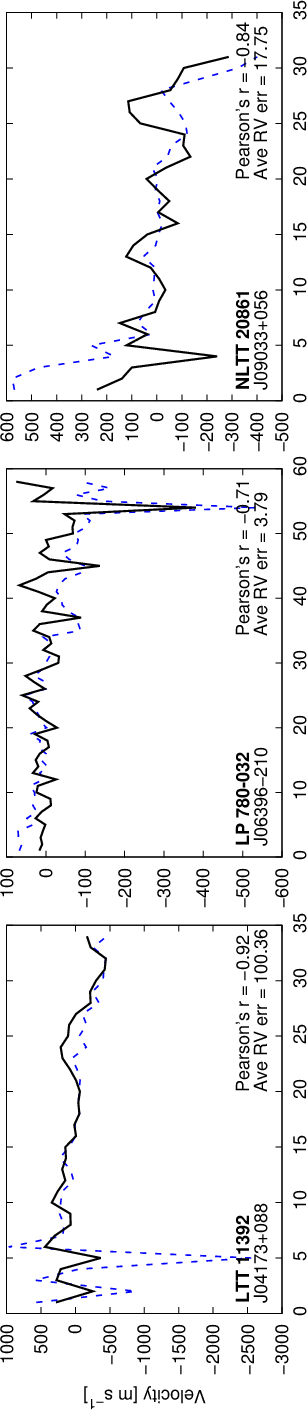}
\includegraphics[angle=270,width=0.95\textwidth]{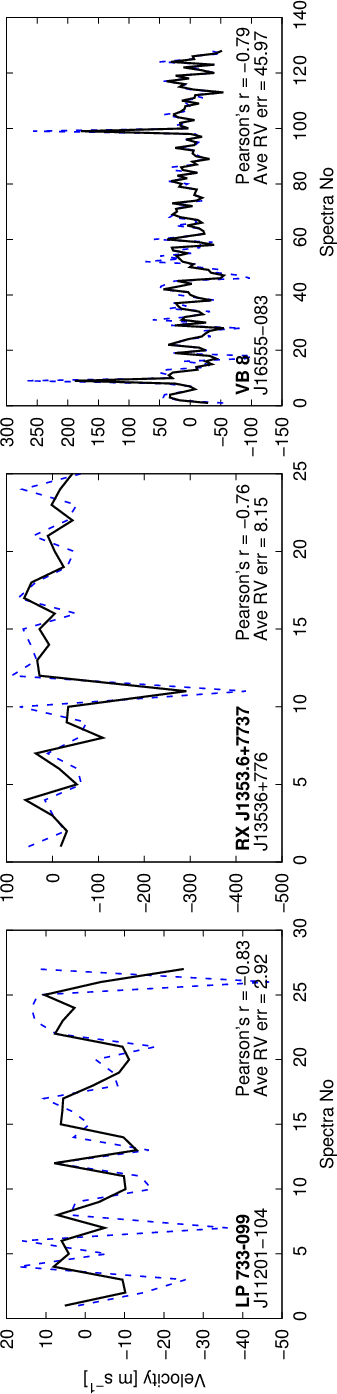}
\includegraphics[angle=270,width=0.33\textwidth]{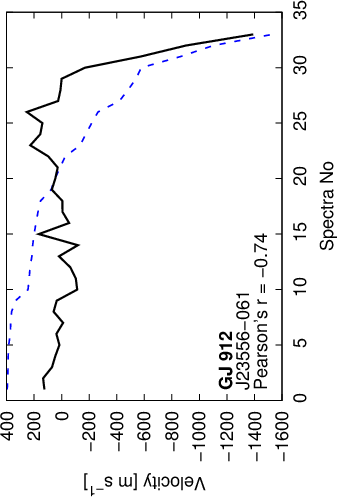}
\caption{Same as Figure~\ref{fig:Appendix_CRX_sub} but for the Pearson sample with strong correlations only and factor of improvement values greater than 1. }
\label{fig:Appendix_Pearson_sub}
\end{figure*}

\begin{figure*}
\centering
\includegraphics[angle=270,width=0.95\textwidth]{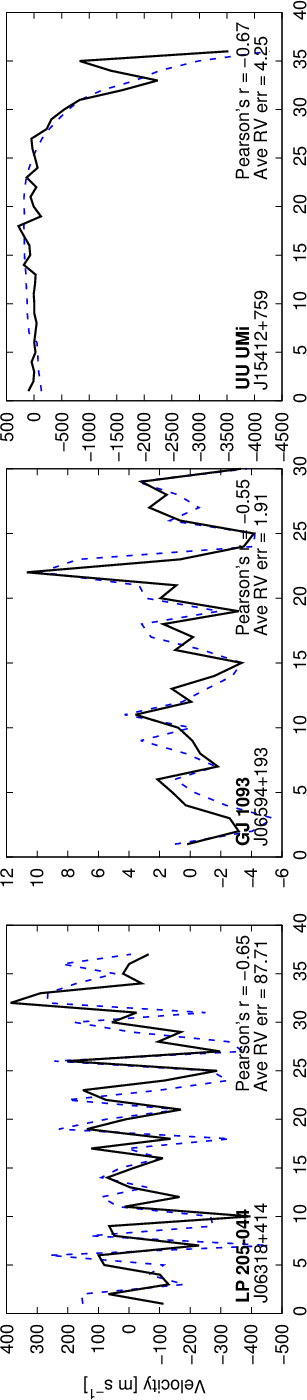}
\includegraphics[angle=270,width=0.95\textwidth]{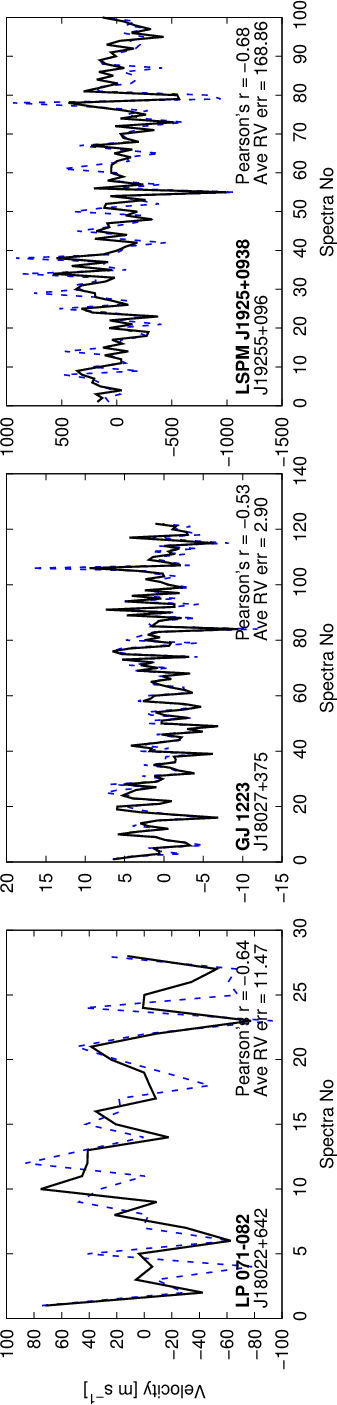}
\caption{Same as Figure~\ref{fig:Appendix_CRX_sub} but for the Pearson sample with moderate correlations only and factor of improvement values greater than 1. }
\label{fig:Appendix_Pearson_sub_mod}
\end{figure*}

\section{Stellar Parameters}

\longtab[1]{
\scriptsize
   \begin{landscape}
     \begin{longtable}{lllccccccccccccc}      \caption{\label{tab:Sample}Table of stellar parameters}\\
       \hline \hline
       \noalign{\smallskip}
       Karmn & Name & Sp. & N$_{\rm Obs}$ & $T_{\rm eff}$ & $R$ & $M$ & $\varv\sin{i}$ & Ref. & pEW(H$\alpha$) &   $B_{\rm field}$   & $V-Ks$ & $\log{\tau}$ & TiO\,7050 & $P_{\rm rot}$ & Ref. \\
       & & type & & (K) & (R$_{\odot}$) & (M$_{\odot}$) &  (km\,s$^{-1}$) & & (\AA) & (G) & (mag) & & & (d) & \\
       
      \noalign{\smallskip}
      \hline
       \noalign{\smallskip}
       \endfirsthead
       \caption{continued.}\\
       \hline \hline
       \noalign{\smallskip}

       Karmn & Name & Sp. & N$_{\rm Obs}$ & $T_{\rm eff}$ & $R$ & $M$ & $\varv\sin{i}$ & Ref. & pEW(H$\alpha$) &   $B_{\rm field}$   & $V-Ks$ & $\log{\tau}$ & TiO\,7050 & $P_{\rm rot}$ & Ref. \\
       & & type & & (K) & (R$_{\odot}$) & (M$_{\odot}$) &  (km\,s$^{-1}$) & & (\AA) & (G) & (mag) & & & (d) & \\
       
      \noalign{\smallskip}
       \hline
       \noalign{\smallskip}
       \endhead
       \hline
       \endfoot
J00051+457 & GJ 2 & M1.0 V & 52 & 3773 & 0.49 & 0.49 & <2 & Rein18 & -0.06 &     & 4.25 & 1.77 & 0.77 & 15.37 & DA19 \\
J00067-075 & GJ 1002 & M5.5 V & 89 & 3169 & 0.12 & 0.11 & <2 & Rein18 & 0.00 &     & 6.40 & 2.37 & 0.35 & 126.00 & SM23 \\
J00183+440 & GX And & M1.0 V & 216 & 3603 & 0.40 & 0.39 & <2 & Rein18 & -0.07 &     & 4.27 & 1.77 & 0.77 & 45.00 & SM18 \\
J00184+440 & GQ And & M3.5 V & 193 & 3318 & 0.18 & 0.16 & <2 & Rein18 & -0.03 &     & 5.14 & 2.02 & 0.55 &  &  \\
J00286-066 & GJ 1012 & M4.0 V & 50 & 3419 & 0.35 & 0.34 & <2 & Rein18 & 0.12 &     & 4.92 & 1.96 & 0.53 &  &  \\
J00389+306 & Wolf 1056 & M2.5 V & 60 & 3551 & 0.41 & 0.41 & <2 & Rein18 & 0.04 &     & 4.39 & 1.81 & 0.66 & 50.20 & DA19 \\
J00403+612 & 2MASS J00402129+6112490 & M2.0 V & 44 & 3709 & 0.47 & 0.47 &  &  &  &     & 4.17 & 1.75 & 0.74 &  &  \\
J01025+716 & BD+70 68 & M3.0 V & 115 & 3524 & 0.46 & 0.47 & <2 & Rein18 & 0.08 &     & 4.60 & 1.87 & 0.60 & 50.52 & Shan23 \\
J01026+623 & BD+61 195 & M1.5 V & 82 & 3791 & 0.51 & 0.51 & <2 & Rein18 & -0.04 &     & 4.39 & 1.81 & 0.74 & 18.82 & Shan23 \\
J01033+623 & V388 Cas & M5.0 V & 26 & 3057 & 0.24 & 0.23 & 10.5 & Rein18 & -9.77 & 4800 & 5.64 & 2.16 & 0.38 & 1.02 & Shan23 \\
J01048-181 & GJ 1028 & M5.0 V & 78 & 3209 & 0.15 & 0.14 & <2 & Rein18 & 0.01 &     & 6.10 & 2.29 & 0.42 & 143.20 & New18 \\
J01066+192 & LSMP J0106+1913 & M3.0 V & 65 & 3496 & 0.34 & 0.34 &  &  &  &     & 4.63 & 1.88 & 0.59 & 70.10 & Shan23 \\
J01125-169 & YZ Cet & M4.5 V & 110 & 3228 & 0.15 & 0.14 & <2 & Rein18 & -1.36 &     & 5.68 & 2.17 & 0.43 &  &  \\
J01433+043 & GJ 70 & M2.0 V & 26 & 3547 & 0.40 & 0.40 & <2 & Rein18 & -0.08 &     & 4.24 & 1.77 & 0.67 & 14.65 & Shan23 \\
J02002+130 & TZ Ari & M3.5 V: & 93 & 3237 & 0.16 & 0.14 & <2 & Rein18 & -1.91 &     & 5.75 & 2.19 & 0.43 & 2.00 & Shan23 \\
J02015+637 & G 244-047 & M3.0 V & 51 & 3579 & 0.42 & 0.42 & <2 & Rein18 & 0.05 &     & 4.63 & 1.88 & 0.62 &  &  \\
J02070+496 & G 173-037 & M3.5 V & 27 & 3455 & 0.33 & 0.32 & <2 & Rein18 & -0.48 &     & 4.69 & 1.89 & 0.61 & 7.20 & DA19 \\
J02123+035 & BD+02 348 & M1.5 V & 65 & 3575 & 0.47 & 0.47 & <2 & Rein18 & 0.07 &     & 4.06 & 1.72 & 0.74 & 17.11 & Shan23 \\
J02222+478 & BD+47 612 & M0.5 V & 53 & 3894 & 0.57 & 0.58 & <2 & Rein18 & 0.06 &     & 3.79 & 1.64 & 0.83 & 29.50 & DA19 \\
J02336+249 & GJ 102 & M4.0 V & 30 & 3313 & 0.23 & 0.22 & 3 & Rein18 & -2.73 &     & 5.35 & 2.08 & 0.48 & 3.00 & Shan23 \\
J02358+202 & BD+19 381 & M2.0 V & 32 & 3703 & 0.48 & 0.49 & <2 & Rein18 & 0.07 &     & 4.31 & 1.79 & 0.71 & 31.90 & DA19 \\
J02362+068 & BX Cet & M4.0 V & 50 & 3335 & 0.27 & 0.26 & <2 & Rein18 & 0.08 &     & 5.09 & 2.01 & 0.51 &  &  \\
J02442+255 & VX Ari & M3.0 V & 51 & 3510 & 0.35 & 0.34 & <2 & Rein18 & 0.01 &     & 4.63 & 1.88 & 0.63 & 38.70 & DA19 \\
J02489-145W & PM J02489-1432W & M0 & 33 & 3655 & 0.45 & 0.45 &  &  &  &     & -8.65 & -1.84 & 0.66 &  &  \\
J02530+168 & Teegarden's Star & M7.0 V & 253 & 3034 & 0.10 & 0.08 & <2 & Rein18 & -0.50 &     & 7.49 & 2.68 & 0.20 & 97.56 & Laf21 \\
J02573+765 & G 245-61 & M3.5 & 57 & 3381 & 0.25 & 0.24 &  &  &  & 197 & 4.54 & 1.85 & 0.63 &  &  \\
J03133+047 & CD Cet & M5.0 V & 107 & 3214 & 0.18 & 0.16 & <2 & Rein18 & 0.04 & 253 & 6.17 & 2.31 & 0.43 & 126.20 & New16 \\
J03181+382 & HD 275122 & M1.5 V & 56 & 3826 & 0.57 & 0.58 & <2 & Rein18 & 0.17 & 181 & 4.12 & 1.73 & 0.75 & 77.20 & DA19 \\
J03213+799 & GJ 133 & M2.0 V & 28 & 3599 & 0.40 & 0.40 & <2 & Rein18 & -0.01 &     & 4.22 & 1.76 & 0.71 & 23.26 & Shan23 \\
J03463+262 & HD 23453 & M0.0 V & 51 & 3949 & 0.57 & 0.58 & 3.3 & Rein18 & -0.09 & 470 & 3.74 & 1.63 & 0.85 & 16.40 & Shan23 \\
J03531+625 & Ross 567 & M3.0 V & 36 & 3501 & 0.35 & 0.35 & <2 & Rein18 & 0.06 & 261 & 4.50 & 1.84 & 0.64 &  &  \\
J04153-076 & omi02 Eri C & M4.5 V & 55 & 3179 & 0.26 & 0.25 & 2.1 & Rein18 & -3.36 & 3018 & 3.53 & 1.57 & 0.42 & 8.56 & BF19 \\
J04167-120 & LP 714-47 & K7.0 V & 34 & 3961 & 0.58 & 0.59 &  &  &  & 645 & 3.87 & 1.66 & 0.80 & 33.00 & Dre20 \\
J04173+088 & LTT 11392 & M4.5 V & 34 & 3100 & 0.28 & 0.27 & 190.3 & Jeff18 & -11.15 &     & 5.69 & 2.17 & 0.68 & 0.37 & Shan23 \\
J04198+425 & LSR J0419+4233 & M8.5 V & 35 & 2400 & 0.13 & 0.11 & 3.6 & Rein18 & -10.82 &     & -9.90 & -2.19 & 0.37 & 0.99 & Shan23 \\
J04290+219 & BD+21 652 & M0.5 V & 164 & 4123 & 0.66 & 0.68 & 3.9 & Rein18 & 0.19 & 380 & 3.62 & 1.59 & 0.90 & 25.40 & DA19 \\
J04343+430 & PM J04343+4302 & M2 & 55 & 3525 & 0.46 & 0.47 &  &  & -0.17 & 187 & 4.62 & 1.87 & 0.62 &  &  \\
J04376-110 & BD-11 916 & M1.5 V & 36 & 3666 & 0.45 & 0.45 & <2 & Rein18 & 0.07 &     & 4.26 & 1.77 & 0.72 &  &  \\
J04376+528 & BD+52 857 & M0.0 V & 126 & 4039 & 0.55 & 0.55 & 3.4 & Rein18 & -0.01 & 428 & 3.82 & 1.65 & 0.91 & 15.47 & Laf21 \\
J04406-128 & TOI-2457 & M0 V & 45 & 4100 & 0.58 & 0.59 &  &  &  &     & 3.33 & 1.51 & 0.94 &  &  \\
J04588+498 & BD+49 1280 & M0.0 V & 56 & 3983 & 0.59 & 0.60 & <2 & Rein18 & -0.05 & 459 & 3.72 & 1.62 & 0.86 & 17.46 & Shan23 \\
J05019+011 & 1RXS J050156.7+010845 & M4.0 V & 29 & 3066 & 0.65 & 0.30 & 6.5 & Rein18 & -6.36 &     & 5.18 & 2.03 & 0.47 & 2.08 & Shan23 \\
J05033-173 & LP 776-046 & M3.0 V & 50 & 3464 & 0.27 & 0.26 & <2 & Rein18 & -0.01 & 266 & 4.82 & 1.93 & 0.61 &  &  \\
J05084-210 & 2M J05082729-2101444 & M5.0 V & 33 & 3115 & 0.68 & 0.33 & 25.2 & Rein18 & -17.39 &     & 5.83 & 2.21 & 0.38 & 0.28 & Shan23 \\
J05127+196 & GJ 192 & M2.0 V & 40 & 3599 & 0.42 & 0.42 & <2 & Rein18 & 0.04 &     & 4.32 & 1.79 & 0.72 & 32.70 & BF19 \\
J05314-036 & HD 36395 & M1.5 V & 96 & 3850 & 0.57 & 0.58 & <2 & Rein18 & 0.10 & 211 & 4.32 & 1.79 & 0.75 & 33.80 & DA19 \\
J05360-076 & Wolf 1457 & M4.0 V & 34 & 3422 & 0.32 & 0.32 & <2 & Rein18 & 0.07 &     & 5.31 & 2.07 & 0.52 &  &  \\
J05365+113 & V2689 Ori & M0.0 V & 128 & 4067 & 0.57 & 0.58 & 3.8 & Rein18 & -0.43 & 1111 & 3.95 & 1.69 & 0.90 & 11.75 & Shan23 \\
J05415+534 & HD 233153 & M1.0 V & 98 & 3825 & 0.54 & 0.55 & <2 & Rein18 & -0.04 & 418 & 4.01 & 1.70 & 0.80 & 17.39 & Shan23 \\
J05421+124 & V1352 Ori & M4.0 V & 50 & 3324 & 0.24 & 0.23 & <2 & Rein18 & 0.07 & 140 & 5.09 & 2.01 & 0.51 &  &  \\
J06011+595 & G 192-013 & M3.5 V & 79 & 3431 & 0.26 & 0.25 & <2 & Rein18 & -0.01 & 208 & 5.10 & 2.01 & 0.55 &  &  \\
J06024+498 & G 192-015 & M5.0 V & 83 & 3213 & 0.15 & 0.13 & <2 & Rein18 & -0.04 & 368 & 6.09 & 2.28 & 0.41 & 105.00 & DA19 \\
J06103+821 & GJ 226 & M2.0 V & 58 & 3554 & 0.41 & 0.41 & <2 & Rein18 & 0.00 & 100 & 4.37 & 1.80 & 0.67 & 44.60 & DA19 \\
J06105-218 & HD 42581 A & M0.5 V & 53 & 3801 & 0.54 & 0.54 & <2 & Rein18 & 0.00 & 375 & 4.92 & 1.96 & 0.78 & 27.30 & DA19 \\
J06318+414 & LP 205-044 & M5.0 V & 37 & 3255 & 0.36 & 0.35 & 58.4 & Rein18 & -11.19 & 2197 & 5.98 & 2.25 & 0.40 & 0.30 & Shan23 \\
J06371+175 & HD 260655 & M0.0 V & 95 & 3803 & 0.44 & 0.44 & <2 & Rein18 & -0.07 & 186 & 3.91 & 1.67 & 0.87 & 37.50 & Luq22 \\
J06396-210 & LP 780-032 & M4.0 V & 57 & 3428 & 0.33 & 0.33 & <2 & Rein18 & -0.13 & 620 & 5.11 & 2.01 & 0.54 & 79.15 & New18 \\
J06548+332 & Wolf 294 & M3.0 V & 364 & 3503 & 0.35 & 0.34 & <2 & Rein18 & 0.00 & 145 & 4.76 & 1.91 & 0.60 & 122.00 & Sto20b \\
J06594+193 & GJ 1093 & M5.0 V & 28 & 3146 & 0.13 & 0.12 & <2 & Rein18 & -0.43 & 392 & -8.23 & -1.72 & 0.36 &  &  \\
J07001-190 & 2M J07000682-1901235 & M5.0 V & 27 & 3000 & 0.64 & 0.65 & 3.8 & Rein18 & -6.99 &     & 5.83 & 2.21 & 0.41 &  &  \\
J07274+052 & Luyten's Star & M3.5 V & 768 & 3380 & 0.29 & 0.28 & <2 & Rein18 & -0.05 & 219 & 5.05 & 1.99 & 0.54 & 93.50 & SM17 \\
J07319+362N & BL Lyn & M3.5 V & 48 & 3384 & 0.40 & 0.41 & <2 & Rein18 & -2.78 & 2529 & 4.98 & 1.98 & 0.51 & 16.40 & DA19 \\
J07361-031 & BD-02 2198 & M1.0 V & 47 & 3825 & 0.56 & 0.57 & 3.1 & Rein18 & -0.86 &     & 4.10 & 1.73 & 0.81 & 12.20 & Shan23 \\
J07393+021 & BD+02 1729 & M0.0 V & 50 & 3980 & 0.58 & 0.59 & <2 & Rein18 & 0.02 & 277 & 3.88 & 1.67 & 0.86 & 11.45 & Shan23 \\
J07403-174 & LP 783-002 & M6.0 V & 52 & 3031 & 0.11 & 0.09 & <2 & Rein18 & 0.00 & 276 & -9.29 & -2.02 & 0.24 &  &  \\
J07446+035 & YZ CMi & M4.5 V & 49 & 2908 & 0.34 & 0.19 & 4 & Rein18 & -7.28 & 4539 & 5.32 & 2.07 & 0.43 & 2.78 & Shan23 \\
J07582+413 & GJ 1105 & M3.5 V & 27 & 3431 & 0.27 & 0.26 & <2 & Rein18 & -0.03 & 119 & 5.14 & 2.02 & 0.54 &  &  \\
J08023+033 & G 050-016 A & M4.0 V & 64 & 3422 & 0.35 & 0.34 &  &  &  & 255 & 4.91 & 1.96 & 0.54 &  &  \\
J08161+013 & GJ 2066 & M2.0 V & 74 & 3570 & 0.44 & 0.44 & <2 & Rein18 & 0.01 & 251 & 4.33 & 1.79 & 0.69 & 40.70 & DA19 \\
J08298+267 & DX Cnc & M6.5 V & 34 & 2997 & 0.11 & 0.09 & 10.5 & Rein18 & -4.55 & 2679 & -7.26 & -1.45 & 0.25 & 0.46 & DA19 \\
J08409-234 & LP 844-008 & M3.5 V & 30 & 3496 & 0.40 & 0.40 & 2.5 & Bro10 & 0.04 &     & 4.95 & 1.97 & 0.55 & 57.50 & Irv23 \\
J08413+594 & LP 090-018 & M5.5 V & 204 & 3141 & 0.14 & 0.12 & <2 & Rein18 & -1.34 & 1276 & 6.50 & 2.40 & 0.35 & 88.65 & Laf21 \\
J08536-034 & LP 666-009 & M9.0 V & 52 & 2400 & 0.10 & 0.08 & 9.3 & Rein18 &  & 4786 & -9.94 & -2.20 & 0.55 & 0.46 & Shan23 \\
J09033+056 & NLTT 20861 & M7.0 V & 32 & 3027 & 0.15 & 0.14 & 9.7 & Rein18 & -3.79 & 3224 & -9.83 & -2.17 & 0.28 &  &  \\
J09143+526 & HD 79210 & M0.0 V & 72 & 4015 & 0.57 & 0.58 & 2.9 & Del98 & -0.01 & 404 & 3.80 & 1.64 & 0.90 & 17.54 & BF19 \\
J09144+526 & HD 79211 & M0.0 V & 159 & 3983 & 0.58 & 0.58 & 2.3 & Rein18 & -0.02 & 335 & 4.38 & 1.81 & 0.89 & 16.88 & BF19 \\
J09411+132 & Ross 85 & M1.5 V & 48 & 3699 & 0.45 & 0.45 & <2 & Rein18 & 0.01 &     & 4.27 & 1.77 & 0.73 &  &  \\
J09425+700 & GJ 360 & M2.0 V & 50 & 3547 & 0.49 & 0.50 & <2 & Rein18 & -0.58 & 1027 & 4.51 & 1.84 & 0.66 & 21.00 & DA19 \\
J09428+700 & GJ 362 & M3.0 V & 50 & 3504 & 0.42 & 0.42 & <2 & Rein18 & -0.80 &     & 4.78 & 1.92 & 0.59 & 24.33 & Shan23 \\
J09561+627 & BD+63 869 & M0.0 V & 71 & 3938 & 0.58 & 0.59 & <2 & Rein18 & -0.06 & 392 & 3.98 & 1.69 & 0.85 & 16.88 & \\
J10088+692 & TYC 4384-1735-1 & M0.5 V & 42 & 3952 & 0.63 & 0.64 &  &  & -0.74 & 207 & 3.60 & 1.59 & 0.85 & 41.20 & Blu20 \\
J10122-037 & AN Sex & M1.5 V & 77 & 3720 & 0.49 & 0.50 & <2 & Rein18 & -0.03 & 449 & 4.58 & 1.86 & 0.72 & 23.00 & Shan23 \\
J10185-117 & LP 729-054 & M4.0 V & 52 & 3434 & 0.37 & 0.37 & 3 & Jeff18 & 0.18 & 141 & 4.94 & 1.96 & 0.54 &  &  \\
J10196+198 & AD Leo & M3.0 V+ & 84 & 3455 & 0.43 & 0.45 & 2.4 & MR14 & -4.52 & 3568 & 4.93 & 1.96 & 0.56 & 2.24 & Mor08 \\
J10251-102 & BD-09 3070 & M1.0 V & 26 & 3759 & 0.51 & 0.51 & <2 & Rein18 & -0.11 &     & 4.19 & 1.75 & 0.75 & 25.05 & Shan23 \\
J10289+008 & BD+01 2447 & M2.0 V & 84 & 3588 & 0.41 & 0.41 & <2 & Rein18 & -0.01 & 273 & 4.39 & 1.81 & 0.70 & 31.72 & Shan23 \\
J10482-113 & LP 731-058 & M6.5 V & 75 & 3029 & 0.09 & 0.08 & 2.1 & Rein18 & -2.27 &     & 7.86 & 2.78 & 0.21 & 1.50 & Mor10 \\
J10504+331 & G 119-037 & M4.0 V & 44 & 3441 & 0.40 & 0.40 & <2 & Rein18 & 0.09 &     & 5.04 & 1.99 & 0.53 &  &  \\
J10508+068 & EE Leo & M4.0 V & 49 & 3410 & 0.26 & 0.25 & <2 & Rein18 & 0.05 & 201 & 5.33 & 2.07 & 0.52 & 64.00 & DA19 \\
J10564+070 & CN Leo & M6.0 V & 78 & 3071 & 0.11 & 0.09 & <2 & Rein18 & -7.22 & 3014 & 7.44 & 2.66 & 0.27 & 2.70 & DA19 \\
J10584-107 & LP 731-076 & M5.0 V & 50 & 3218 & 0.21 & 0.19 & 3.2 & Rein18 & -0.62 & 4098 & -8.64 & -1.84 & 0.40 &  &  \\
J11000+228 & Ross 104 & M2.5 V & 62 & 3532 & 0.37 & 0.37 & <2 & Rein18 & -0.01 & 361 & 4.58 & 1.86 & 0.66 & 53.17 & BF19 \\
J11026+219 & DS Leo & M1.0 V & 53 & 3846 & 0.54 & 0.54 & 2.6 & Rein18 & -0.49 & 1041 & 3.90 & 1.67 & 0.82 & 14.26 & Shan23 \\
J11033+359 & Lalande 21185 & M1.5 V & 504 & 3557 & 0.40 & 0.39 & <2 & Rein18 & -0.04 & 100 & 4.24 & 1.77 & 0.72 & 56.15 & Dia19 \\
J11044+304 & LSPM J1104+3027 & M3 & 75 & 3534 & 0.35 & 0.35 &  &  &  &     & 5.18 & 2.03 & 0.52 &  &  \\
J11054+435 & BD+44 2051A & M1.0 V & 117 & 3628 & 0.38 & 0.38 & <2 & Rein18 & -0.09 & 100 & 4.05 & 1.71 & 0.80 & 100.90 & SM18 \\
J11055+435 & WX UMa & M5.5 V & 47 & 3278 &  & 0.11 & 8.2 & Rein18 & -9.69 & 6878 & -7.84 & -1.61 & 0.22 & 0.78 & Mor10 \\
J11110+304W & HD 97101 B & M2.0 V & 51 & 3730 & 0.54 & 0.55 & <2 & Rein18 & 0.05 & 279 & 4.25 & 1.77 & 0.72 &  &  \\
J11201-104 & LP 733-099 & M2.0 V & 28 & 3661 & 0.51 & 0.53 & 3.6 & Rein18 & -1.88 & 2170 & 4.28 & 1.78 & 0.70 & 5.64 & Shan23 \\
J11302+076 & K2-18 & M2.5 V & 61 & 3563 & 0.43 & 0.43 & <2 & Rein18 & 0.04 & 270 & 4.60 & 1.87 & 0.64 & 36.40 & DA19 \\
J11417+427 & Ross 1003 & M4.0 V & 76 & 3387 & 0.35 & 0.35 & <2 & Rein18 & 0.11 & 122 & 5.14 & 2.02 & 0.52 & 71.50 & DA19 \\
J11421+267 & Ross 905 & M2.5 V & 124 & 3533 & 0.41 & 0.41 & <2 & Rein18 & 0.00 & 100 & 4.54 & 1.85 & 0.64 & 44.60 & DA19 \\
J11423+230 & NLTT 28297 & M0.5 V & 83 & 3879 & 0.53 & 0.53 &  &  &  &     & 3.76 & 1.63 & 0.84 &  &  \\
J11474+667 & 1RXS J114728.8+664405 & M5.0 V & 50 & 3171 & 0.30 & 0.29 & 2.7 & Rein18 & -7.99 & 4773 & 5.84 & 2.22 & 0.38 & 13.30 & Shan23 \\
J11476+786 & GJ 445 & M3.5 V & 66 & 3361 & 0.26 & 0.25 & <2 & Rein18 & 0.00 & 158 & 4.82 & 1.93 & 0.55 &  &  \\
J11477+008 & FI Vir & M4.0 V & 56 & 3325 & 0.19 & 0.17 & <2 & Rein18 & 0.00 & 279 & 5.46 & 2.11 & 0.49 & 112.83 & Ste16 \\
J11509+483 & GJ 1151 & M4.5 V & 108 & 3282 & 0.18 & 0.16 & <2 & Rein18 & -0.10 & 340 & 5.60 & 2.15 & 0.46 & 125.00 & DA19 \\
J11511+352 & BD+36 2219 & M1.5 V & 112 & 3692 & 0.45 & 0.45 & <2 & Rein18 & -0.07 & 658 & 4.20 & 1.75 & 0.76 & 22.80 & DA19 \\
J12100-150 & LP 734-032 & M3.5 V & 64 & 3415 & 0.36 & 0.36 & <2 & Rein18 & -0.02 & 162 & 5.22 & 2.04 & 0.52 & 79.30 & Shan23 \\
J12123+544S & HD 238090 & M0.0 V & 109 & 3902 & 0.58 & 0.59 & <2 & Rein18 & 0.04 & 207 & 3.64 & 1.60 & 0.85 & 96.70 & Sto20b \\
J12230+640 & Ross 690 & M3.0 V & 149 & 3563 & 0.48 & 0.49 & <2 & Rein18 & 0.12 & 111 & 4.47 & 1.83 & 0.62 & 32.90 & DA19 \\
J12312+086 & BD+09 2636 & M0.5 V & 50 & 3896 & 0.55 & 0.56 & <2 & Rein18 & 0.01 & 363 & 3.86 & 1.66 & 0.86 &  &  \\
J12479+097 & Wolf 437 & M3.5 V & 85 & 3408 & 0.32 & 0.31 & <2 & Rein18 & -0.01 & 118 & 5.03 & 1.99 & 0.54 & 49.90 & Cab22 \\
J13102+477 & G 177-025 & M5.0 V & 36 & 3221 & 0.17 & 0.16 & <2 & Rein18 & -3.02 & 2293 & 5.90 & 2.23 & 0.41 & 29.06 & Shan23 \\
J13229+244 & Ross 1020 & M4.0 V & 104 & 3327 & 0.28 & 0.27 & <2 & Rein18 & 0.06 & 438 & 5.05 & 1.99 & 0.52 & 95.00 & Luq18 \\
J13255+688 & 2MASS J13253177+6850106 & M0 & 64 & 4049 & 0.58 & 0.59 &  &  &  & 363 & 3.61 & 1.59 & 0.84 &  &  \\
J13299+102 & BD+11 2576 & M0.5 V & 401 & 3765 & 0.48 & 0.48 & <2 & Rein18 & -0.02 & 159 & 4.32 & 1.79 & 0.79 & 30.00 & SM17 \\
J13450+176 & BD+18 2776 & M0.0 V & 29 & 3829 & 0.49 & 0.49 & 2.3 & Rein18 & -0.03 & 100 & 3.59 & 1.58 & 0.90 &  &  \\
J13457+148 & HD 119850 & M1.5 V & 248 & 3620 & 0.49 & 0.49 & <2 & Rein18 & 0.13 & 179 & -4.42 & -0.66 & 0.74 & 52.30 & SM15 \\
J14010-026 & HD 122303 & M1.0 V & 28 & 3719 & 0.50 & 0.51 & <2 & Rein18 & 0.01 & 229 & 4.03 & 1.71 & 0.76 & 43.90 & SM17 \\
J14082+805 & BD+81 465 & M1.0 V & 33 & 3803 & 0.59 & 0.60 & <2 & Rein18 & 0.05 & 402 & 3.96 & 1.69 & 0.75 &  &  \\
J14257+236E & BD+24 2733B & M0.5 V & 50 & 3933 & 0.59 & 0.60 & <2 & Rein18 & 0.03 & 268 & 3.83 & 1.65 & 0.80 & 17.60 & DA19 \\
J14257+236W & BD+24 2733A & M0.0 V & 64 & 3985 & 0.61 & 0.61 & <2 & Rein18 & 0.00 & 226 & 3.73 & 1.62 & 0.83 & 111.00 & DA19 \\
J14307-086 & BD-07 3856 & M0.5 V & 94 & 4037 & 0.63 & 0.64 & 2.4 & Rein18 & 0.17 & 211 & 3.97 & 1.69 & 0.87 &  &  \\
J14321+081 & LP 560-035 & M6.0 V & 58 & 3161 & 0.14 & 0.13 & 6.3 & Rein18 & -5.11 & 2842 & 6.52 & 2.41 & 0.35 & 1.12 & Shan23 \\
J14342-125 & HN Lib & M4.0 V & 94 & 3347 & 0.30 & 0.29 & <2 & Rein18 & 0.02 & 109 & 5.36 & 2.08 & 0.50 & 94.77 & Shan23 \\
J14524+123 & G 066-037 & M2.0 V & 26 & 3703 & 0.50 & 0.50 & <2 & Rein18 & -0.05 &     & 4.46 & 1.83 & 0.66 & 26.03 & Shan23 \\
J14544+355 & Ross 1041 & M3.5 V & 33 & 3433 & 0.35 & 0.35 & <2 & Rein18 & 0.00 & 254 & 4.85 & 1.94 & 0.54 &  &  \\
J15194-077 & HO Lib & M3.0 V & 54 & 3500 & 0.30 & 0.30 & <2 & Rein18 & 0.00 & 156 & 4.72 & 1.90 & 0.62 & 132.50 & SM15 \\
J15218+209 & OT Ser & M1.5 V & 52 & 3706 & 0.52 & 0.52 & 4.3 & Rein18 & -2.88 & 3357 & 4.27 & 1.78 & 0.70 & 3.35 & Shan23 \\
J15412+759 & UU UMi & M3.0 V+ & 33 & 3300 & 0.35 & 0.34 & 3 & Jeff18 & -0.11 &     & 4.63 & 1.88 & 0.62 & 83.40 & DA19 \\
J15474-108 & LP 743-031 & M2.0 V & 37 & 3481 & 0.51 & 0.52 & 3 & Jeff18 & -0.03 &     & 4.52 & 1.84 & 0.66 & 46.71 & Shan23 \\
J15583+354 & G 180-018 & M3.5 V & 77 & 3478 & 0.30 & 0.29 & 3 & Jeff18 & 0.15 & 153 & 4.81 & 1.93 & 0.58 &  &  \\
J16102-193 & K2-33 & M3.0 V & 27 & 3365 & 0.94 & 0.51 & 7.3 & Rein18 & -1.43 &     & 5.69 & 2.17 & 0.62 & 6.26 & DA19 \\
J16167+672N & EW Dra & M3.0 V & 105 & 3569 & 0.45 & 0.45 & <2 & Rein18 & 0.06 & 243 & 4.55 & 1.85 & 0.61 &  &  \\
J16167+672S & HD 147379 & M0.0 V & 186 & 4034 & 0.63 & 0.65 & 2.7 & Rein18 & 0.07 & 507 & 3.94 & 1.68 & 0.87 & 22.00 & Laf21 \\
J16254+543 & GJ 625 & M1.5 V & 33 & 3595 & 0.31 & 0.30 & <2 & Rein18 & 0.00 & 290 & 4.23 & 1.77 & 0.72 & 76.79 & DA19 \\
J16303-126 & V2306 Oph & M3.5 V & 93 & 3426 & 0.30 & 0.29 & <2 & Rein18 & 0.01 & 112 & 5.03 & 1.99 & 0.55 & 119.00 & DA19 \\
J16343+571 & CM Dra Aab & M4.5 V+ & 43 & 3200 & 0.34 & 0.33 & 10.22 & Mora09 & -5.00 &     & -7.80 & -1.60 & 0.50 &  &  \\
J16554-083N & GJ 643 & M3.5 V & 30 & 3397 & 0.22 & 0.21 & <2 & Rein18 & -0.01 & 257 & 5.04 & 1.99 & 0.55 & 6.52 & DA19 \\
J16555-083 & vB 8 & M7.0 V & 122 & 3005 & 0.09 & 0.07 & 5.4 & Rein18 & -4.54 & 3022 & -8.82 & -1.89 & 0.21 & 1.09 & Laf21 \\
J16581+257 & BD+25 3173 & M1.0 V & 55 & 3782 & 0.50 & 0.50 & <2 & Rein18 & -0.05 & 536 & 4.21 & 1.76 & 0.78 & 23.80 & DA19 \\
J17033+514 & G 203-042 & M4.5 V & 37 & 3273 & 0.20 & 0.18 & <2 & Rein18 & 0.01 & 258 & 5.71 & 2.18 & 0.46 &  &  \\
J17052-050 & Wolf 636 & M1.5 V & 50 & 3587 & 0.48 & 0.48 & <2 & Rein18 & 0.11 & 156 & 4.22 & 1.76 & 0.71 & 50.20 & DA19 \\
J17115+384 & Wolf 654 & M3.5 V & 53 & 3472 & 0.37 & 0.37 & <2 & Rein18 & 0.08 & 100 & 4.84 & 1.93 & 0.58 & 62.60 & DA19 \\
J17303+055 & BD+05 3409 & M0.0 V & 55 & 3821 & 0.52 & 0.52 & <2 & Rein18 & -0.02 & 268 & 4.01 & 1.70 & 0.83 & 34.60 & Shan23 \\
J17364+683 & BD+68 946 AB & M3.0 V+ & 41 & 3389 & 0.43 & 0.43 & 2.5 & Bro10 & 0.00 & 249 & 4.84 & 1.94 & 0.59 &  &  \\
J17378+185 & BD+18 3421 & M1.0 V & 102 & 3663 & 0.43 & 0.43 & <2 & Rein18 & -0.07 & 105 & 4.07 & 1.72 & 0.79 & 35.02 & BF19 \\
J17578+046 & Barnard's Star & M3.5 V & 664 & 3254 & 0.19 & 0.17 & <2 & Rein18 & 0.00 & 433 & 5.02 & 1.98 & 0.53 & 145.00 & TP19 \\
J18022+642 & LP 071-082 & M5.0 V & 26 & 3213 & 0.18 & 0.16 & 11.3 & Rein18 & -5.13 & 4931 & 5.72 & 2.18 & 0.39 & 0.28 & Shan23 \\
J18027+375 & GJ 1223 & M5.0 V & 118 & 3210 & 0.16 & 0.14 & <2 & Rein18 & 0.02 & 271 & 6.10 & 2.29 & 0.41 & 123.80 & New16 \\
J18051-030 & HD 165222 & M1.0 V & 55 & 3706 & 0.44 & 0.44 & <2 & Rein18 & 0.00 & 266 & 4.07 & 1.72 & 0.79 & 127.80 & SM15 \\
J18165+048 & G 140-051 & M5.0 V & 46 & 3240 & 0.19 & 0.18 & <2 & Rein18 & 0.05 & 132 & 5.80 & 2.20 & 0.45 &  &  \\
J18174+483 & TYC 3529-1437-1 & M2.0 V & 71 & 3692 & 0.52 & 0.52 & <2 & Rein18 & -1.58 & 1905 & 4.42 & 1.82 & 0.67 & 15.83 & Shan23 \\
J18198-019 & HD 168442 & K7 V & 145 & 4155 & 0.59 & 0.59 & 6.42 & LS10 & -0.60 & 274 & 3.64 & 1.60 & 0.94 & 31.69 & BF19 \\
J18224+620 & GJ 1227 & M4.0 V & 60 & 3294 & 0.18 & 0.16 & <2 & Rein18 & -0.07 & 496 & 5.60 & 2.15 & 0.46 &  &  \\
J18346+401 & LP 229-017 & M3.5 V & 78 & 3447 & 0.41 & 0.41 & <2 & Rein18 & -0.05 & 240 & 4.36 & 1.80 & 0.52 & 40.20 & DA19 \\
J18356+329 & LSR J1835+3259 & M8.5 V & 39 & 2500 & 0.09 & 0.07 & 49.2 & Rein18 & -1.65 &     & -9.17 & -1.99 & 0.35 & 0.12 & Shan23 \\
J18409-133 & BD-13 5069 & M1.0 V & 53 & 3799 & 0.52 & 0.53 & <2 & Rein18 & 0.05 & 307 & 4.08 & 1.72 & 0.76 & 28.23 & Shan23 \\
J18427+596N & HD 173739 & M3.0 V & 72 & 3473 & 0.34 & 0.34 & 2.5 & Bro10 & 0.00 & 118 & 4.62 & 1.87 & 0.62 &  &  \\
J18427+596S & HD 173740 & M3.5 V & 78 & 3393 & 0.27 & 0.26 & 2.5 & Bro10 & 0.00 & 113 & 4.88 & 1.95 & 0.58 &  &  \\
J18482+076 & G 141-036 & M5.0 V & 52 & 3175 & 0.16 & 0.14 & 2.4 & Rein18 & -3.37 & 1189 & 6.34 & 2.35 & 0.38 & 2.76 & DA19 \\
J18498-238 & V1216 Sgr & M3.5V & 53 & 3340 & 0.19 & 0.18 & 3 & Rein18 & -2.22 & 2083 & 5.06 & 2.00 & 0.53 & 2.87 & DA19 \\
J18580+059 & BD+05 3993 & M0.5 V & 33 & 3857 & 0.58 & 0.58 & <2 & Rein18 & 0.00 & 557 & 4.01 & 1.70 & 0.82 & 35.20 & DA19 \\
J19025+754 & LSPM J1902+7525 & M2.5 V & 44 & 3759 & 0.45 & 0.45 &  &  &  &     & 3.69 & 1.61 & 0.82 &  &  \\
J19070+208 & Ross 730 & M2.0 V & 37 & 3543 & 0.32 & 0.31 & <2 & Rein18 & -0.11 & 111 & -6.52 & -1.25 & 0.72 &  &  \\
J19072+208 & HD 349726 & M2.0 V & 42 & 3558 & 0.31 & 0.31 & <2 & Rein18 & -0.10 & 111 & -6.52 & -1.24 & 0.71 & 3.80 & DA19 \\
J19169+051N & V1428 Aql & M2.5 V & 123 & 3575 & 0.47 & 0.48 & <2 & Rein18 & 0.06 & 233 & 4.47 & 1.83 & 0.65 & 46.00 & DA19 \\
J19169+051S & V1298 Aql (vB 10) & M8.0 V & 50 & 2600 & 0.10 & 0.08 & 2.7 & Rein18 & -5.24 & 1207 & -8.77 & -1.87 & 0.30 & 23.60 & DA19 \\
J19242+755 & GJ 1238 & M5.5 V & 149 & 3044 & 0.14 & 0.13 &  &  & -0.28 &     & -8.96 & -1.93 & 0.35 &  &  \\
J19255+096 & LSPM J1925+0938 & M8.0 V & 101 & 2500 & 0.17 & 0.16 & 34.7 & Rein18 & -1.60 &     & -10.02 & -2.23 & 0.32 &  &  \\
J19346+045 & BD+04 4157 & M0.0 V & 52 & 4093 & 0.55 & 0.56 & 3.9 & Rein18 & 0.10 & 216 & 3.37 & 1.52 & 0.93 & 21.79 & BF19 \\
J19422-207 & 2M J19421282-2045477 & M5.1 V & 26 & 3241 & 0.22 & 0.21 & 6.2 & Rein18 & -5.80 & 3838 & 5.52 & 2.13 & 0.43 & 1.34 & Shan23 \\
J20260+585 & Wolf 1069 & M5.0 V & 268 & 3223 & 0.17 & 0.16 & <2 & Rein18 & -0.07 & 307 & 5.90 & 2.23 & 0.42 & 57.70 & DA19 \\
J20305+654 & GJ 793 & M2.5 V & 53 & 3514 & 0.38 & 0.37 & <2 & Rein18 & -0.24 & 217 & 4.58 & 1.86 & 0.64 & 32.80 & DA19 \\
J20336+617 & GJ 1254 & M4.0 V & 52 & 3406 & 0.38 & 0.38 & <2 & Rein18 & 0.06 & 261 & -7.40 & -1.49 & 0.52 & 12.60 & DA19 \\
J20450+444 & BD+44 3567 & M1.5 V & 70 & 3600 & 0.41 & 0.41 & <2 & Rein18 & 0.07 & 171 & 4.17 & 1.75 & 0.73 & 39.12 & BF19 \\
J20451-313 & AU Mic & M0.5 V & 100 & 3485 & 0.86 & 0.61 & 9.3 & Tor06 & -2.34 & 3007 & 4.36 & 1.80 & 0.75 & 4.89 & Shan23 \\
J20525-169 & LP 816-060 & M4.0 V & 40 & 3382 & 0.23 & 0.22 & <2 & Rein18 & 0.02 & 378 & 5.32 & 2.07 & 0.52 & 67.60 & DA19 \\
J20533+621 & HD 199305 & M1.0 V & 157 & 3815 & 0.52 & 0.53 & <2 & Rein18 & 0.00 & 345 & 4.14 & 1.74 & 0.80 &  &  \\
J20556-140N & GJ 810 A & M4.0 V+ & 27 & 3200 & 0.33 & 0.33 & 3 & Karmn & -0.02 &     & -7.37 & -1.48 & 0.51 & 137.37 & New18 \\
J20556-140S & GJ 810 B & M5.0 V & 45 & 3193 & 0.16 & 0.15 & 2 & Marf21 & 0.08 & 372 & 5.65 & 2.16 & 0.43 & 134.63 & New18 \\
J21019-063 & Wolf 906 & M2.5 V & 67 & 3568 & 0.45 & 0.45 & <2 & Rein18 & 0.04 & 238 & 4.58 & 1.86 & 0.64 & 49.29 & Shan23 \\
J21164+025 & LSPM J2116+0234 & M3.0 V & 83 & 3550 & 0.41 & 0.40 & <2 & Rein18 & 0.10 & 158 & 4.64 & 1.88 & 0.62 & 42.68 & BF19 \\
J21221+229 & TYC 2187-512-1 & M1.0 V & 94 & 3761 & 0.48 & 0.49 & <2 & Rein18 & -0.03 & 515 & 3.96 & 1.69 & 0.79 & 38.40 & Shan23 \\
J21348+515 & Wolf 926 & M3.0 V & 70 & 3525 & 0.44 & 0.44 & <2 & Rein18 & 0.06 & 226 & 4.45 & 1.83 & 0.61 & 54.30 & DA19 \\
J21463+382 & LSPM J2146+3813 & M4.0 V & 41 & 3247 & 0.21 & 0.20 & <2 & Rein18 & 0.10 & 309 & 5.01 & 1.98 & 0.53 &  &  \\
J21466+668 & G 264-012 & M4.0 V & 159 & 3409 & 0.29 & 0.29 & <2 & Rein18 & 0.08 & 215 & 5.09 & 2.01 & 0.53 & 100.00 & Ama21 \\
J21474+627 & TYC 4266-736-1 & M0 V & 58 & 4029 & 0.61 & 0.62 &  &  &  & 294 & -7.93 & -1.64 & 0.88 & 29.50 & DA19 \\
J22021+014 & BD+00 4810 & M0.5 V & 79 & 3882 & 0.56 & 0.57 & <2 & Rein18 & 0.02 & 403 & 4.10 & 1.73 & 0.83 & 120.47 & Laf21 \\
J22057+656 & G 264-018 A & M1.5 V & 91 & 3707 & 0.47 & 0.47 & <2 & Rein18 & 0.11 & 285 & 4.01 & 1.70 & 0.76 &  &  \\
J22096-046 & BD-05 5715 & M3.5 V & 60 & 3540 & 0.45 & 0.45 & <2 & Rein18 & 0.15 & 189 & 4.77 & 1.92 & 0.59 & 39.20 & SM15 \\
J22102+587 & TOI-2285 & M2 & 82 & 3616 & 0.48 & 0.48 &  &  &  &     & 4.36 & 1.80 & 0.70 &  &  \\
J22114+409 & 1RXS J221124.3+410000 & M5.5 V & 54 & 3132 & 0.18 & 0.16 & <2 & Rein18 & -5.74 & 1880 & 6.35 & 2.36 & 0.36 & 30.00 & DA19 \\
J22115+184 & Ross 271 & M2.0 V & 67 & 3704 & 0.50 & 0.51 & <2 & Rein18 & 0.03 & 257 & 4.40 & 1.81 & 0.67 & 36.30 & DA19 \\
J22125+085 & Wolf 1014 & M3.0 V & 111 & 3528 & 0.36 & 0.36 & <2 & Rein18 & 0.04 & 123 & 4.48 & 1.84 & 0.64 &  &  \\
J22137-176 & LP 819-052 & M4.5 V & 87 & 3256 & 0.19 & 0.18 & <2 & Rein18 & 0.06 & 318 & 5.56 & 2.14 & 0.46 & 116.40 & Ste16 \\
J22252+594 & G 232-070 & M4.0 V & 101 & 3448 & 0.40 & 0.40 & <2 & Rein18 & 0.08 & 210 & 5.06 & 2.00 & 0.53 & 64.60 & DA19 \\
J22330+093 & BD+08 4887 & M1.0 V & 67 & 3722 & 0.45 & 0.45 & <2 & Rein18 & -0.03 & 319 & 4.01 & 1.70 & 0.79 & 37.80 & Shan23 \\
J22468+443 & EV Lac & M3.5 V & 107 & 3306 & 0.34 & 0.34 & 3.5 & Rein18 & -4.98 & 4324 & 4.96 & 1.97 & 0.51 & 4.35 & Shan23 \\
J22503-070 & BD-07 5871 & M0.5 V & 53 & 3900 & 0.54 & 0.55 & <2 & Rein18 & -0.03 & 441 & 3.82 & 1.65 & 0.87 &  &  \\
J22532-142 & IL Aqr & M4.0 V & 70 & 3421 & 0.32 & 0.32 & <2 & Rein18 & 0.06 & 197 & 5.19 & 2.03 & 0.53 & 81.00 & DA19 \\
J22565+165 & HD 216899 & M1.5 V & 569 & 3798 & 0.52 & 0.53 & <2 & Rein18 & 0.07 & 332 & 4.62 & 1.87 & 0.74 & 39.50 & DA19 \\
J23113+085 & NLTT 56083 & M3.5 V & 93 & 3477 & 0.43 & 0.43 & <2 & Rein18 & 0.03 & 223 & 5.04 & 1.99 & 0.55 & 44.82 & Shan23 \\
J23216+172 & LP 462-027 & M4.0 V & 66 & 3398 & 0.38 & 0.37 & <2 & Rein18 & 0.02 & 249 & 5.21 & 2.04 & 0.51 & 74.70 & DA19 \\
J23245+578 & BD+57 2735 & M1.0 V & 60 & 3811 & 0.54 & 0.54 & <2 & Rein18 & 0.02 & 433 & 4.24 & 1.77 & 0.76 & 36.48 & BF19 \\
J23340+001 & GJ 899 & M2.5 V & 33 & 3530 & 0.43 & 0.43 & <2 & Rein18 & 0.08 & 189 & 4.21 & 1.76 & 0.66 &  &  \\
J23351-023 & GJ 1286 & M5.5 V & 71 & 3136 & 0.13 & 0.11 & <2 & Rein18 & -0.82 & 907 & -8.18 & -1.71 & 0.35 & 88.92 & New18 \\
J23381-162 & G 273-093 & M2.0 V & 56 & 3570 & 0.36 & 0.36 & <2 & Rein18 & 0.01 & 159 & 4.29 & 1.78 & 0.70 & 61.66 & Wat06 \\
J23419+441 & HH And & M5.0 V & 99 & 3186 & 0.16 & 0.14 & <2 & Rein18 & -0.38 & 413 & -5.93 & -1.08 & 0.40 & 106.00 & DA19 \\
J23492+024 & BR Psc & M1.0 V & 378 & 3573 & 0.42 & 0.42 & <2 & Rein18 & -0.02 & 147 & 4.37 & 1.80 & 0.76 & 49.90 & SM18 \\
J23505-095 & LP 763-012 & M4.0 V & 71 & 3377 & 0.30 & 0.29 & <2 & Rein18 & 0.01 & 252 & -8.04 & -1.67 & 0.50 &  &  \\
J23556-061 & GJ 912 & M2.5 V & 30 & 3639 & 0.54 & 0.55 & <2 & Rein18 & 0.07 &     & 4.46 & 1.83 & 0.66 &  &  \\
J23585+076 & Wolf 1051 & M3.0 V & 28 & 3496 & 0.47 & 0.47 & <2 & Rein18 & 0.04 &     & 4.66 & 1.88 & 0.60 & 55.40 & Shan23 \\

       \noalign{\smallskip}
       \footnote{References:\\
         Shan23  --        Sahn et al. (2023 submitted) ;
         DA19    --        \citet{DiezAlonso2019A&A...621A.126D};
         SM15    --        \citet{SuarezM2015MNRAS.452.2745S};
         SM17b   --        \citet{SuarezM2017MNRAS.468.4772S};
         SM18    --        \citet{SuarezM2018A&A...612A..89S};
         SM23    --        \citet{SuarezM2023A&A...670A...5S};
         New18   --        \citet{Newton2018AJ....156..217N};
         New16a  --        \citet{2016ApJ...821...93N};
         Laf21    --        \citet{Lafarga2021A&A...652A..28L};
         Luq22    --       \citet{Luque2022A&A...664A.199L};
         Mori08  --        \citet{Morin2008MNRAS.390..567M};
         Mor10   --         \citet{Morin2010MNRAS.407.2269M};
         Cab22   --         \citet{Caballero2022A&A...665A.120C};
         Jef18   --    \citet{Jeffers2018A&A...614A..76J};
         Rei18   --     \citet{Reiners2018A&A...612A..49R};
         BF  -- \citet{Fuhrmeister2019A&A...623A..24F}
       }
     \end{longtable}
   \end{landscape}
\normalsize
}


\end{appendix}

\end{document}